\documentclass[structabstract]{aa} %, draft onecolumn referee for the abstract without structuration 
\usepackage{graphicx}
\usepackage{lscape}
\usepackage{natbib}
\usepackage{xcolor}
\usepackage[colorlinks, citecolor=blue]{hyperref} 

\usepackage{etoolbox}
\makeatletter
\patchcmd\@combinedblfloats{\box\@outputbox}{\unvbox\@outputbox}{}{%
   \errmessage{\noexpand\@combinedblfloats could not be patched}%
}%
 \makeatother
\usepackage{txfonts}
\usepackage{longtable}
\begin{document}
\newcommand{\kms}{km~s$^{-1}$}
\newcommand{\Msun}{M$_{\odot}$}
\newcommand{\Teff}{$T_{\rm eff}$}
\newcommand{\FeH}{[Fe/H]}
\newcommand{\doe}{{\scriptsize DOE}}
\newcommand{\ha}{H$\alpha$}

\title{The \emph{Gaia}-ESO Survey\thanks{Tables 3 and 4 are only available in electronic form at the CDS via anonymous ftp to cdsarc.u-strasbg.fr (130.79.128.5) or via \url{http://cdsweb.u-strasbg.fr/cgi-bin/qcat?J/A+A/}.
Based on data products from observations made with ESO Telescopes at the La Silla Paranal Observatory under programme IDs 188.B-3002 and 193.B-0936. These data products have been processed by the Cambridge Astronomy Survey Unit (CASU) at the Institute of Astronomy, University of Cambridge, and by the FLAMES/UVES reduction team at INAF/Osservatorio Astrofisico di Arcetri. These data have been obtained from the \emph{Gaia}-ESO Survey Data Archive, prepared and hosted by the Wide Field Astronomy Unit, Institute for Astronomy, University of Edinburgh, which is funded by the UK Science and Technology Facilities Council.}: detection and characterisation of single-line spectroscopic binaries}

\author{T. Merle\inst{1}
          \and M. Van der Swaelmen\inst{1}
          \and S. Van Eck\inst{1}          
          \and A. Jorissen\inst{1}
          \and R.~J. Jackson\inst{2}
          \and G. Traven\inst{3}
          \and T. Zwitter\inst{4}
          \and D. Pourbaix\inst{1}
          \and A. Klutsch\inst{5}
          \and G. Sacco\inst{6}
          \and R. Blomme\inst{7}
          \and T. Masseron\inst{8}
          \and G. Gilmore\inst{9}
          \and S. Randich\inst{6}
          \and C. Badenes\inst{10}
          \and A. Bayo\inst{11}
          \and T. Bensby\inst{3}
          \and M. Bergemann\inst{12}
          \and K. Biazzo\inst{13}          
          \and F. Damiani\inst{14}
          \and D. Feuillet\inst{12}    
          \and A. Frasca\inst{13}          
          \and A. Gonneau\inst{9}
          \and R.~D. Jeffries\inst{2}
          \and P. Jofr\'e\inst{15}
          \and L. Morbidelli\inst{6}
          \and N. Mowlavi\inst{16}         
          \and E. Pancino\inst{6}
          \and L. Prisinzano\inst{14}
          }

\institute{$^1$ Institut d'Astronomie et d'Astrophysique, Universit\'e Libre de Bruxelles, CP. 226, Boulevard du Triomphe, 1050 Brussels, Belgium \\
\email{tmerle@ulb.ac.be} \\        
$^2$ Astrophysics Group, Keele University, Keele, Staffordshire ST5 5BG, UK\\
$^3$ Lund Observatory, Department of Astronomy and Theoretical Physics, Box 43, SE-221 00 Lund, Sweden\\
$^4$  Faculty of Mathematics and Physics, University of Ljubljana, Jadranska 19, 1000, Ljubljana, Slovenia\\
$^5$ Institut f\"ur Astronomie und Astrophysik, Eberhard Karls Universit\"at, Sand 1, 72076 T\"ubingen, Germany\\
$^6$  INAF – Osservatorio Astrofisico di Arcetri, Largo E. Fermi 5, 50125 Firenze, Italy\\
$^7$  Royal Observatory of Belgium, Ringlaan 3, 1180, Brussels, Belgium\\
$^8$  Instituto de Astrof\'{\i}sica de Canarias, E-38205 La Laguna, Tenerife, Spain\\
$^9$  Institute of Astronomy, University of Cambridge,  Madingley Road, Cambridge CB3 0HA, UK\\
$^{10}$ PITT PACC, University of Pittsburgh, 3941 O’Hara Street, Pittsburgh, PA 15260, USA\\
$^{11}$  Instituto de F\'isica y Astronomi\'ia, Universidad de Valpara\'iso, Chile\\
$^{12}$  Max-Planck Institut f\"{u}r Astronomie, K\"{o}nigstuhl 17, 69117 Heidelberg, Germany\\
$^{13}$ INAF – Osservatorio Astrofisico di Catania, via S. Sofia 78, 95123, Catania, Italy\\
$^{14}$ INAF – Osservatorio Astronomico di Palermo, Piazza del Parlamento 1, 90134, Palermo, Italy\\
$^{15}$  N\'ucleo de Astronom\'{i}a, Facultad de Ingenier\'{i}a, Universidad Diego Portales, Av. Ej\'ercito 441, Santiago, Chile\\
$^{16}$ Department of Astronomy, University of Geneva, Ch. des Maillettes 51, 1290 Versoix, Switzerland\\
}

\date{Received ...; accepted ...}

\abstract 
%Context
{Multiple stellar systems  play a fundamental role in the formation and evolution of stellar populations in galaxies. Recent and ongoing large ground-based multi-object spectroscopic surveys significantly increase the sample of spectroscopic binaries (SBs) allowing analyses of their statistical properties.
}
%Aims
{We investigate the repeated spectral observations of the \emph{Gaia}-ESO Survey internal data release 5 (GES iDR5) to identify and characterise SBs with one visible component (SB1s) in fields covering mainly the discs, the bulge, the CoRot fields, and some stellar clusters and associations.}
%Methods
{A statistical $\chi^2$-test is performed on spectra of the iDR5 subsample of approximately 43\,500 stars characterised by at least two observations and a signal-to-noise ratio larger than three.  In the GES iDR5, most stars have four~observations generally split into two~epochs. A careful estimation of the radial velocity (RV) uncertainties is performed. Our sample of RV variables is cleaned from contamination by pulsation- and/or convection-induced variables using \emph{Gaia} DR2 parallaxes and photometry. 
Monte-Carlo simulations using the SB9 catalogue of spectroscopic orbits allow to estimate our detection efficiency and to correct the SB1 rate to evaluate the GES SB1 binary fraction and its relation to effective temperature and metallicity.}
%Results
{We find 641 (resp., 803) FGK SB1 candidates at the $5\sigma$ (resp., $3\sigma$) level. The maximum RV differences range from 2.2~\kms\ at the $5\sigma$  confidence level (1.6~\kms\ at $3\sigma$) to 133~\kms\ (in both cases). Among them a quarter of the primaries are giant stars and can be located as far as 10~kpc. The orbital-period distribution is estimated from the RV standard-deviation distribution and reveals that the detected SB1s probe binaries with $\log{P[\text{d}]}\lessapprox 4$. We show that SB1s with dwarf primaries tend to have shorter orbital periods than SB1s with giant primaries. This is consistent with binary interactions removing shorter period systems as the primary ascends the red giant branch. For two systems, tentative orbital solutions with periods of 4 and 6~d are provided. After correcting for detection efficiency, selection biases, and the present-day mass function, we estimate the global GES SB1 fraction to be in the range 7-14\% with a typical uncertainty of 4\%. A small increase of the SB1 frequency is observed from K- towards F-type stars, in agreement with previous studies. The GES SB1 frequency decreases with metallicity at a rate of $(-9\pm3)\%$~dex$^{-1}$ in the metallicity range $-2.7\le$\FeH$\le+0.6$. This anticorrelation is obtained with a confidence level higher than 93\% on a homogeneous sample covering spectral types FGK and a large range of metallicities. When the present-day mass function is accounted for, this rate turns to $(-4\pm2)\%$~dex$^{-1}$ with a confidence level higher than 88\%. In addition we provide the variation of the SB1 fraction with metallicity separately for F, G, and K spectral types, as well as for dwarf and giant primaries.}
%Conclusions
{}


\titlerunning{GES: SB1 catalogue}
\maketitle

\section{Introduction}
Binary stars are now recognised as playing a fundamental role in the evolution of stellar populations in galaxies (\emph{e.g.} \mbox{\citealt{hurley2002}}, \mbox{\citealt{widmark2018}}, \mbox{\citealt{dorn2018}}).
Statistical properties  linked to stellar multiplicity provide fundamental information for both stellar formation and evolution theories \citep{duquennoy1991, raghavan2010, duchene2013, moe2017, breivik2019}. The impact of companions on stellar evolution \mbox{\citep[\emph{e.g.}][]{demarco2017}} is indeed very diverse. In particular, binarity is responsible for stars  with photometric \citep[\emph{e.g.} heartbeat stars:][]{thompson2012,pigulski2018} and/or chemical peculiarities: barium stars \citep[\emph{e.g.}][]{mcclure84a,mcclure90, merle2016,Jorissen2019}, CH and many other carbon-enriched metal-poor stars \citep[\emph{e.g.}][]{mcclure84b, mcclure90, jorissen16}, extrinsic S stars \mbox{\citep[\emph{e.g.}][]{vaneck1998, shetye2018}}, blue stragglers \citep[\emph{e.g.}][]{bailyn1995, mathieu2009, jofre2016}, and so on. Binary stars are also the best benchmarks to constrain stellar evolution models because radii and masses of binary-star components can be measured with a precision of a few percent \citep[\emph{e.g.}][]{torres2010, eker2018}. Among binary systems, spectroscopic binaries (SBs) probe short orbital periods, that is, from tens of hours to tens of years, according to simulations \citep[\emph{e.g.}][]{soderhjelm2004}. The main advantage of SBs compared to astrometric or visual binaries is that their detectability is independent of distance until the limiting magnitude is reached. They can be detected over large volumes in our Galaxy or even in other galaxies. Spectroscopic binaries are therefore prime targets to study binarity in large surveys with multi-epoch spectroscopy. 

Recent studies on SBs were conducted in large multi-object spectroscopic surveys such as LAMOST \citep[with a resolution of $R\sim1800$;][]{luo15} and SDSS \citep[with a resolution of $R\sim2000$;][]{gunn06}. From these surveys, \citet{gao2014} estimated the frequency of SBs with periods shorter than 10$^3$~d using Bayesian statistics, and found $30\pm8$\% in the LEGUE subsample of LAMOST (that operated from October 2011 to June 2013) reaching $43\pm2$\% in the SEGUE subsample of SDSS \citep[that operated from 2000 to 2008;][]{yanny09} for solar-type stars. Moreover, \citet{gao2017} found that the binary fraction increases monotonically from 20\% (among stars with $T_{\rm eff} \sim 4000$~K) to 50\% (at 7500~K). These latter authors also found a statistically significant anti-correlation of the binary fraction with metallicity.
 
 Within the APOGEE survey  \citep[that operated from 2011 to 2014 at $R\sim22\,000$;][]{majewski16},  the analysis of early observations for 14\,000 stars  provided a catalogue of new SBs, including binaries with substellar companions \citep{troup2016}, while new SB2s were identified in open clusters and star-forming regions \citep{fernandez2017}. In addition, analysing a sample of 90\,000 APOGEE red-giant stars with sparse radial-velocity (RV) measurements, \citet{badenes2018} found a correlation between the maximum RV span and their surface gravity: the most evolved stars (on the red giant branch, RGB) have the smallest RV spans, as expected, since their large radii impose large orbital separations, and therefore small velocity amplitudes. These latter authors also measured an excess by a factor of two for the frequency of SBs with [Fe/H]~$\sim-0.5$ as compared to SBs with solar metallicity on the main sequence. This excess reaches a factor of three at the tip of the RGB. 
 
 Finally, new developments involving machine-learning techniques should be mentioned:  these allow SBs with a virtually null RV shift to be identified by fitting synthetic composite spectra \citep{gullikson2016,el-badry2018}. Such objects consist in binaries seen almost pole-on and with long periods, typically larger than $10^4$~d (\emph{i.e.} 27.4~yr). A total of 20\,000 APOGEE main sequence stars were analysed and 2500 SBs with no detectable RV changes (called `unresolved' binaries in the spectroscopic sense) were discovered by fitting synthetic composite spectra. This technique allowed to reach a detection efficiency close to 100\% for mass ratios $q$ ranging from 0.4 to 0.8. This method, although not easy to implement and rather time-consuming, has great potential. 

The spectroscopic ground-based \emph{Gaia}-ESO survey \citep[GES][]{gilmore2012, randich2013} is almost completed. With its sample of $10^5$ stars probing all stellar populations of the Galaxy down to magnitude $V=19$, this survey unravels, with exquisite detail, the kinematics and dynamical structures in the Galaxy  \citep{jeffreson2017}, and the chemical compositions \citep[\emph{e.g.}][]{smiljanic2014} and histories  of stellar clusters, associations, and field stars \citep[\emph{e.g.}][]{bergemann2014}. The observing strategy of this survey was neither designed to discover SBs nor to study them. Nevertheless, \citet{merle2017, merle2018a} investigated SBs with two or more components (\emph{i.e.} SB$n$, with $n \ge 2$) in the GES internal data release 4 (iDR4) and identified 342 SB2, 11 SB3, and one SB4 candidates, among which only two were previously known. These latter authors also confirmed two of them by providing orbital solutions. Investigation of the GES iDR5 with new cross-correlation functions \citep[CCFs,][]{vdswaelmen2017,vdswaelmen2018a,vdswaelmen2018b} provides an additional 30\% of new SB2 and SB3 candidates (Van der Swaelmen et al., in prep.). 

\begin{figure}
 \includegraphics[width=\linewidth]{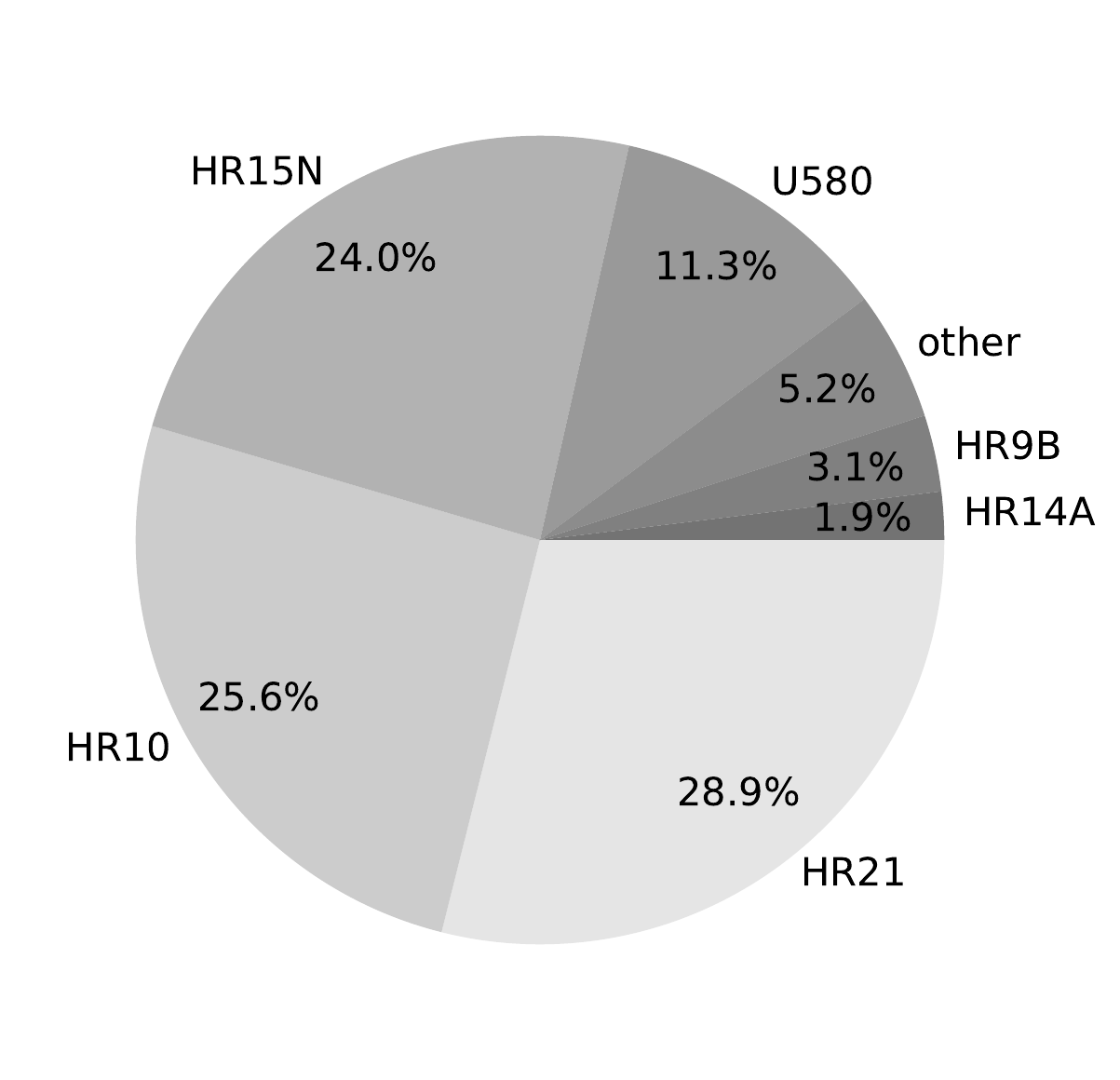}

 \includegraphics[width=\linewidth]{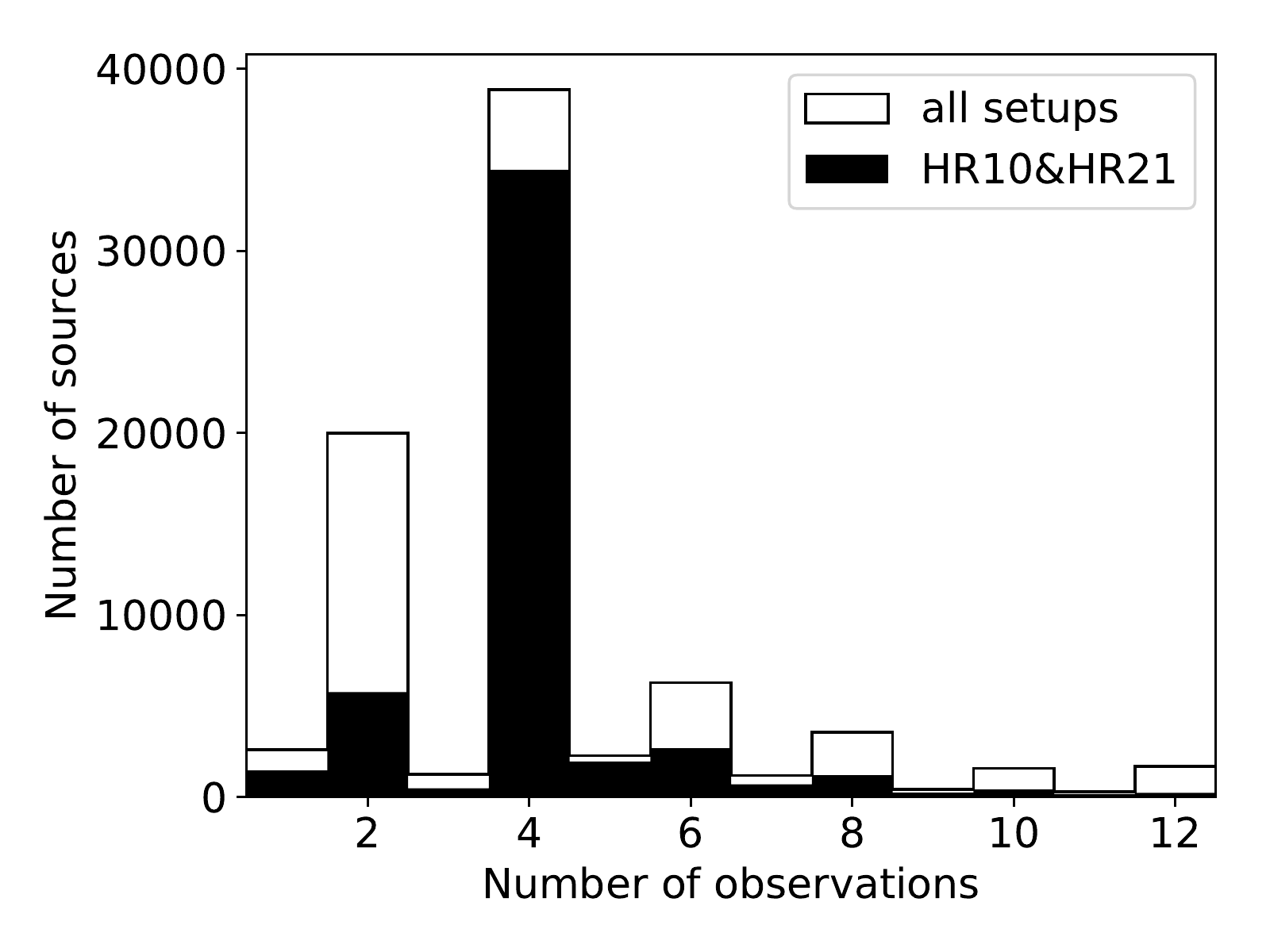}
 \caption{Top panel: Pie chart of single GES iDR5 exposures per instrumental setup. The part labelled `other' contains GIRAFFE setups HR3, HR5A, HR6, and HR15, as well as the U520 UVES setup. Bottom panel: Distribution of sources per number of observations (single exposures) within GES iDR5. Even numbers are favoured because sources are generally observed in both HR10 and HR21. In the case of four observations, we generally have two consecutive observations in HR10 and in HR21, both close in time.}
 \label{fig:pie_histo_nobs}
\end{figure}

The present study addresses the detection and characterisation of SB1s in the GES iDR5, the preliminary results of which were presented in \mbox{\citet{merle2018b}}. Many spectra are required in order to obtain time series for each SB1 showing RV variations. The GES observing strategy was designed so as to achieve an accurate determination of atmospheric parameters and elemental abundances. Repeated observations of the same targets were not planned, in order to maximise binary detection. We restrict our analysis to spectra acquired with the medium-resolution spectrograph FLAMES/GIRAFFE ($R\sim20\,000$) and setups HR10 and HR21, covering the wavelength ranges [$5340 - 5610$] and [$8490-9000$]~\AA, respectively. Our sample comprises more than 50\% of the iDR5 data, corresponding to almost 50\,000 stars. %#49557

\begin{figure*}[h]
 \includegraphics[width=0.49\linewidth]{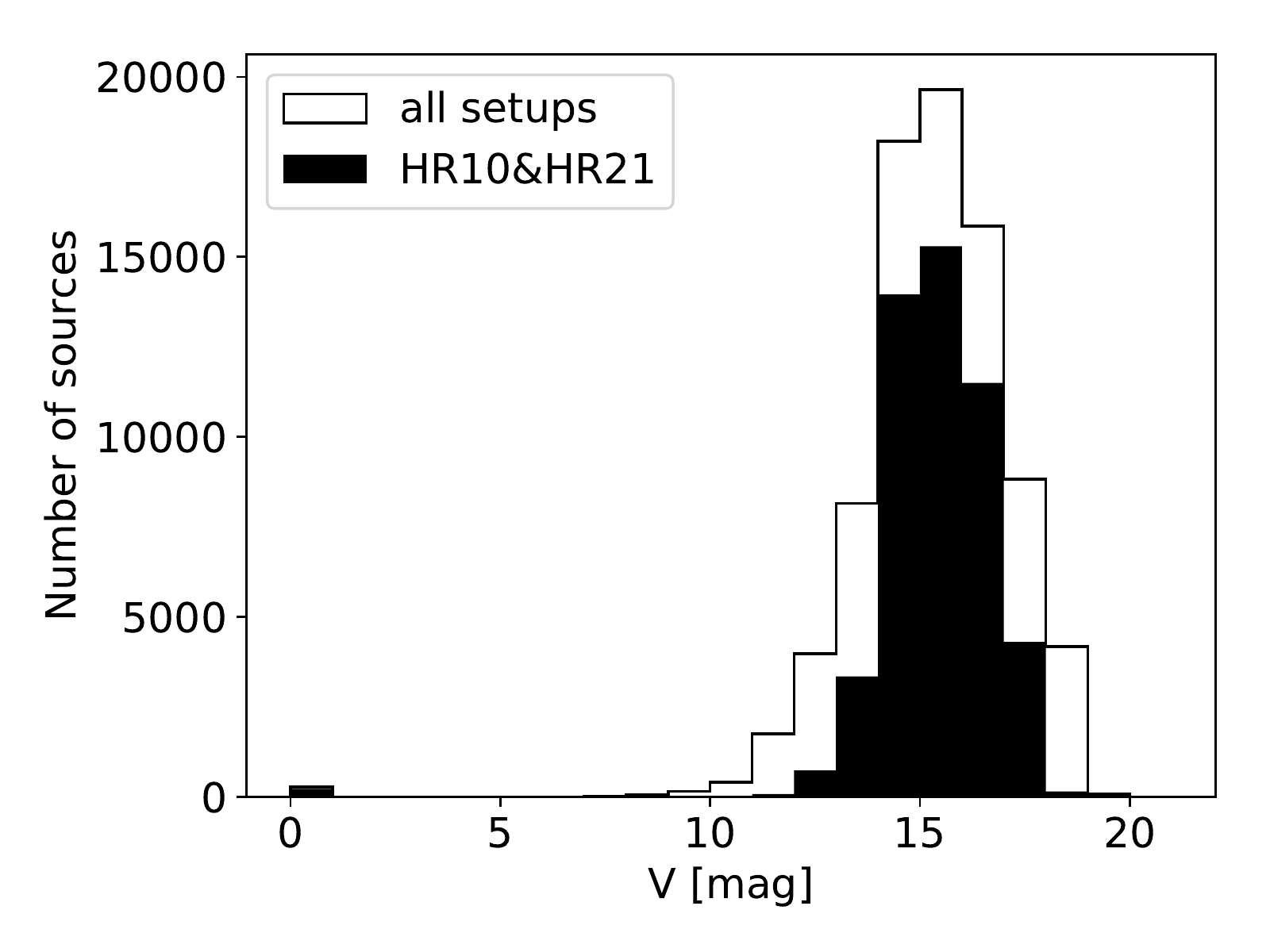}
 \includegraphics[width=0.49\linewidth]{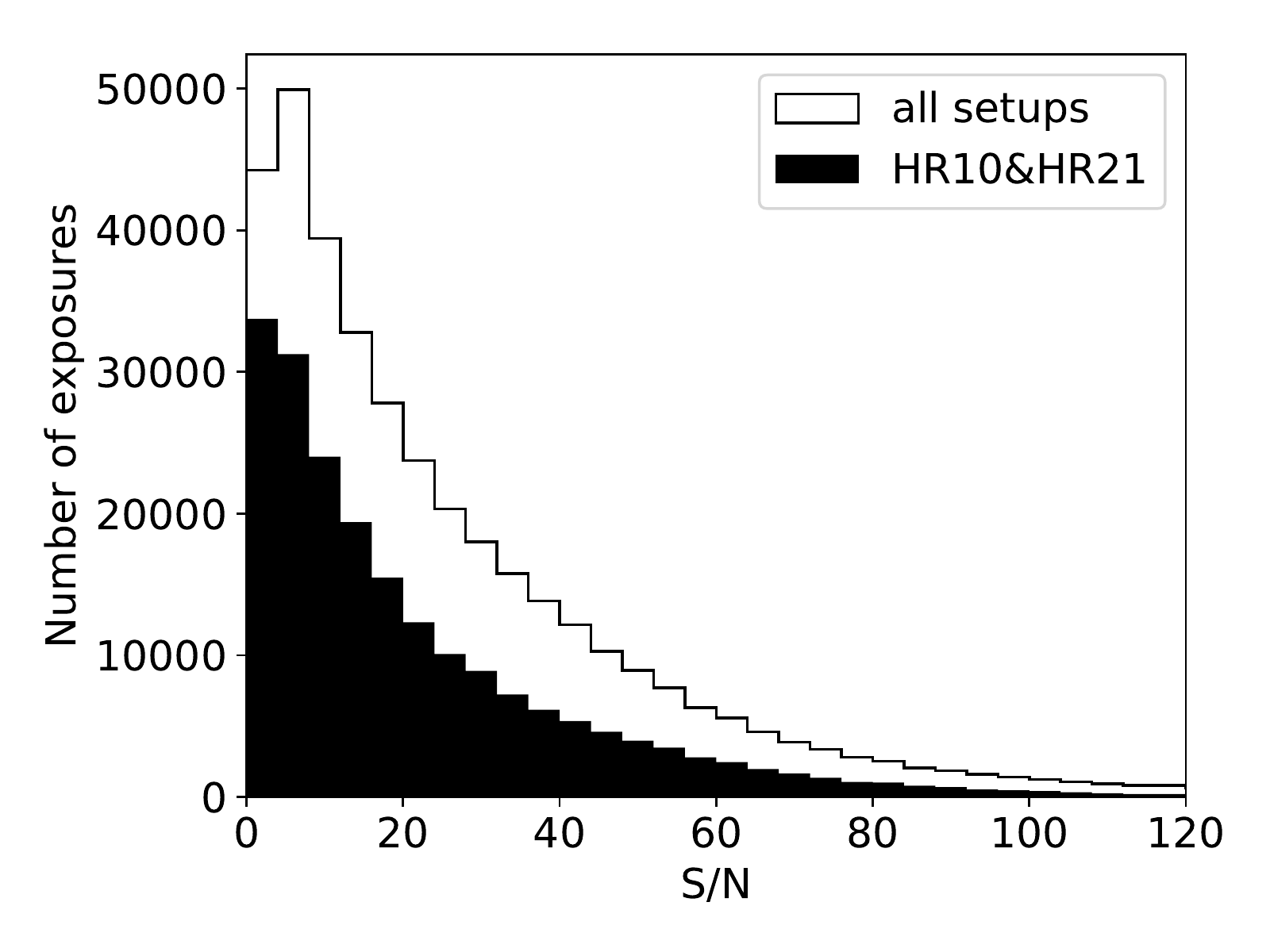}
 \caption{Left: Histogram of the visual apparent magnitude of GES iDR5 sources. Right: Histogram of the $S\!/N$ of GES iDR5 single exposures.}
 \label{fig:histo_mag_snr}
 \end{figure*}

\begin{figure*}[h]
 \includegraphics[width=0.33\linewidth]{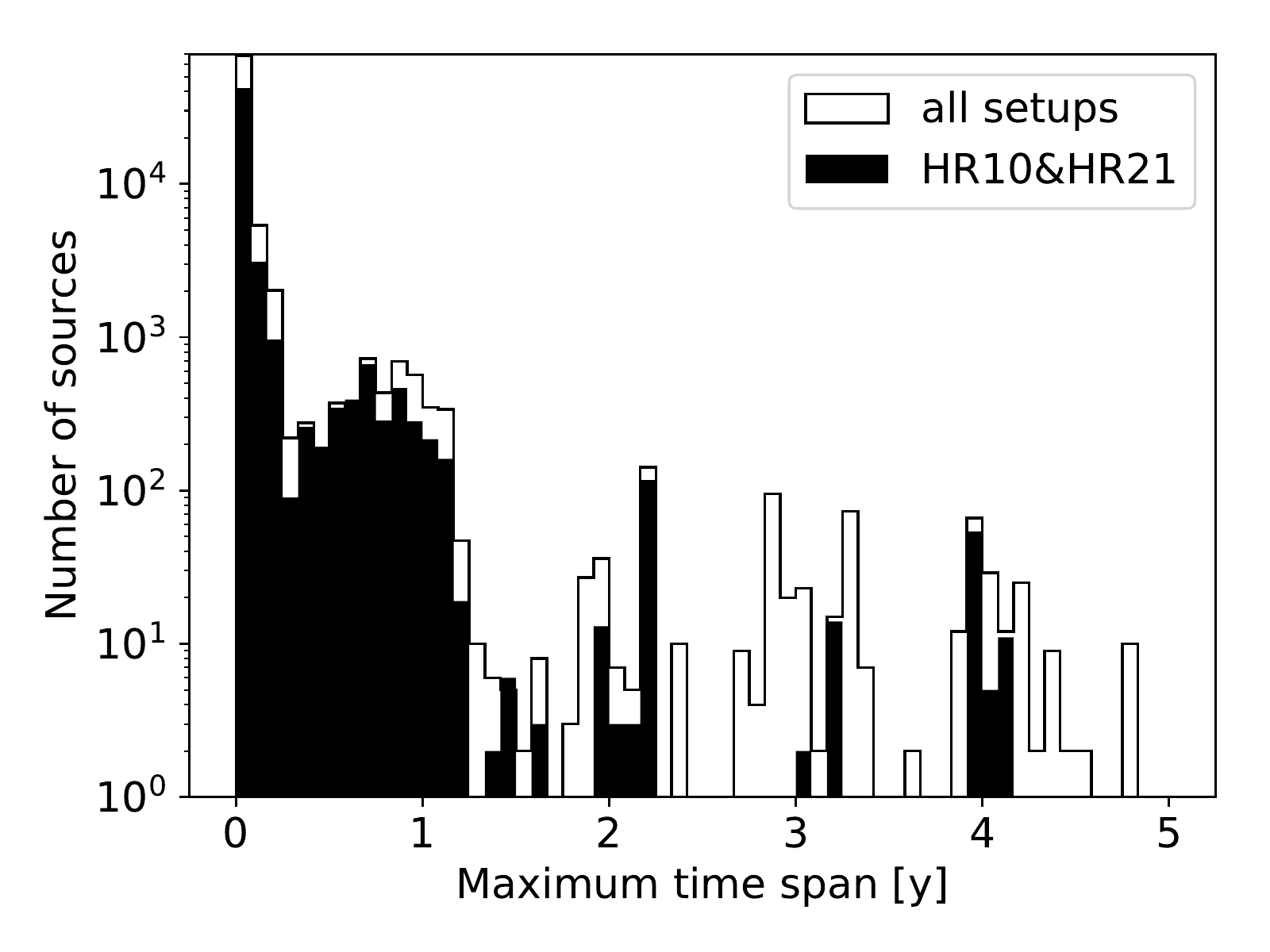}
 \includegraphics[width=0.33\linewidth]{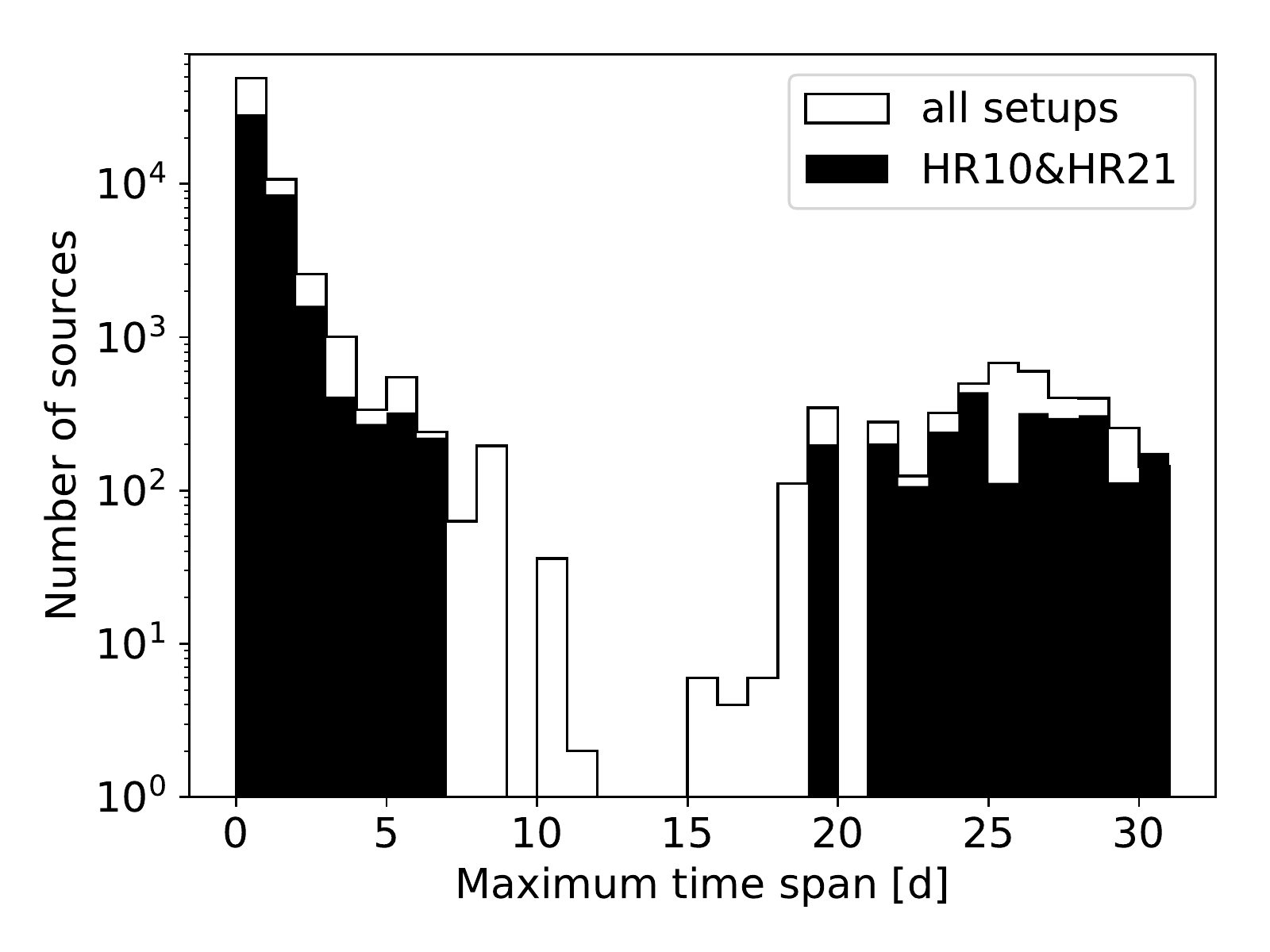}
 \includegraphics[width=0.33\linewidth]{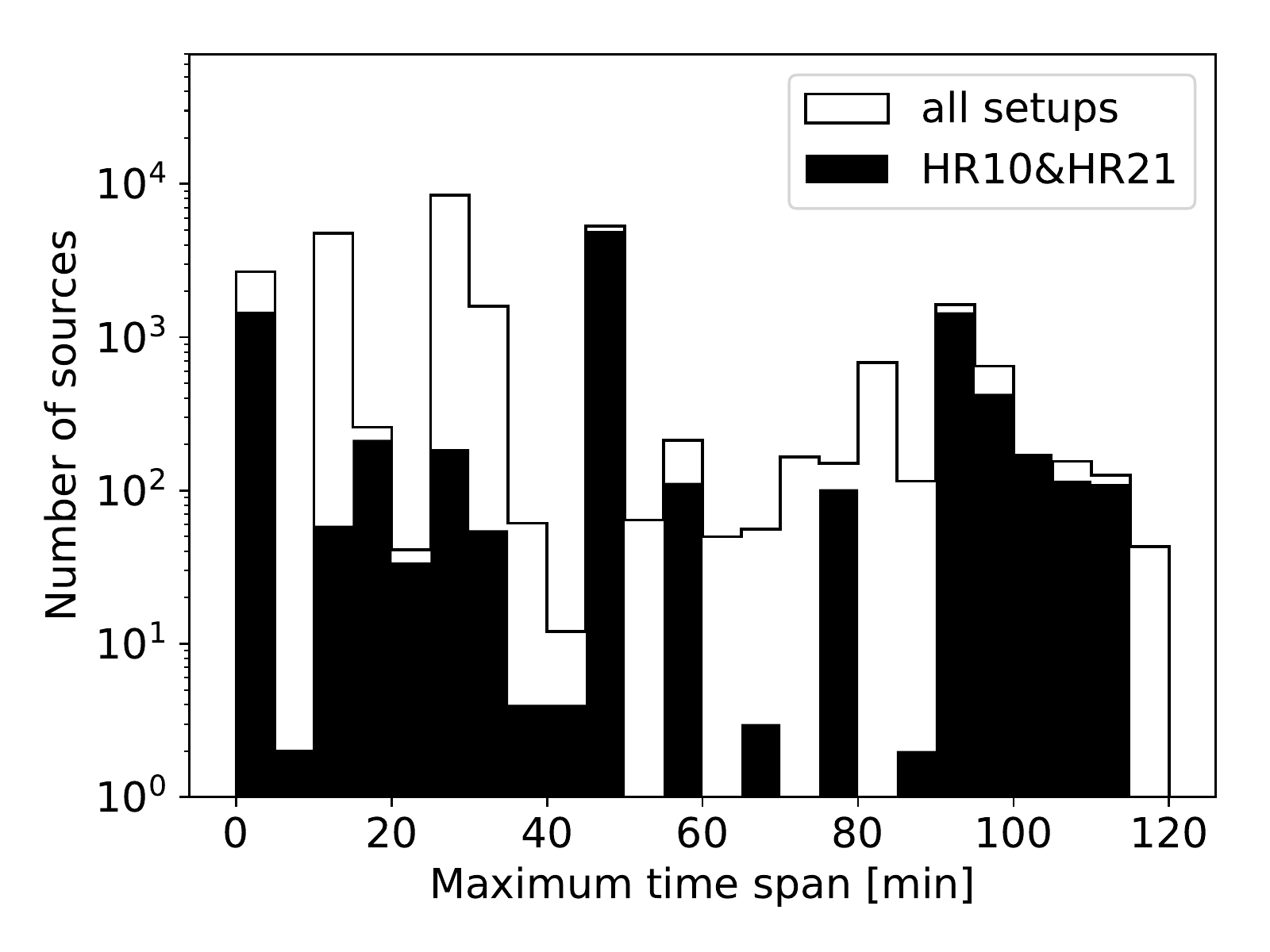}
 \caption{Distribution of the maximum time span per source. Left panel: - Over the five years of the survey; middle panel: - Over one month; right panel: - Over two hours.}
 \label{fig:dtmax}
\end{figure*}

\section{Data}
\subsection{Spectral sample selection}
The number of science targets within GES iDR5 amounts to more than 82\,000 targets corresponding to almost 380\,000 single exposures. These exposures were taken with the FLAMES (UVES and GIRAFFE) optical spectrographs \citep{pasquini2002}. The distribution of single exposures among the spectral setups is displayed in the top panel of Fig.~\ref{fig:pie_histo_nobs}. More than 75\% of the observations were obtained with GIRAFFE setups HR21, HR10, and HR15N. The GIRAFFE HR10 and HR21 setups were mainly dedicated to disc, halo, and bulge stars, whereas HR15N and the remaining setups were used for stars in stellar clusters, associations, and specific targets such as massive stars. The distribution of the target visual magnitudes is displayed in the left panel of Fig.~\ref{fig:histo_mag_snr}, with a median magnitude of $V=15.4$ (which holds true for the subsample of targets with at least one observation in HR10 and/or HR21 considered in the present study). The signal-to-noise ratio ($S\!/N$) of GES iDR5 data is shown in the right panel of Fig.~\ref{fig:histo_mag_snr}. The median $S\!/N$ is 19 but decreases to 15 when only relying on the HR10 and HR21 setups. This is due to the fact that the UVES setups have a higher median of about 29, but are restricted to a much smaller set of observations than HR10 and HR21. 

The number of exposures per target is crucial to discovering SB1s because the method relies on the analysis of RVs taken at least at two different epochs. As the GES was not thought of as a monitoring survey, the number of visits per target is minimal. The bottom panel of Fig.~\ref{fig:pie_histo_nobs} shows the number of targets per number of exposures. Even numbers of observations per target are the most frequent because field stars, the most numerous in the iDR5 sample, were observed with HR10 and HR21; these observations were generally consecutive, possibly with one or more repeats over the following nights. The most frequent (more than 70\%) numbers of exposures per target are two and four (white histogram on Fig.~\ref{fig:pie_histo_nobs}). The number of sources for which HR10 or HR21 observations are available amounts to about $50\,000$ out of the 82\,000 individual targets in iDR5.

The time coverage (baseline) of the observations per target is another crucial point to assess the kind of SB1s detectable within the GES. The distribution of the maximum time span per target is represented in Fig.~\ref{fig:dtmax}. Two types of repeats are present within the GES. Short-term repeats involve targets observed within an observing block. The time delay between these exposures was shorter than one hour, and the same fibre with the same configuration and wavelength calibration were used. On the contrary, long-term repeats involve targets observed within different observing blocks, and in such cases, different fibres and configurations were used, thus increasing the uncertainty on the RV determination. The maximum time span for 80\% of the iDR5 sources was shorter than or equal to one week.  For one-third of the targets, the maximum time span was shorter than two hours. At the other end of the exposure time span distribution, for less than 3\% of the sources the maximum time span was larger than one year, and this was mainly due to the inclusion of ESO archive spectra for benchmark, cluster and association stars. These numbers clearly reveal that the search for SB1s within the GES was biased toward short-period binaries, from a few hours to a few weeks, with possible exceptional cases of a few years.

With these facts in mind, we decided to focus our investigation of SB1 on the two main GIRAFFE setups HR21 ($\sim$29\% of the iDR5 single exposures) and HR10 (26\%), that is, involving mainly field (halo, bulge, disc) stars. This restriction is justified by the number of exposures per target (four in general) covering at least two, sometimes very close, epochs. Another argument in favour of restricting the study to the HR10 and HR21 setups is given by the control of the inter-setup bias correction and of the cleanliness of the RV measurements. Neither HR10 nor HR21 include the strong H$\alpha$ line, which is sometimes in emission or exhibits asymmetrical profiles \citep{traven2015} that make a precise RV measurement challenging (Klutsch et al., in prep.). As we do not want to deteriorate the SB1 detection process with such disturbances, we decided to exclude all setups dedicated to hot and/or pre-main sequence/cluster stars, thus keeping only HR10 and HR21. Some tests were performed with the UVES setups, which confer the advantage of having a higher resolution than the GIRAFFE setups, but it turned out that troublesome issues appeared with the sky line correction for the brightest targets. The UVES exposures involve only a very small fraction (6\%) of iDR5. Moreover, only 0.3\% among the targets have observations in both GIRAFFE HR10 and HR21 and in  UVES 520 and 580. This is why we did not consider UVES observations in the present analysis.

\subsection{Atmospheric parameters of the sample stars}
\begin{figure*}
    \centering
    \includegraphics[width=\linewidth]{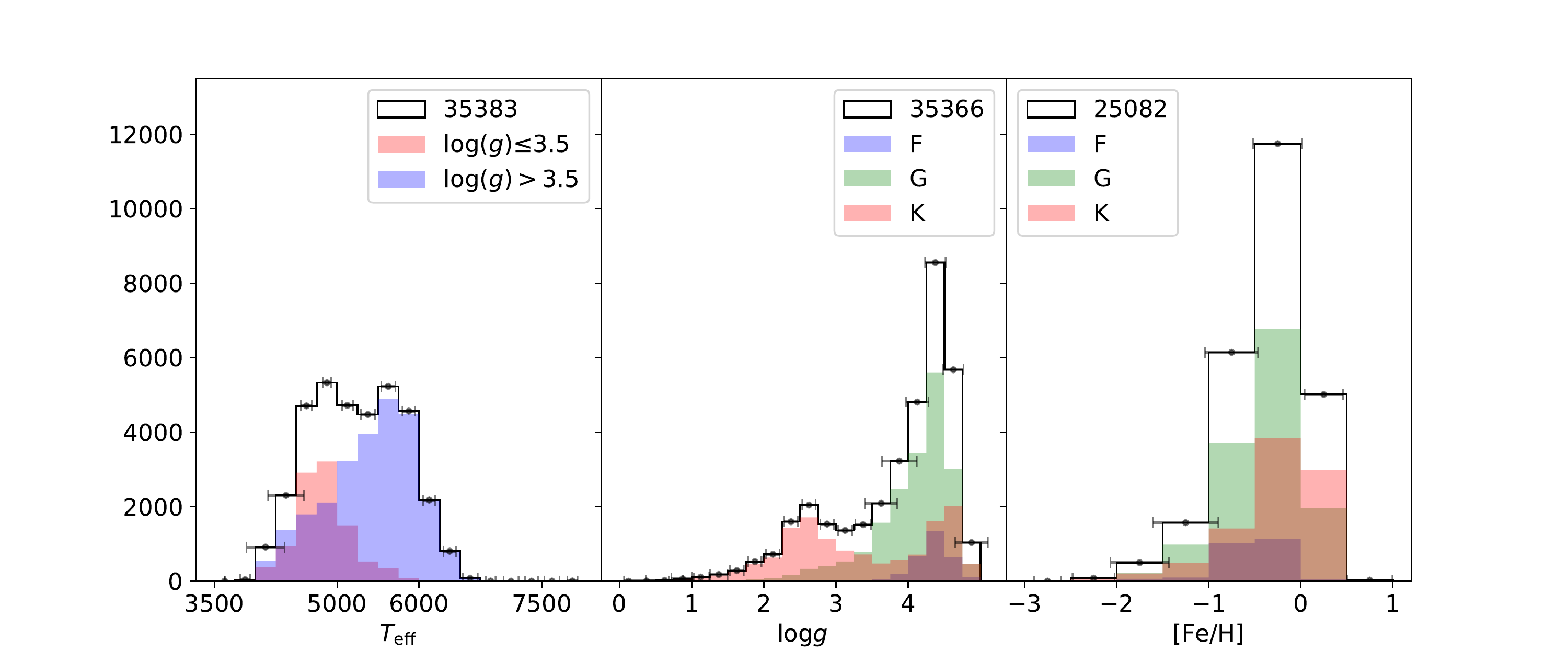}
    \caption{Distribution of the GES iDR5 recommended atmospheric parameters for the analysed stellar sample. The \Teff, $\log{g}$ and [Fe/H] bin size are 250~K, 0.25 and 0.5 dex, respectively. Horizontal error bars represent the median error in each bin.}
    \label{fig:idr5_ap}
\end{figure*}
We present in Fig.~\ref{fig:idr5_ap} the distributions of the GES-recommended atmospheric parameters for the subsample of stars analysed in this paper (\emph{i.e.} those stars with at least two observations with HR10 and HR21, as fully described in Sect.~\ref{sect:filter}). The GES-recommended atmospheric parameters (effective temperatures \Teff, gravities $\log{g}$, and metallicities [Fe/H]) are obtained by a weighted average of the astrophysical parameters obtained by different groups within the GES consortium. These groups used different state-of-the-art methods to eliminate systematics and to derive reliable uncertainties. For details about this procedure for GES UVES spectra, see \citet{sacco2014}. The sample covers FGK stars with a few additional A and M stars. The two peaks in the \Teff\ distribution apparent in Fig.~\ref{fig:idr5_ap} are attributable to dwarf and giant stars (hot peak at $5750$~K and cool peak at $4750$~K, respectively). Two peaks are also visible in the gravity distribution, at $\log{g}=4.5$ and $\log{g}=2.5$  for dwarfs and giants, respectively. The distribution of $\log{g}$ per spectral type (middle panel of Fig.~\ref{fig:idr5_ap}) clearly shows that all F-type stars are on the main sequence, that K-type stars include main sequence and red giant objects whereas G-type stars are mainly main sequence and turn-off stars.  The metallicity distribution peaks at [Fe/H]~$\sim -0.25$ with a tail toward metal-poor stars. The shape of the metallicity distribution is almost independent of the spectral type.

\subsection{New computation of cross-correlation functions}
\label{Sect:CCFrecomputations}

Figure~2 of \citet{merle2017} illustrates in the case of the Sun how the cross-correlation FWHM\footnote{Full Width at Half Maximum} varies among the different setups used by GES. The FWHM for the CCF obtained with the HR10 setup is of the order of 40~km~s$^{-1}$ while it reaches $\sim$120~km~s$^{-1}$ for HR21. The difference in instrumental resolutions between HR10 and HR21 cannot explain the observed difference in the FWHMs. Rather, this FWHM change comes from the differing properties of the spectral lines present in the two setups. Whereas a forest of weak lines is present in HR10, the HR21 setup is dominated by the strong \ion{Ca}{II} triplet. This triplet is visible all the way from A- to M-type stars and it may be used as a proxy for metallicity. This is one of the reasons why the \emph{Gaia} spectrograph operates in this wavelength range. Nevertheless, the \ion{Ca}{II} triplet is responsible for the large width of the  CCF peaks, which in turn makes the RV measurement less precise. 

\citet{vdswaelmen2017} showed that it was possible to get narrower cross-correlation functions by designing cross-correlation masks that include only mildly blended and unsaturated lines. Our findings motivated the recomputation of the CCFs for both HR10 and HR21 spectra. In the following, the new set of CCFs are referred to as `NACRE\footnote{NArrow CRoss-correlation Enterprise} CCFs' (in contrast to `GES CCF', which only applies to the CCFs released together with iDR5 spectra).

In the context of the detection of SB$n$ (to be described in Van der Swaelmen et al., in prep.), it was necessary to reduce the CCF FWHM as much as possible, in order (i) to detect new SB$n$ (with $n\ge2$) or to confirm SB$n$ detected with the HR10 setup alone \citep{merle2017}, and (ii) to obtain more precise RVs from the HR21 setup. We summarise hereafter the main steps of the NACRE CCF computation (Van der Swaelmen et al., in prep.). First, we designed new masks from synthetic spectra using a selection of weak lines in the spectral ranges corresponding to the HR10 and HR21 setups, and in particular, we excluded the strong \ion{Ca}{II} triplet from the HR21 masks. A dozen masks were built for both HR10 and HR21 spectral ranges in order to sample the spectral diversity of FGK dwarfs and giants. Second, we computed the NACRE CCFs (\emph{i.e.} a dozen CCFs per single GES exposure) over the velocity range $[+500, -500]$~\kms\ with a velocity step of 1~\kms\ (to be compared to 2.74~\kms\ for HR10 and 1.72~\kms\ for HR21 in the GES CCFs), which corresponds to one-tenth of the velocity resolution element. A comparison between GES and NACRE CCFs is presented in Sect.~\ref{sec:rv_comp}. Third, we selected the best NACRE CCF based on a quality score (measure of the contrast by comparing the correlation noise in the feet of the CCF to the height of the highest peak of the CCF) and discarded the CCFs failing the quality test. Fourth, for each observation, we ran an updated version of \doe\ \citep[`detection of extrema' as described by][see also Sect.~\ref{sect:sigma_int}]{merle2017} on the dozen CCFs in order to derive the radial velocities (by Gaussian fitting of the CCF core). In the end, for a given observation, Van der Swaelmen et al. deliver a series of RV estimates (up to 12): one of them, being marked as `best', is used as the nominal RV;  its uncertainty is estimated using the standard deviation of the whole series of velocities derived from the dozen masks and used as the physical precision $e_\mathrm{phy}$ of the RV as described in the following section.

The RV series is selected only if the star does not show evidence for multiply peaked CCFs, since it then qualifies for testing its possible SB1 nature, following the method outlined in Sect.~\ref{sec:methods}; otherwise, it joins the sample of SB$n$ ($n \ge 2$) discussed by Van der Swaelmen et al. (in prep.). The test for detecting multiply peaked CCFs, fully described in this latter paper, may be summarised as follows: we estimate the typical cross-correlation noise over a velocity range far from the absolute CCF maximum and then consider any peak whose height is larger than $\sim 1.5$ times the typical cross-correlation noise as the possible signature of a stellar component. If there are $n$ such peaks, the star is flagged as SB$n$.  

\begin{figure*}
\includegraphics[width=0.333\linewidth]{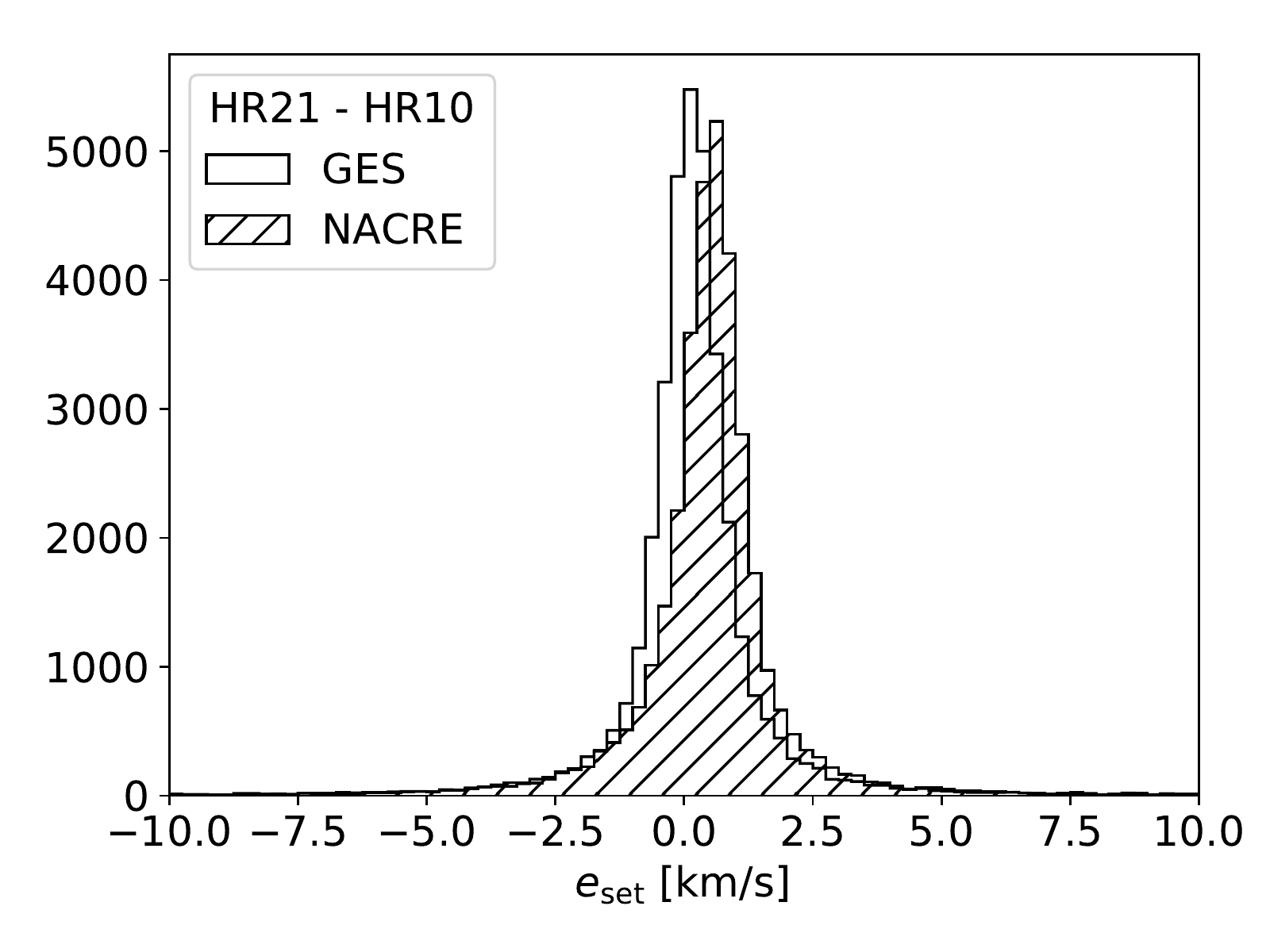}
\includegraphics[width=0.333\linewidth]{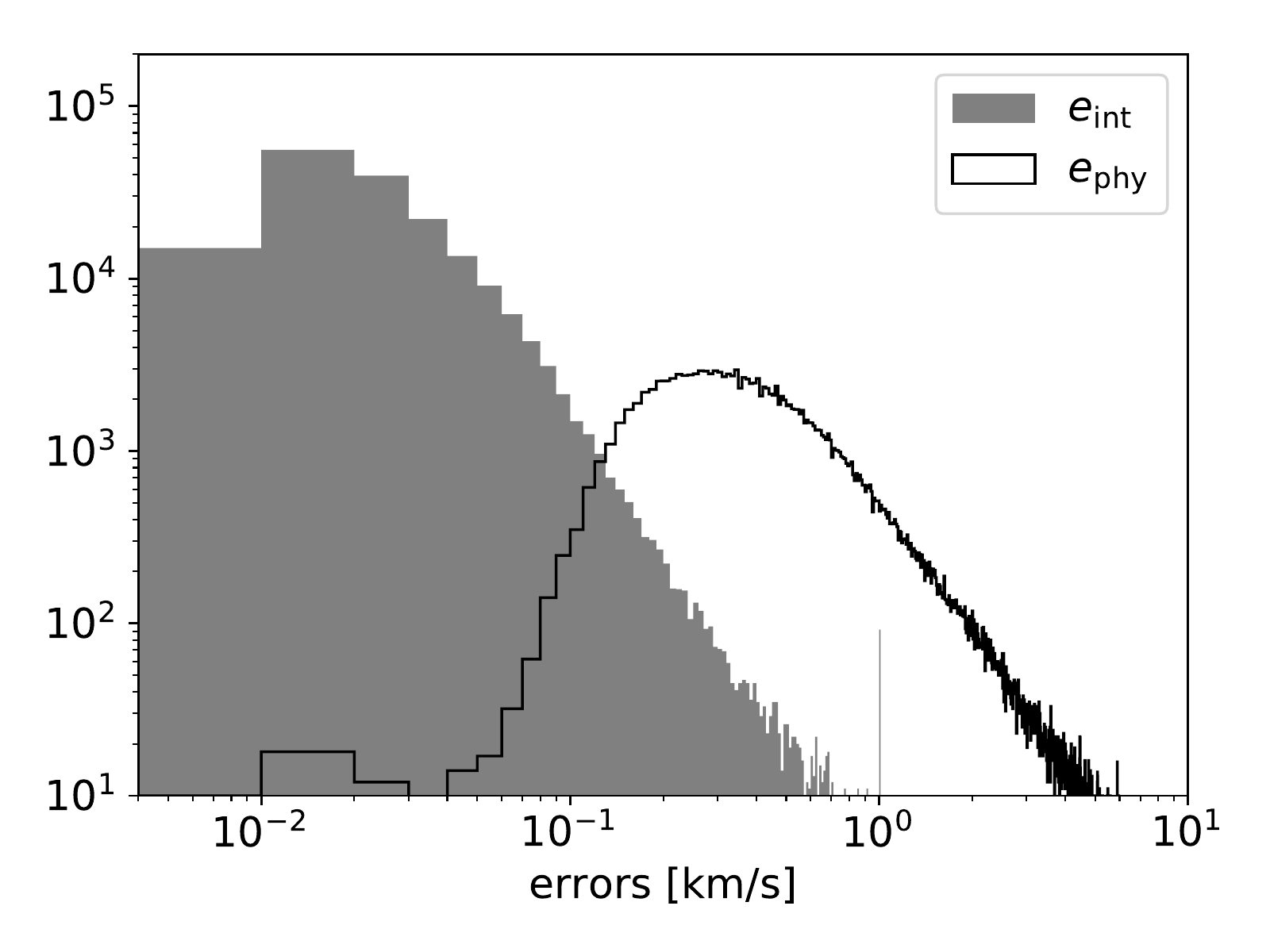}
\includegraphics[width=0.333\linewidth]{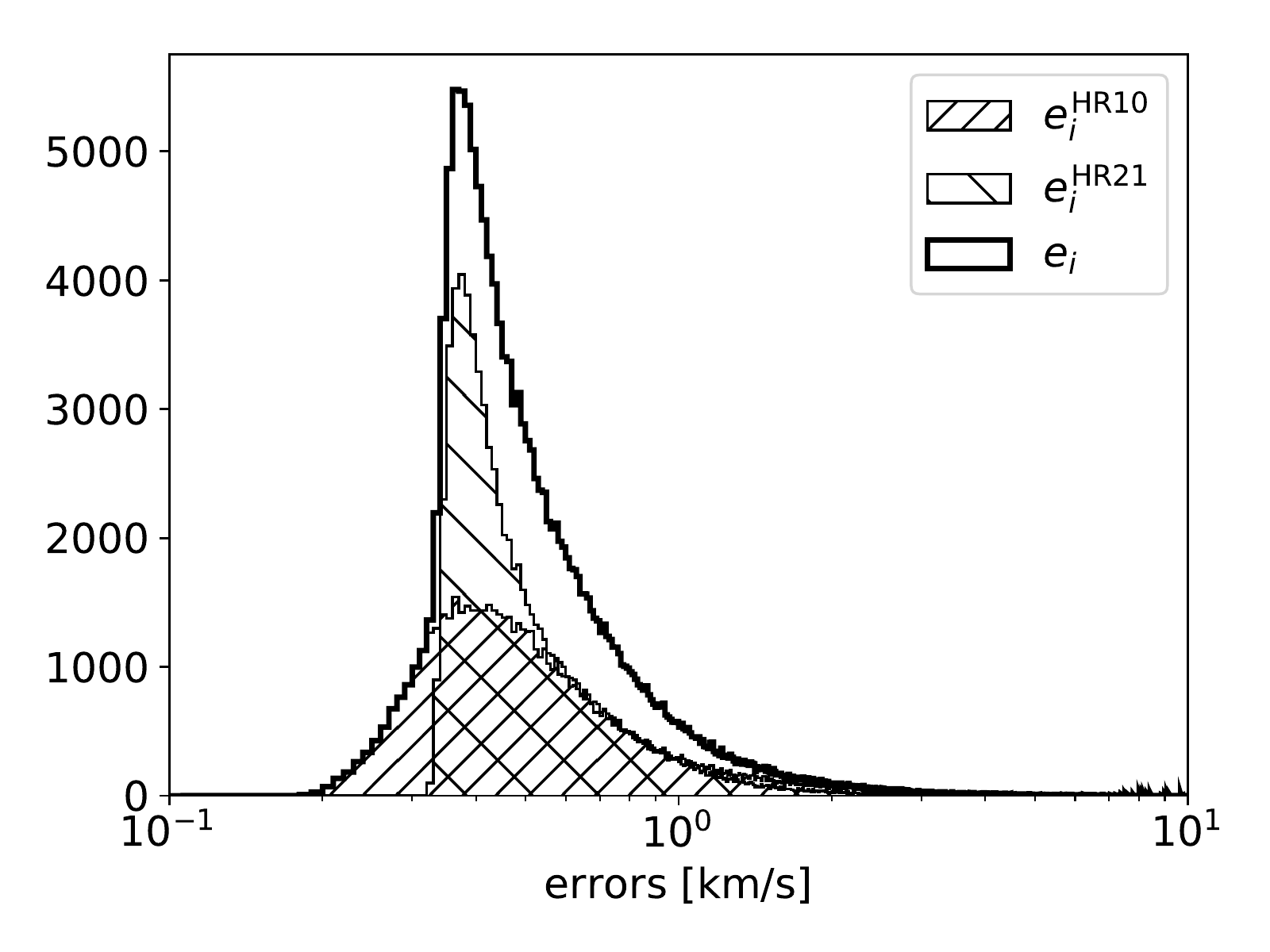}
\caption{Left: Histograms of the RV biases between the HR21 and the HR10 setups per target. The `GES' and `NACRE' labels refer to RVs as measured by the `VRAD' GES keyword, and as re-computed in this work and in Van der Swaelmen et al. (in prep.), respectively. The bin width of histograms is 0.25 \kms. Middle: Intrinsic uncertainty on the RV ($e_\mathrm{int}$) and physical uncertainty ($e_\mathrm{phy}$) measured on HR10 and HR21 single exposures. Right: RV errors according to Eq.~\ref{eq:err} per setup. Middle and right histograms have a bin width of 10~m~s$^{-1}$.}
\label{fig:drvs_errs}
\end{figure*}

\subsection{Radial velocities and their uncertainties}
\label{sec:rv}
For the analysis presented below we used the set of HR10 and HR21 RVs obtained from the NACRE CCFs (Sect.~\ref{Sect:CCFrecomputations}). As stated above, we measured the RVs using an updated version of \doe. While \cite{merle2017} used the location of the zeros of the third CCF derivative to obtain the RV, we performed a Gaussian fit of the CCF core instead. This Gaussian fit generally involves several tens of data points and the error on the parameters derived from the fit is generally very small (see Sect.~\ref{sect:sigma_int}). Larger uncertainties are encountered when multiple peaks are present, especially when the RV separation between the components decreases. However, here, we exclusively deal with single-component CCFs while the companion paper by Van der Swaelmen et al. (in prep.) is dedicated to the analysis of multiple-component CCFs.

SB1s are detected by comparing the standard deviation of the RV time series to the  uncertainties on the corresponding measurements. A careful analysis of the uncertainty sources is therefore necessary. They are discussed in turn in the following sections. The global uncertainty attached to each RV measurement is the root mean square of three terms:
\begin{equation}
 e_i = \sqrt{e_\mathrm{int}^2+e_\mathrm{phy}^2+e^2_\mathrm{cfg}},
 \label{eq:err}
\end{equation}	
where $e_\mathrm{int}$ is the intrinsic precision of the Gaussian fit (Sect.~\ref{sect:sigma_int}), and $e_\mathrm{phy}$ is the precision linked to the physical parameters of the source (spectral type, rotational velocity) and to the measurement ($S\!/N$ and wavelength range), as discussed in Sect.~\ref{sect:sigma_phys}. Finally, $e_\mathrm{cfg}$ is an additional uncertainty (changing with time) associated with the spectrograph configuration, as discussed in Sect.~\ref{sect:sigma_cfg}. The histograms of the uncertainties $e_\mathrm{int}$, $e_\mathrm{phy}$, and $e_i$  are shown in the middle and right panels of Fig.~\ref{fig:drvs_errs}. The global uncertainty histogram is obtained by adding the global uncertainties per setup ($e_i^\mathrm{HR10}$ and $e_i^\mathrm{HR21}$). The global RV uncertainty is  $0.55\pm0.24$~\kms\ (median and interquartile values). This median value is very similar for HR10 and HR21, but with a higher dispersion for HR10 (right panel of Fig.~\ref{fig:drvs_errs}). 

\subsubsection{Inter-setup bias -- $e_\mathrm{set}$}
\label{sect:sigma_setup}
The inter-setup bias is defined as follows:
\begin{equation}
e_\mathrm{set}= v_\mathrm{HR21} - v_\mathrm{HR10},
\end{equation}
where $v_\mathrm{HR21}$ and $v_\mathrm{HR10}$ are the RV measurements in the HR21 and HR10 setups, respectively. The GIRAFFE HR10 instrumental setup is used as the reference one \citep{pancino2017} because it shows the smallest offset compared to the \emph{Gaia} RV standard stars \citep{chubak2012, soubiran2013}. For each target with both HR10 and HR21 exposures, we calculate the median RV in HR10 and HR21 separately, and take the difference. The distribution of these differences is shown in the left panel of Fig.~\ref{fig:drvs_errs}, separately for the GES RVs and for the newly computed NACRE RVs (Sect.~\ref{Sect:CCFrecomputations}). We define the inter-setup bias as the median of these RV differences, wich for NACRE RVs amounts to
$e_\mathrm{set} = 0.521$~\kms\ with an interquartile of 0.520~\kms (For GES RVs, $e_\mathrm{set} = 0.100$~\kms). This inter-setup bias is consistent with the value reported by  \citet{pancino2017} for the GES iDR4, and was substracted from all the RVs measured in the HR21 setup to obtain RV on the scale of the HR10 setup.

\subsubsection{\doe\ intrinsic precision -- $e_\mathrm{int}$}
\label{sect:sigma_int}

In \citet{merle2017}, the RV components were measured using the ascending zeros of the third derivatives. This method turned out to be unbiased but not very precise because some \doe\ parameters can slightly affect the position of the zeros. Larger differences would appear in cases of multiple peaks, especially when the RV separation between the components decreases (for more details, see Van der Swaelmen  al. in prep.). To avoid complications related to this way of measuring RV components (which would produce a non-negligible intrinsic uncertainty), we perform a Gaussian fit of the core of each component based on $22\pm3$ points (median and interquartile values) of each selected CCF for a given exposure. We take the intrinsic uncertainty as the square root of the variance of the optimised RV mean parameters of the CCF fit. The middle panel of Fig.~\ref{fig:drvs_errs} shows the histogram of the intrinsic uncertainty on a log-log scale with a bin width of 10~m~s$^{-1}$ for all single exposures of HR10 and HR21 with a single-component CCF. As a characteristic value, the intrinsic error is $0.024\pm0.013$~\kms (median and interquartile values).

\subsubsection{Physical precision -- $e_\mathrm{phy}$}
\label{sect:sigma_phys}
The physical precision is given by the dispersion of the RV measurements from the NACRE CCFs computed with different masks for a given single exposure (Sect.~\ref{Sect:CCFrecomputations}). Because the CCFs are computed using synthetic masks reproducing weak spectral lines of different FGK dwarfs and giants, the RV measurements are possibly affected  by the template mismatch due to an incorrect match between the effective temperature of the target and of the mask, and to a lesser extent by a gravity or metallicity mismatch. The effect of the projected rotational velocity of the target is mostly included in the intrinsic error that obviously increases for peaks broadened by rotation. These mismatches translate into a  dispersion of the RV measurements for a given exposure. Similarly, the higher the $S\!/N$, the lower the dispersion of the RV measurements for a single exposure. This is investigated in detail in Van der Swaelmen et al. (in prep.). 

The physical precision $e_\mathrm{phy}$ is taken at 3\textit{$\sigma$} with the appropriate Student's coefficient. The histogram of the physical precision for single-component CCFs is shown in the middle panel of Fig.~\ref{fig:drvs_errs}. As a characteristic value, the physical uncertainty is $0.486\pm0.268$~\kms (median and interquartile values). If the analysis is performed separately for HR10 and HR21, we find a physical error of $0.584\pm0.331$~\kms\ for HR10 and $0.389\pm0.221$~\kms\ for HR21. We conclude  that the physical precision is better for HR21 than for HR10. When the histograms of $e_\mathrm{phy}$ and $e_\mathrm{int}$ are compared (middle panel of Fig.~\ref{fig:drvs_errs}), the physical errors generally dominate over the intrinsic ones.

\subsubsection{Spectrograph configuration precision -- $e_\mathrm{cfg}$}
\label{sect:sigma_cfg}
The last source of uncertainty to be considered  comes from the random changes in the wavelength calibration with time and from the changes in fibre allocation.  
We apply constant values of $e_\mathrm{cfg}=0.17$~\kms\ for HR10 and  $e_\mathrm{cfg}=0.32$~\kms\ for HR21 as given in Table~\ref{tab:err_phys}, irrespective of the short- or long-term repeats, using the same procedure designed for GES iDR4 data \citep{Jackson2015}. The latter authors have shown that $e_\mathrm{cfg}$ is not very sensitive to the time span between the observations.

\section{Searching for SB1s}
\label{sec:methods}

\subsection{Data-cleaning filters}
\label{sect:filter}

Before applying a statistical test to identify targets with variable RVs (Sect.~\ref{Sect:chi2}), we first restrict the sample using three different filters that reduce the initial sample of 49557 stars to 43421 stars (Table~\ref{tab:selection}).

First, we decided to consider only RV measurements coming from exposures with $S\!/N \ge 3$. This threshold appear to be very low, but it has been demonstrated on simulated data that RV measurements are still possible at  $S\!/N=2$ with the HR10 setup and at $S\!/N=5$ with HR21 using the  GES CCF (see Fig.~13 in \citealt{merle2017}). The combination of tens or hundreds of lines entering the CCF  makes a  RV measurement possible even with spectra of poor quality, though at the cost of a larger RV uncertainty. 

\begin{table}
 \caption{Fitting parameters for the computation of GES RV uncertainties following \cite{Jackson2015} for iDR5 data.}
 \centering
 \begin{tabular}{lrrr}
\hline\hline
Setup & HR10 & HR21 \\
\hline
$A$ [km/s] & 0.17 & 0.32 \\
$C$ [km/s] & 13.6 & 14.9 \\
$b_0$ [km/s] & 1.21 & 4.84 \\
$b_1$ [km/s] & 120.3 & 103.1\\
median($v_\mathrm{cor}$) [km/s]& 4.7 & 1.5 \\ 
\hline
 \end{tabular}
 \label{tab:err_phys}
\end{table}

\begin{figure*}
 \includegraphics[width=0.5\linewidth]{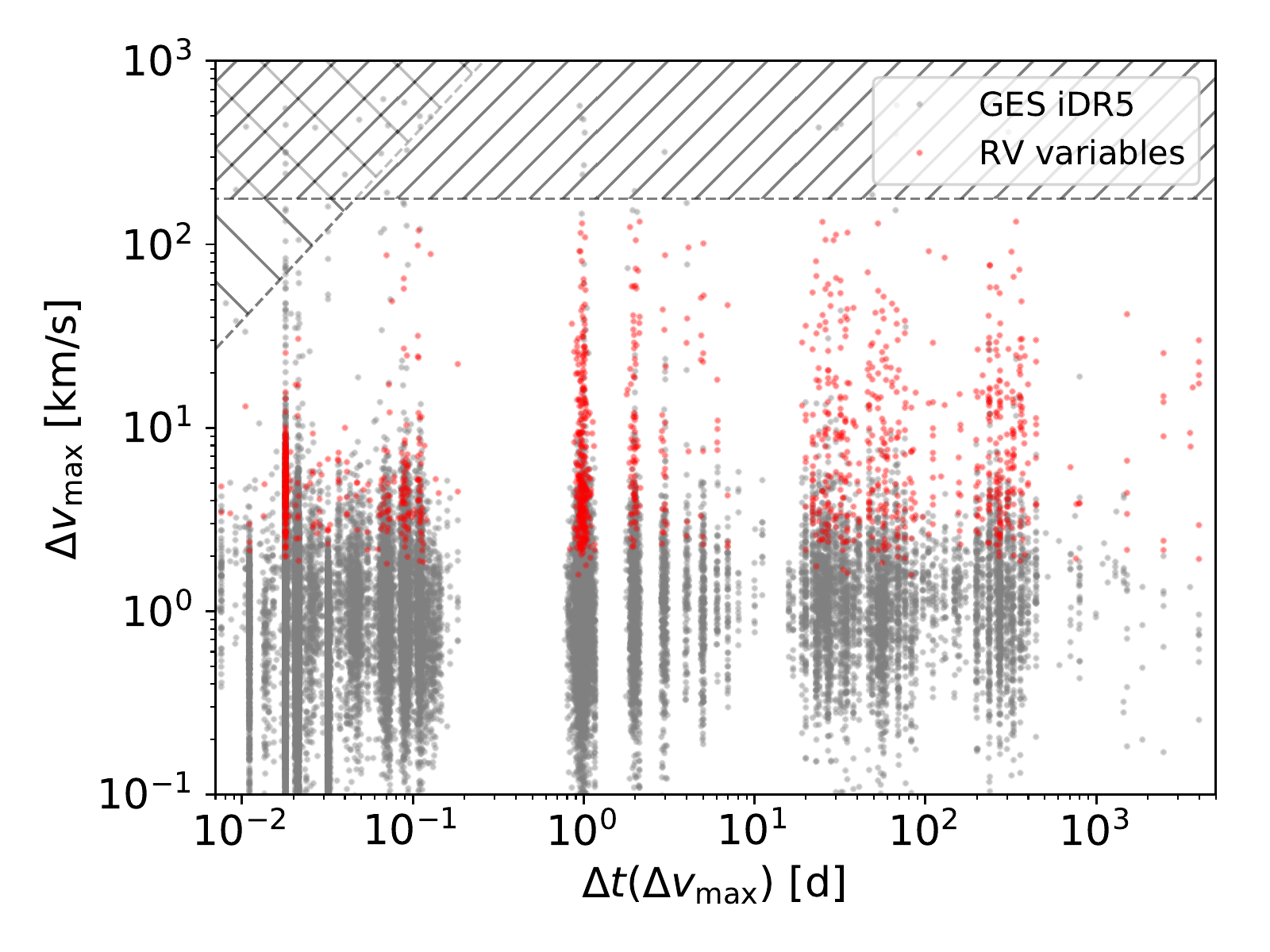}
 \includegraphics[width=0.5\linewidth]{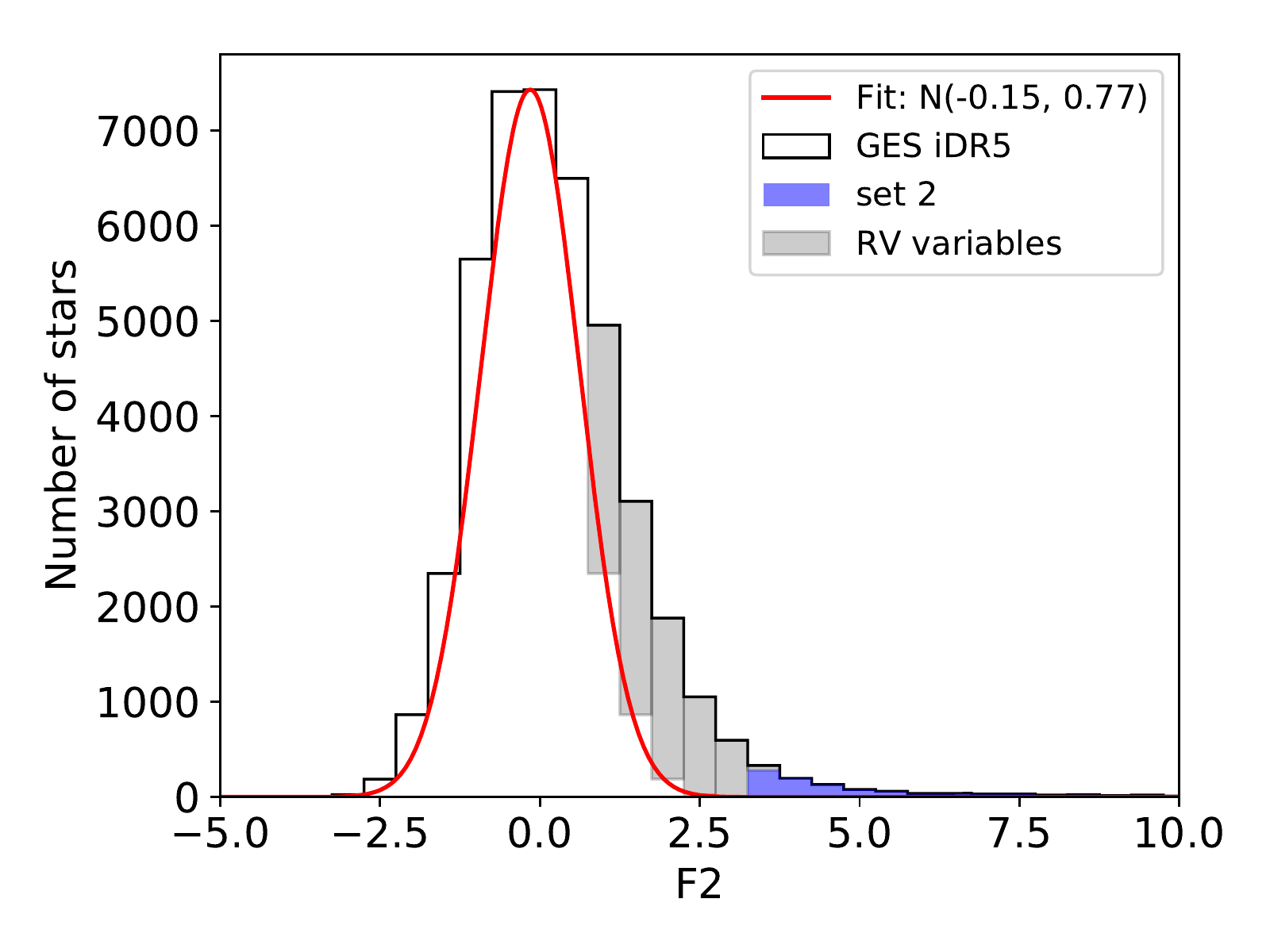}
 \caption{Left: The RV range per star as a function of the time difference between the RV measurements for GES iDR5 (grey dots) and for RV variables of set~1 (red dots). The horizontal and inclined hatched areas excluded stars showing $\Delta v_{\max}$ larger than 177~\kms, and stars showing $(\Delta v/\Delta t)_{\max} > 160 $~\kms/h. Right: The $F2$ distribution of GES iDR5 stars (empty histogram), RV variables (filled grey) and set~2 (filled blue). The red line is the fit of the total GES  iDR5 histogram.}
 \label{fig:chi2_test}
\end{figure*}

Second, we applied an acceleration criterion based on the RV rate of variation (${\rm d}v/{\rm d}t$, where $v$ is the RV) to identify those stars requiring a visual evaluation of the CCFs to identify CCFs with technical issues leading to unreliable velocities. The acceleration criterion is obtained by deriving the expression of the velocity for a SB1 system with respect to time, and using the Kepler equation and the relation between the mean and eccentric anomalies. This leads to
\begin{equation}
\label{Eq:acceleration}
\frac{{\rm d}v}{{\rm d}t} = K_1 \frac{a}{r} \frac{2\pi
}{P} \sin(\omega+\phi) \left(\left(\frac{1+e}{1-e}\right)^{1/2}\cos^2\frac{\phi}{2}\;+\;\left(\frac{1-e}{1+e}\right)^{1/2}\sin^2\frac{\phi}{2}\right),
\end{equation}
where $a$ is the semi-major axis of the orbit, $r$ is the radius-vector for the  true anomaly $\phi$, $\omega$ is the argument of periastron,
\begin{equation}
    K_1 = \frac{2\pi}{P}\;\frac{a_1\,\sin i}{(1-e^2)^{1/2}}
\end{equation}
is the velocity amplitude of  the visible component, and
\begin{equation} 
a_1 = \frac{q}{1+q}\;a
\end{equation}
 is the semi-major axis of the visible component of mass $M_1$ around the centre of mass of the system (with $q=M_2/M_1$, where $M_2$ is the mass of the invisible component).

Large rates of RV variation may be encountered in two situations:  (i) for eccentric systems at periastron passage, and (ii) for close semi-detached systems possibly involving a compact and possibly massive companion like a neutron star. 

For case (i), a large eccentricity and a short-period should be adopted. However, these two parameters are constrained by the eccentricity - period ($e - P$) diagram \citep[see, \emph{e.g.}][for an $e - P$ diagram involving  mainly pre-mass-transfer giants, or  \citealt{pourbaix2004} for the $e - P$ diagram from the SB9 catalogue]{mermilliod2007}. In SB9, the largest eccentricity at a given period obeys the relation $(1-e_{\max})^3\;P  \sim 0.3$ (with the orbital period $P$ expressed in days). The maximum acceleration is obtained by imposing $\phi = 0$ (and thus $r = (1-e)\;a$; periastron passage) in Eq.~\ref{Eq:acceleration}, leading to 
\begin{equation}
\frac{{\rm d}v}{{\rm d}t}  = K_1\; \sin\,\omega\;\frac{2\pi}{P}\;\frac{1}{1-e}\;\left(\frac{1+e}{1-e}\right)^{1/2},   
\end{equation}
or 
\begin{equation}
\frac{{\rm d}v}{{\rm d}t}  = K_1\; \;\frac{2\pi}{0.3}\;(1+e)^{1/2}(1-e)^{3/2},   
\end{equation}
after inserting the expression for the $e - P$ upper envelope and taking into account the fact that the resulting expression  is itself maximum for $\omega = \pi/2$. This expression (with ${\rm d}v/{\rm d}t$ in units of $K_1$~d$^{-1}$) is clearly maximum when $e = 0$. The period of 0.3~d appearing on the denominator of the above equation  corresponds to the minimum period of orbits with non-zero eccentricities observed in the SB9 catalogue. However, there are systems with even shorter periods and zero eccentricities that will lead to even larger accelerations. These systems are cataclysmic variables consisting of a red dwarf on the main sequence and a white dwarf star in a semi-detached system with periods of the order of 3 or 4~h. 

As a consequence, case (i) described immediately above turns out to be identical to case (ii) -- the semi-detached systems--, meaning that we may conclude that the acceleration criterion reduces in all cases to  
\begin{equation}
\label{Eq:caseii}
\left(\frac{{\rm d}v}{{\rm d}t}\right)_{\rm max} = \frac{2\pi\;K_1 }{P}.
\end{equation}
Adopting the values of one of the prototypical cataclysmic variables for $P$  and $K_1$, namely WW~Cet \citep[$P=0.1758$~d, $K_1 = 108$~\kms;][]{Thorstensen1985}, we find 160~\kms~h$^{-1}$ as the typical threshold for the acceleration criterion.  We stress that this criterion does not constitute a hard threshold to reject targets, but is rather a flag identifying targets requiring a visual inspection of their CCFs.

The statistical test described in Sect.~\ref{Sect:chi2} that we apply on the RV data to identify SBs, relies on at least two single exposures. These two exposures may have been obtained as close as a few minutes apart (see the right panel of Fig.~\ref{fig:dtmax}). The criterion based on ${\rm d}v/{\rm d}t$ that we just described is especially designed to prevent  false positives under such circumstances.

Third, we flagged targets that show RV amplitudes larger than 177~\kms\ for visual inspection of their CCFs. This value was selected on a similar basis as for the acceleration criterion described above. Clearly, large velocity amplitudes will be obtained in massive short-period systems. Since the GES HR10 and HR21 setups were mainly used for FGK spectral types,  implying $0.5 \le M_1 \le 3$~M$_\odot$ for main sequence primaries, we base our amplitude filter on a semi-detached system consisting of a 3~M$_\odot$ primary star (with a radius of 2~R$_{\odot}$ according to the usual mass -- radius relationship for main sequence stars) in a $q = 0.5$ system. In a semi-detached system, the primary star fills its Roche lobe. Therefore, the orbit must be circular since a system evolves towards such a state when it dissipates energy at constant angular momentum. Based on the simple \citet{Paczynski1971} formula for the Roche radius, equating the latter with the stellar radius leads to an orbital separation of 6.25~R$_{\odot}$ or $2.9\; 10^{-2}$~au for the semi-detached system, corresponding to an orbital period of 0.8~d and a velocity amplitude $K_1$ of 125~\kms, and a maximum RV range potentially as large as twice that value or 250~\kms. A more realistic statistical estimate would be $\sqrt{2}\;K_1$, or 177~\kms, which corresponds to the standard deviation expected for measurements randomly sampling a sinusoidal RV curve (i.e., circular orbit). A 1.5 + 0.5~M$_\odot$ main sequence pair would correspond to a semi-detached orbital separation of 3.2~R$_{\odot}$, a period of 0.52~d, and a semi RV amplitude $K_1$ of 101~\kms, thus smaller than the former case which is therefore retained as threshold value. As we show below in Fig.~\ref{fig:histo_drvmax}, this threshold is realistic given the observed fall-off of the observed $\Delta v_{\max}$ ranges for SB1 systems involving dwarf stars.  The acceleration and amplitude criteria are shown in the left panel of Fig.~\ref{fig:chi2_test}. The number of false positives with $\Delta v_{\max} > 177$~\kms\ and $(\Delta v/\Delta t)_{\max} > 160 $~\kms/h due to technical problems with the CCF computation amount to about 100, which is relatively marginal compared to the 43\,500 sources analysed (Table~\ref{tab:selection}). 

\begin{figure*}
\includegraphics[width=0.49\linewidth]{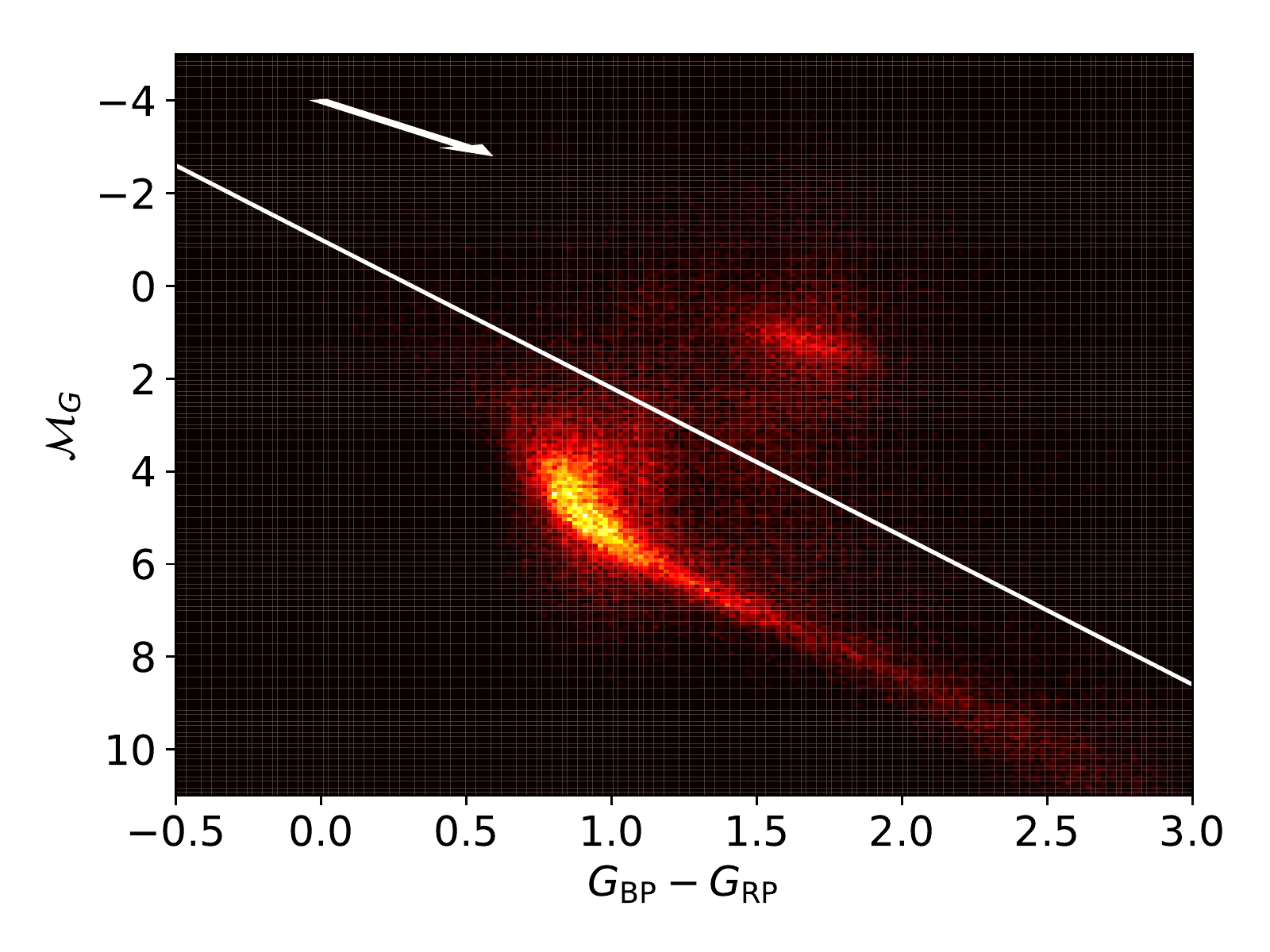}
\includegraphics[width=0.49\linewidth]{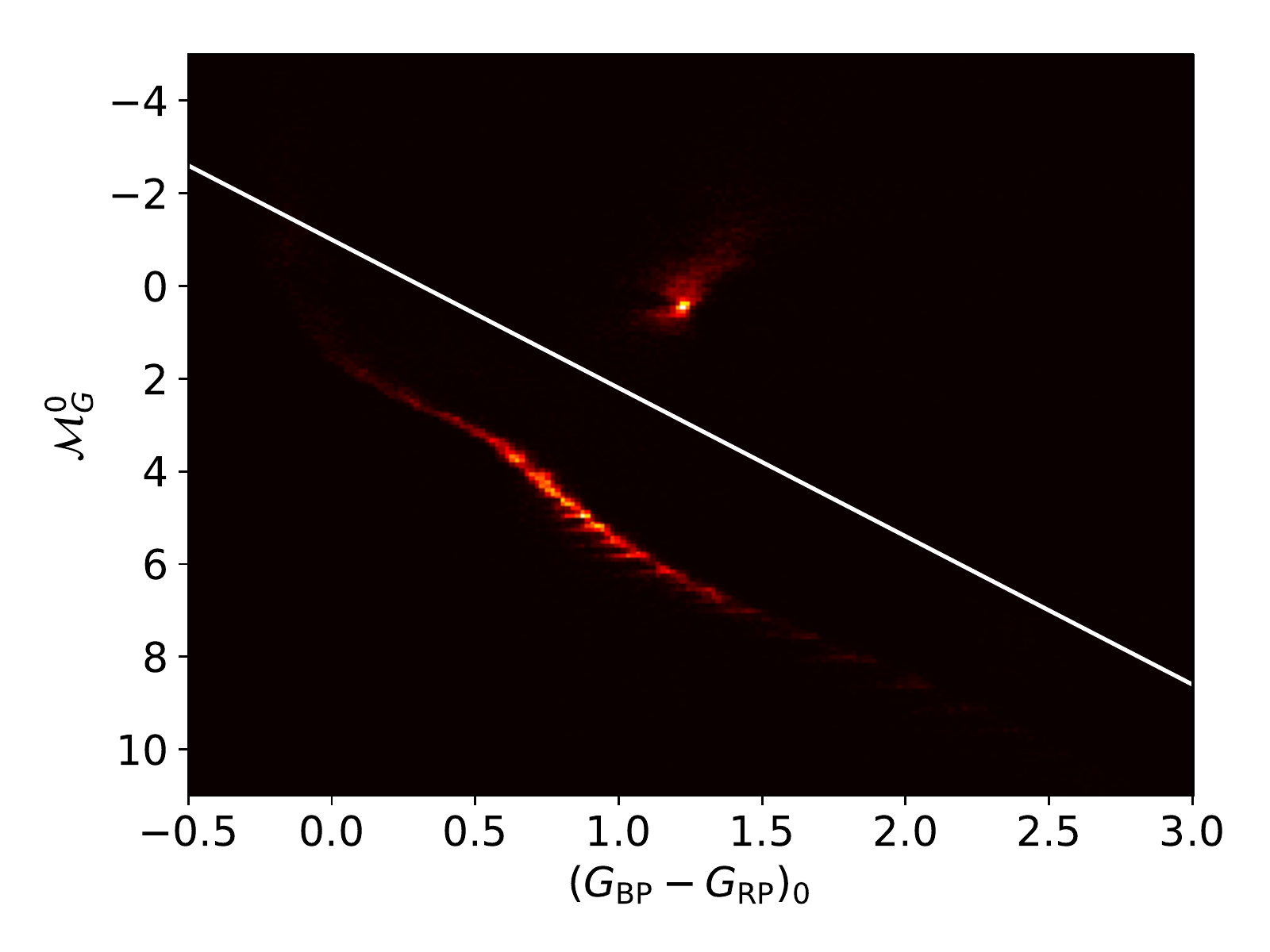}
\caption{Absolute magnitude vs. colour diagram for GES iDR5 stars using \emph{Gaia} DR2 parallaxes and photometry without (left) and with (right) correction for extinction and reddening taken from \citet{andrae2018}. We included all sources irrespective of their parallax and flux uncertainties. Almost 40\% of the stars in the left diagram have a \emph{Gaia}~DR2 extinction and colour excess values. The white line marks the separation between dwarfs and giants  using Eq.~\ref{eq:lc}. The white arrow in the left panel represents the reddening vector.}
\label{fig:cmd_ges}
\end{figure*}

\subsection{Statistical $\chi^2$-test}
\label{Sect:chi2}

To assess whether the dispersion of the measured RVs of a given star calls for an external cause of variation (binarity or pulsations), we use a simple $\chi^2$ test, defined as follows:
\begin{equation}
\chi^2_{N-1} = \sum_{i=1}^{N} \left(\frac{v_i- \bar{v}}{e_i}\right)^2,
\end{equation} 
where $v_i$ is the $i^\mathrm{th}$ RV measurement (corrected from the inter-setup bias $e_\mathrm{set}$ if needed) in the time series of $N$ single exposures and  $\bar{v}$ the weighted RV mean, while $e_i$ is the total uncertainty on $v_i$ as defined by Eq.~\ref{eq:err}. Due to the small number of exposures for each target, we require a high confidence level of 99.9\% (those stars matching this confidence level constitute set~2, as denoted in the remainder of this paper), or even 99.9999\% (denoted as set~1). For $N=4$ (\emph{i.e.}, three degrees of freedom), the null hypothesis that the star has a constant velocity may be rejected at such confidence levels when $\chi^2$ is larger than 16.3 or 30.7,  respectively.

\subsection{F2 statistics}
The $\chi^2$ distribution (which depends on the number of degrees of freedom -- dof) can be transformed into a dof-independent  distribution, called $F2$, as shown by \citet{wilson1931}.
The $F2$ distribution behaves like a normal distribution with zero mean and unit standard deviation, and is defined as: 
\begin{equation}
\label{eq:F2}
F2(\chi^2, N) = \sqrt{\frac{9(N-1)}{2}}\left[\left(\frac{\chi^2}{N-1}\right)^{1/3} +\frac{2}{9(N-1)} -1 \right],
\end{equation} 
where $N-1$ are the dof and $\chi^2/(N-1)$ is the reduced $\chi^2$. If the RV uncertainties $e_i$ were overestimated, $\chi^2$ and $F2$ would be too small, with the $F2$ distribution peaking at negative values. Confidence levels of 99.9\% and 99.9999\%  for the $\chi^2$-test with $N=4$ translate into $F2$ values of 3.1 and 4.6. Sets 1 and 2 may therefore be considered as collecting stars with RV variations at the $\sim 5\sigma$ and $\sim 3\sigma$ confidence levels, respectively. According to the procedure described in \cite{matijevic2011}, which makes use of the logarithm of the $p$-value, these confidence levels are equivalent to $\log p =-3$ and  $\log p=-6$, respectively. 

For the sake of completeness, we also consider set~3 which contains the number of stars with $F2 \ge 1$ located above the theoretical $F2$ curve in the right panel of Fig.~\ref{fig:chi2_test}, which correspond to the excess stars with RV variability over the normal statistical distribution. Individual RV variables cannot be identified for this set, since they are mixed among a large number of normal stars; therefore only the binary frequency can be provided for set~3 (see Sect.~\ref{Sect:RVvariables}).

\subsection{Comparison between GES and NACRE RVs}
\label{sec:rv_comp}
For the detection of SBs, accurate RV measurements and a correct evaluation of their associated uncertainties are crucial. This is why new CCFs have been computed (Sect.~\ref{Sect:CCFrecomputations}), and the number of components and their RVs re-evaluated with the \doe\ tool \citep{merle2017}. We compared the efficiency of SB1 detection when using the NACRE RVs or the GES ones, for a set of single exposures for which both NACRE and GES RVs are available, along with their uncertainties. This common set consists of about 42\,800 stars corresponding to 162\,000 single exposures. The GES RV uncertainties $e_i^\mathrm{GES}$ result from the quadratic sum of (i) an uncertainty computed as described in \citet{Jackson2015} (their Eq. 2) but applied to the iDR5 data with the fitting parameters given in Table~\ref{tab:err_phys}, and (ii) an internal error $e_\mathrm{int}^\mathrm{GES}$ equal to half the velocity step chosen for the GES CCF. For HR10 and HR21,  $e_\mathrm{int}^\mathrm{GES}$ amounts to 1.370 and 0.867~\kms, respectively. Applying the filters on the $S\!/N$, the $(\Delta v/ \Delta t)_{\max}$, and the $\Delta v_{\max}$ as described in Sect.~\ref{sect:filter}, the $\chi^2$ test delivers more RV variables with NACRE RVs than with GES RVs (878 against 709) for a confidence level of 99.9999\%. This is explained by the fact that, on average, GES uncertainties are larger than NACRE ones, leading to a lower detection efficiency for a given confidence level. However, stars detected as variable in both datasets amount to only 435, or about half the number of RV variables in either the NACRE or the GES samples. A careful look at the GES RV-variable sample reveals that 
most of them are false positives whereas the NACRE sample appears to contain mostly  true RV variables.

\section{Results and discussion}
\label{Sect:Results}

From the analysis of the GES iDR5 subsample (limited to GIRAFFE HR10 and HR21 setups), consisting of 43\,421 stars totalling about 165\,000 single exposures, from which about 100 stars were rejected after visual inspection of their CCFs, as  required by the criteria defined in Sect.~\ref{sect:filter}, we obtain 772, 1395, and 11316 RV variables in sets~1, 2, and 3, respectively, as defined in Sect.~\ref{Sect:chi2} and listed in Table~\ref{tab:selection}.

\subsection{Radial velocity variables from the $\chi^2$ test}
\label{Sect:RVvariables}
We present in the left panel of Fig.~\ref{fig:chi2_test} the RV range per star as a function of the time interval between the RV extrema (\emph{i.e.} the absolute time difference between the highest and the lowest RVs). The minimum range for the RV variables is 2.2~\kms\ and 1.6~\kms\ for sets~1 and 2 respectively, while the maximum one is 133~\kms\ for both sets, for a time span ranging from 7.6~min to 10.9~yr. The latter is larger than the duration of the entire survey (5~yr) because $\sim3$\% of the GES iDR5 spectra are recovered from the ESO archive. The right panel of Fig.~\ref{fig:chi2_test} shows the $F2$ distribution as defined by Eq.~\ref{eq:F2}. The $F2$ distribution for set~2 (detections  at the $3\sigma$ confidence level) is displayed in blue. The observed distribution (black histogram) can be fitted (by a maximum-likelihood estimator) with a Gaussian (red line) with a mean of $-0.15$ and a standard deviation of $0.77$, close to the expected normal-reduced distribution centred on zero and with a standard deviation of unity. The fact that the left side of the $F2$ distribution  ($F2<0$) almost follows the normal distribution indicates that the uncertainties were correctly estimated. However, the right side of the $F2$ distribution shows an expanded tail corresponding to a population with significant RV variations. Indeed, for $F2 > 1$, the observed distribution lies well above the $\mathcal{N}(-0.15, 0.77)$ distribution. We can estimate the excess of RV variables in the sample by counting the number of objects above the Gaussian distribution starting from the bin $F2=1$, where the asymmetry is apparent. This defines set 3, where the number of RV variables amounts to $11\,316$ (26\% of the full sample). Set~3 obviously contains more SB1 candidates than those selected by sets~1 and 2 (as defined in Sect.~\ref{Sect:chi2}) but the former cannot be  identified individually: the $F2$ statistics only allow us to detect an excess of RV variables (with respect to the normal distribution expected if the sample was purely composed of single, non-photometrically variable objects). All cleaning processes described in Sect.~\ref{sec:photo} (for removing photometric variables) are applied to set~3 as well, below, in order to derive a cleaned SB1 fraction at the 1$\sigma$ level. 

\begin{table}
\caption{Number of SB1 detected among the GES iDR5, for  sets 1 ($5\sigma$), 2 ($3\sigma$) and 3 ($1\sigma$).}
\centering
\begin{tabular}{lrrrr}
\hline\hline
 &set~1  &set~2  & set 3 & iDR5\\
 & ($5\sigma$)    & ($3\sigma$)  & ($1\sigma$)& GES\\
\hline\\
& \multicolumn{3}{c}{RV variables} & 
\medskip\\
\cline{2-4}\\
Gaia DR2 cross-matches & 772 &  1\,395 & 11\,316 & 43\,421\\
with photometry   & 738 & 1\,310  & 10\,479 & 40\,551\\
~~dwarfs           & 591 & 1\,070  & 8\,378 & 31\,240\\
~~giants           & 147 & 240     & 2\,101 &  9\,311\\
without photometry & 34 & 85 &  837 & 2\,628\\
\hline\\
& \multicolumn{3}{c}{SB1} & 
\medskip\\
\cline{2-4}
\\
After photometric cleaning& 607& 718 & 975 \\
~~dwarfs         & 464 & 519 & 577\\
~~giants         & 143 & 199 & 398\\

\hline
with \Teff\         & 466 & 605 && 35\,383\\
~~dwarfs            & 314 & 359 && 25\,131\\
~~giants            & 128 & 175 &&  8\,033\\
with [Fe/H] (and \Teff) & 166 & 256 && 25\,082\\
~~dwarfs            &  49 &  75 && 16\,140\\
~~giants            & 104 & 146 &&  7\,491\\
\hline
\end{tabular}
\label{tab:selection}
\end{table} 

\subsection{\emph{Gaia}-ESO Survey benchmark stars and RV standards}
\label{sect:benchmark}

The GES iDR5 contains 39 benchmark stars  of spectral types FGKM \citep{jofre2014,heiter2015} and 30~RV standards \citep[selected from][]{chubak2012,soubiran2013,soubiran2018}. Nineteen of these benchmark stars were analysed in this study. One star (the supergiant $\beta$~Ara)  is selected as an SB1 candidate in both sets~1 and 2. However this is clearly a false positive because a close inspection shows that there is only one discrepant RV measurement (HR21,  MJD\footnote{Modified Julian Date} = 56172.97408, RV=$31.765\pm0.796$~\kms) whereas the other measurements are consistent with RV averages of $-0.193\pm0.305~$ (7 observations) and $-0.445\pm0.279~$~\kms (12 observations) at MJD of 56172.97 and 56195.99, respectively.

Among the 30 GES RV standards, 26 stars had appropriate data to search for RV variations. No SB1 candidates were found in set~1 but one (the dwarf HIP~31415) emerged in set~2. For this star, 12 observations in HR10 and HR21 are available with $S\!/N\sim375$. These 12 observations sampled two epochs of six observations each, separated by two years with RV averages of $-8.600\pm0.799$ and $-7.870\pm1.402$~\kms\ at MJD of  56223.38 and 56967.37. There are no discrepant values in this case, but a closer look at the data shows a systematic bias between HR10 and HR21 RV values of $\sim1.6$~\kms~ at the first epoch and $\sim2.7$~\kms~ at the second epoch, causing the large dispersions of the global RV averages. This GES RV standard appears to be a false positive due to the relatively large inter-setup bias of its observations (see inter-setup bias distribution in the left panel of Fig.~\ref{fig:drvs_errs}). 

These false positives found among benchmark and RV standard stars allow us to derive a an approximate contamination fraction,  of $1/45\sim2.2\%$ for set~1 and $2/45\sim4.4\%$ for set~2. Clearly, these values are not very precise because of the small-number statistics, but are broadly consistent with our SB1 fraction error estimate given in Table~\ref{tab:sb1_freq} (Sect.~\ref{sec:2d}).

\begin{figure*}
 \includegraphics[width=0.5\linewidth]{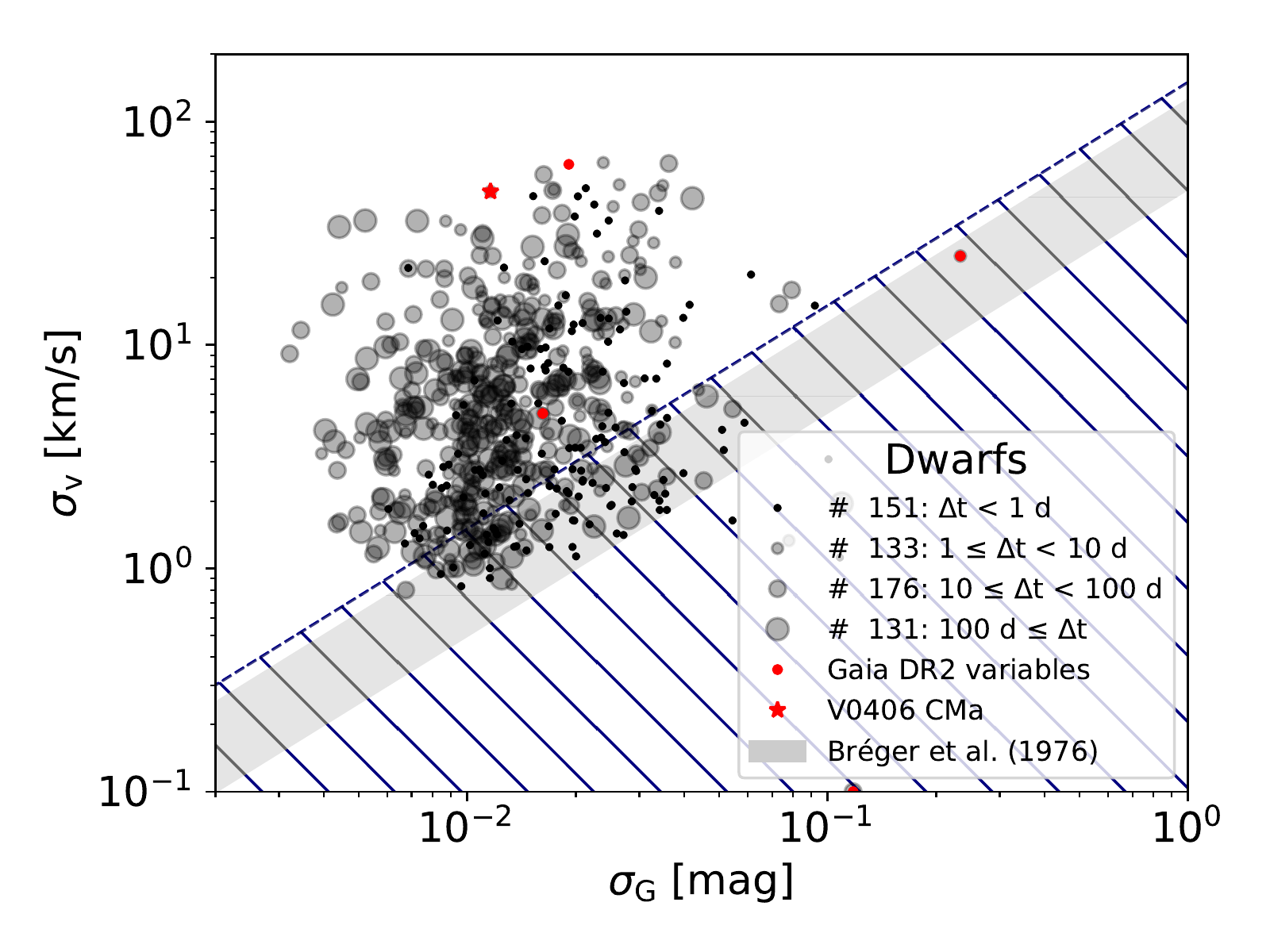}
 \includegraphics[width=0.5\linewidth]{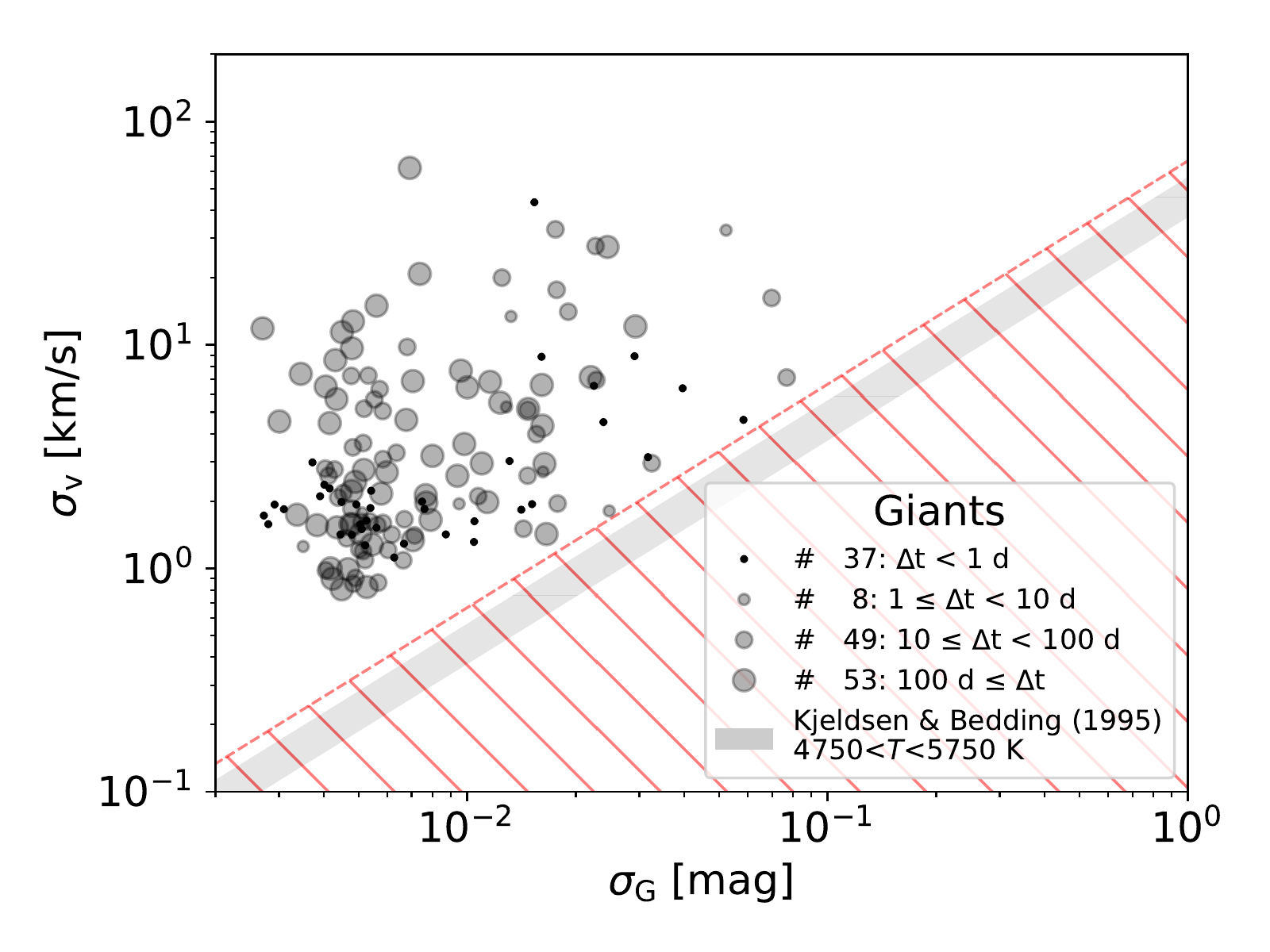}
 \caption{Intrinsic RV standard deviation vs. $G$-magnitude standard deviation for dwarf (left) and giant (right) GES iDR5 RV variables for set 1 (confidence level at $\sim5\sigma$).}
 \label{fig:sigmav_sigmam}
\end{figure*}

\subsection{False positives}
\label{sec:photo}
Stellar envelope pulsations, atmospheric convection, and the presence of spots at the stellar surface can all produce photometric and RV variations. Therefore, a fraction of the detected RV variables are not due to binarity but to jitter (pulsation and/or convection) and/or to the presence of spots.

One way to filter them out is to compare the RV dispersion with the photometric variability (in a given photometric band). The intrinsic RV dispersion is defined as:
\begin{equation}
\label{eq:sigma_v}
\sigma_v = \sqrt{\sigma_{\bar{v}}^2 - \left(\frac{\sum_i^N e_i}{N}\right)^2},
\end{equation}
where $\sigma_{\bar{v}}$ is the standard deviation of the $N$ RV measurements and $e_i$ is the uncertainty attached to the $i^\mathrm{th}$ RV measurement as defined in Eq.~\ref{eq:err}.
If $\left(\sum_i^N e_i/N\right)^2$ is larger than $\sigma_{\bar{v}}^2$, then it means that some of the individual uncertainties attached to RV measurements are overestimated.
This happens in very rare cases.

To estimate the photometric dispersion, we used data from \emph{Gaia} DR2 \citep{gaia2018} which provides mean $G$ magnitudes for more than 1.69 billion sources with precisions varying from around 1 milli-magnitude at the bright end ($G<13$) to around 20 milli-magnitude at $G=20$. No crossmatches between \emph{Gaia} DR2 and GES were available on the advanced query of the \emph{Gaia} archive\footnote{\url{https://gea.esac.esa.int/archive}} at the time of this work. Therefore, we performed a cross-match using a cone radius of two arcseconds. If several cross-matches occur within 2 arcsec for a GES target, we always keep the closest one (more specifically, we found 89.8, 8.5, and 1.7\% with, respectively, one, two, three or more cross-matches of GES iDR5 targets with \emph{Gaia} DR2). We retrieved the \emph{Gaia} identifications of all  iDR5 GES targets, and extracted their photometric data from \emph{Gaia} DR2. We used the mean flux $G$-band $F$, the error on the $G$-band mean flux $e_F$ computed as the standard deviation of the $G$-band fluxes divided by the square root of the number of CCD transits $N_t$ (about 230 on average), and the $G$-band mean magnitude $\bar{G}$. In addition, we need to know the instrumental error on the $G$ magnitude. We used the error distribution as a function of the $G$-band magnitude from Fig.~11 of \cite{riello2018} for sources with photometry produced by the full calibration process and denote it $\sigma_{\mathrm{ins}}(G)$. The uncertainty on the $G$-band magnitude is not provided by \emph{Gaia} DR2; nevertheless, we are able to estimate it from the mean flux $G$-band $F$ and its error $e_F$:
\begin{equation}
\sigma_{\bar{G}} = -2.5 \log{\left(1+\frac{\sqrt{N_t}e_F}{F}\right)}.
\end{equation}
From these quantities, we defined the intrinsic photometric variability as:
\begin{equation}
\label{eq:sigma_G}
\sigma_G = \sqrt{\sigma_{\bar{G}}^2 - \sigma_{\mathrm{ins}}^2(G)}.
\end{equation}
There are no values of $\sigma_{\mathrm{ins}}^2(G)$ larger than $\sigma_{\bar{G}}^2$, meaning that the instrumental photometric error is well estimated. 

Because the origin of photometric variability is different for dwarf and giant stars, we split the sample into these two categories. We use the \emph{Gaia} DR2 parallaxes, the $G$-band photometry, and the colour index $G_\mathrm{BP}-G_\mathrm{RP}$ from the $BP$ and $RP$ passbands. Figure~\ref{fig:cmd_ges} shows the colour -- absolute magnitude diagram obtained without (left panel) and with (right panel) the reddening correction adopted from \citet{andrae2018}. From the \emph{Gaia} DR2 extinction and colour excess of the GES iDR5 targets, we derived an average ratio of total to  selective absorption of $R_G = A_G / E(G_\mathrm{BP} - G_\mathrm{RP}) = 1.994 \pm 0.001$, as expected from the value of $\sim2$  given by \cite{andrae2018}. This correction applies the extinction to the absolute magnitude and the colour excess on the $G_\mathrm{BP}-G_\mathrm{RP}$ colour index. The reddening correction produces narrow main sequence and RGB along with a compact red clump. Nevertheless, for late-type stars, there is a strong degeneracy between colour excess and effective temperature  when they are derived from broad-band photometry \citep{bailer-jones2011,andrae2018}. In addition, there are spurious horizontal stripes on the reddening-corrected main sequence (right panel of Fig.~\ref{fig:cmd_ges}). These artifacts originate from the sparse sampling of PARSEC 1.2S evolutionary tracks \citep{bressan2012} used without further interpolation or smoothing \citep[see in][their Sect. 3.3 and Figs. 19 and 20]{andrae2018}. We classified GES iDR5 targets as dwarfs or giants by comparing their intrinsic absolute magnitude $\mathcal{M}_{\mathrm G}^0$ to the threshold  $\mathcal{M}_{G,\mathrm{lim}}^0$:  
\begin{equation}
\label{eq:lc}
\mathcal{M}_{G,\mathrm{lim}}^0=3.2(G_\mathrm{BP} - G_\mathrm{RP})_0-1.0
\end{equation}
represented as the white line on both panels of Fig.~\ref{fig:cmd_ges}.  Dwarf stars are defined as those with $\mathcal{M}_{\mathrm G} > \mathcal{M}_{G,\mathrm{lim}}^0$. This relation has been applied to the undereddened GES stars (left panel of Fig.~ \ref{fig:cmd_ges}). Given the fact that the straight line defined by this relation is almost parallel to the reddening vector (Fig.~\ref{fig:cmd_ges}), the number of stars misclassified as giants that in reality are reddened dwarfs should be very small. We note that the reddening correction (suffering from strong associated artifacts, as described above) is used solely to define relation \ref{eq:lc} and is not used for other purposes.

\begin{figure*}
 \includegraphics[width=0.5\linewidth]{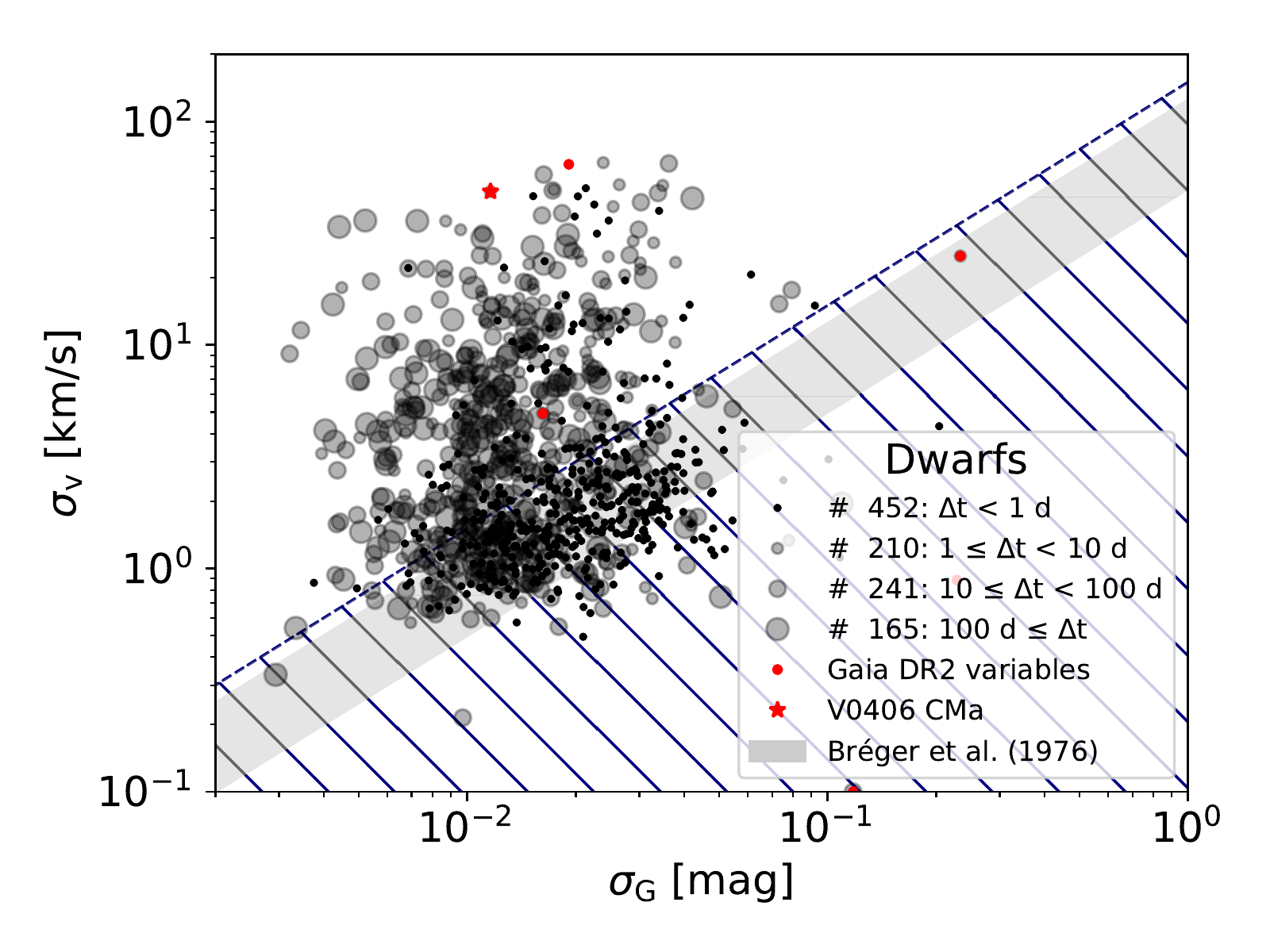}
 \includegraphics[width=0.5\linewidth]{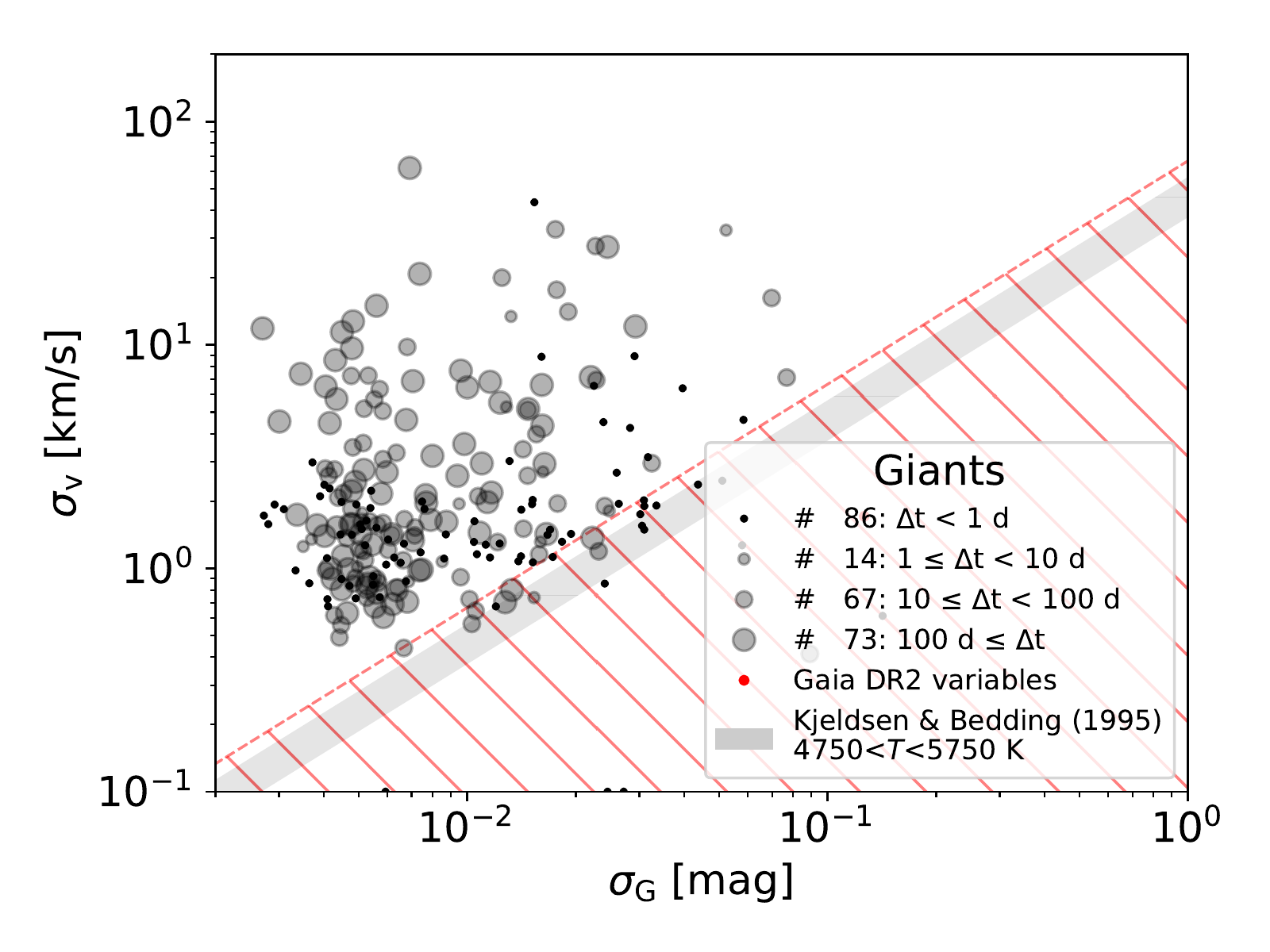}
 \caption{Same as Fig.~\ref {fig:sigmav_sigmam} but for set 2 (confidence level at $\sim3\sigma$).}
 \label{fig:sigmav_sigmam_set2}
\end{figure*}

\begin{figure}
    \centering
    \includegraphics[width=\linewidth]{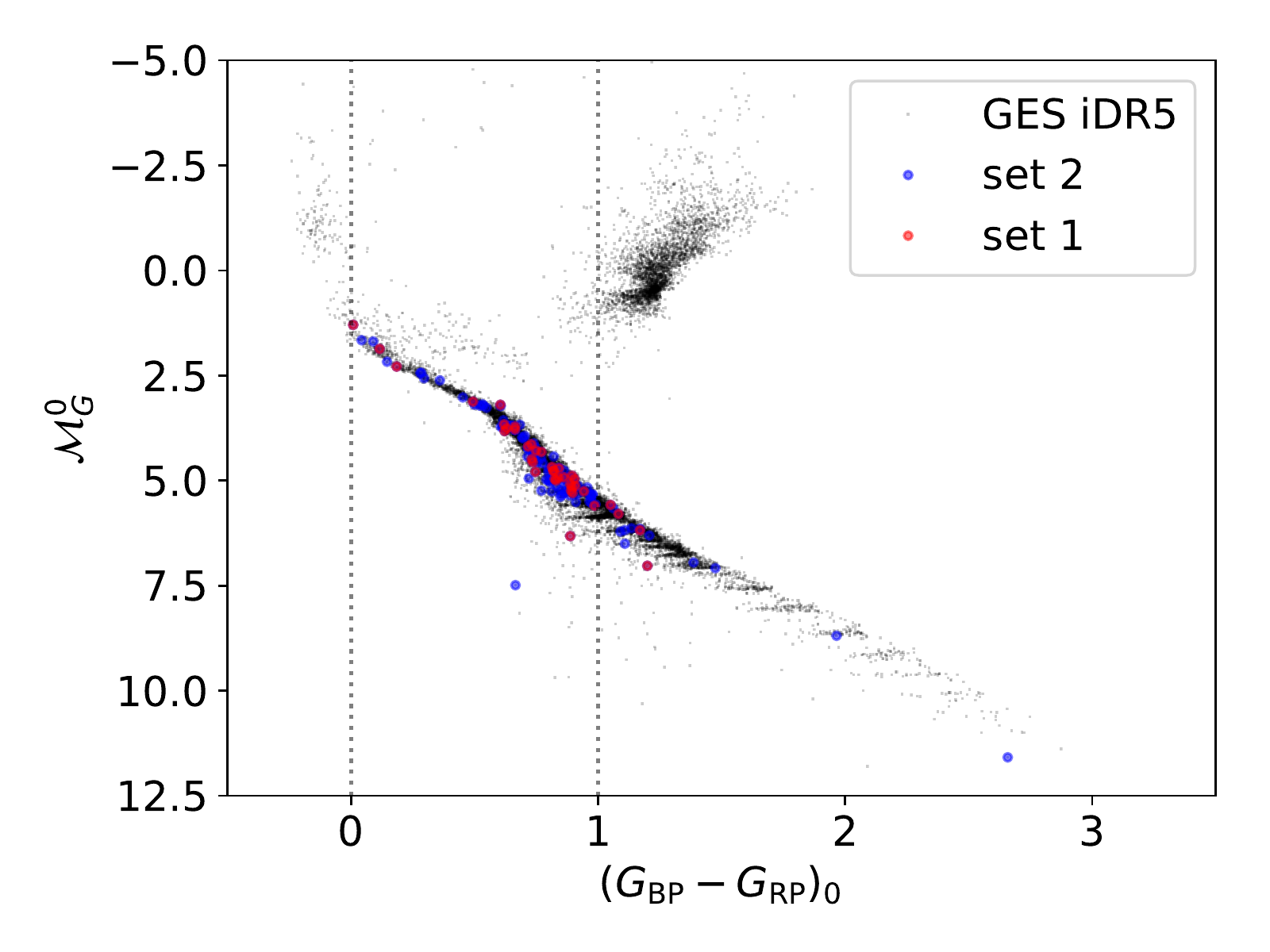}
    \caption{Dereddened colour--absolute magnitude diagram for dwarf RV variables that fall below the Br\'{e}ger criterion of Figs.~\ref{fig:sigmav_sigmam} and \ref{fig:sigmav_sigmam_set2}. All but a few fall in the $\delta$~Sct region defined on the main sequence with $0 \lessapprox(G_\mathrm{BP}-G_\mathrm{RP})_0 \lessapprox 1$ }
    \label{fig:dsct_hrd}
\end{figure}

Once the RV variables are split into these two subsamples (namely dwarfs and giants), we may represent their RV intrinsic dispersions $\sigma_v$ (Eq.~\ref{eq:sigma_v}) as a function of the photometric ones $\sigma_G$ (Eq.~\ref{eq:sigma_G}) as performed by \citet{famaey2009}; see their Fig.~3 for Hipparcos M giants. These graphs are presented in Figs.~\ref{fig:sigmav_sigmam} (set~1) and \ref{fig:sigmav_sigmam_set2} (set~2) for dwarfs (left panels) and giants (right panels). Among FGK main sequence  stars as considered in this paper, $\delta$~Sct pulsations are the major cause of photometrically induced RV variations detectable with the GES data. In fact, $\delta$~Sct variables are restricted to A and early-F dwarfs, that is, with $6\,500 \le T_{\rm eff} \mathrm{(K)} \le 10\,000$ \citep{murphy2019}, while late-F/G/K dwarfs undergo solar-like oscillations and spot-modulated rotational variability with substantially less RV jitter, which is mainly due to granulation. There is relatively little RV jitter associated with the latter physical processes occurring in late-F/G/K  dwarfs  \citep[from $\sim1$~m~s$^{-1}$ for quiet stars like our Sun to $\sim 500$~m~s$^{-1}$ for the youngest and most active solar-like main sequence stars;][]{lanza2016,meunier2017} as compared to the GES RV uncertainty (Sect.~\ref{sec:rv}). Therefore, these effects may be neglected (see the left panels of Figs.~\ref{fig:sigmav_sigmam}  and \ref{fig:sigmav_sigmam_set2}, which show that the lowest observed $\sigma_V$ values are still well above the above-mentioned jitter values for late-F/G/K dwarfs). On the contrary, in short-period $\delta$~Sct stars, \cite{breger1976} found that the ratio between the full amplitude RV variations and the $V$-band photometric variations  ranges from 50 to 125 \kms\ mag$^{-1}$, and must therefore be considered as possibly causing false positives in the search for SBs. These limits are indicated in the left panels of Figs.~\ref{fig:sigmav_sigmam} and \ref{fig:sigmav_sigmam_set2} as the lower and upper limits of the grey area, assuming that $\Delta G = \Delta V$. In the following we use a (conservative) threshold of 150~\kms\ mag$^{-1}$ below which RV variations are supposed to be associated with envelope pulsations. This means that all  RV variables located in the hatched area in Figs.~\ref{fig:sigmav_sigmam} and \ref{fig:sigmav_sigmam_set2} are not considered as SB1. We verified that the majority of these rejected stars are indeed located in the $\delta$~Sct instability strip  \citep[approximately $0 \le G_\mathrm{BP} - G_\mathrm{RP} \le 1$;][]{eyer2018}. We note that  \citet{andrae2018} are a little more conservative, as they locate the $\delta$~Sct instability strip in the range $0.1 \le G_\mathrm{BP} - G_\mathrm{RP} \le 0.6$, which corresponds to $6\,000 \le T_{\rm eff} \mathrm{(K)} \le 10\,000$, in good agreement with the results of  \citet{murphy2019}. Some of them have $ G_\mathrm{BP} - G_\mathrm{RP} > 1$  but turn out to be highly reddened;  applying dereddening (as mentioned above) brings most of them back into the $\delta$~Sct strip (see Fig.~\ref{fig:dsct_hrd}).
 
The few photometric variable stars not falling in the $\delta$~Sct strip could be ellipsoidal or eclipsing variables. Their photometric variability being associated with their binarity, they therefore ought not to be removed from the binary statistics. However, we believe that there are very few -- if any --  eclipsing binaries among the false SB  positives falling in the hatched area of Figs.~\ref{fig:sigmav_sigmam} and \ref{fig:sigmav_sigmam_set2}. First, eclipsing binaries must be close binaries, thus with large velocity amplitudes, which is not the case in the hatched area. Second, none of the six rejected SB candidates falling in the bulge were flagged as eclipsing binaries or ellipsoidal variables by the OGLE survey, although many were detected in their vicinity \citep{soszynski2016}.

Similarly, to identify RV variability arising from (low-amplitude) envelope pulsations in giant stars, we use a simple prediction from the linear theory of adiabatic acoustic oscillations \citep{kjeldsen1995}. This links RV variability and  photometric variability as follows (adapted from Eq.~6 of \citealt{jorissen1997}):
\begin{equation}
\sigma_v = 45.9\; \times \frac{673}{550}\left[\frac{T_\mathrm{eff}}{T_\mathrm{eff, \odot}}\right]^2 \sigma_G,
\label{eq:photoVr}
\end{equation}
where $\sigma_G$ must be expressed in mag and $\sigma_v$ in \kms. The factor 673/550 corresponds to the ratio between the effective wavelength of the \emph{Gaia} $G$ band  \citep[673~nm;][]{Jordi2010} and a reference wavelength (550 nm), \Teff\ is the effective temperature, and $T_{\rm eff, \odot}=5777$~K for the Sun. In the right panel of Figs.~\ref{fig:sigmav_sigmam} and \ref{fig:sigmav_sigmam_set2}, we define the grey shaded area using Eq.~\ref{eq:photoVr} with  \Teff\ =~5750~K and 4750~K for G- and K-type giants, respectively. As for the dwarfs, we define as a conservative limit (red dashed line) 1.2 times the upper limit of the shaded area. All stars falling below this limit are considered as pulsation-induced RV variables and are not counted as SB1.  

In all panels of Figs.~\ref{fig:sigmav_sigmam} and \ref{fig:sigmav_sigmam_set2}, the symbol size is related to the observation sampling: the larger the symbol, the larger the time span corresponding to the largest RV difference. Among dwarf stars, about 20\%  (resp. 50\%) of RV-variables are considered as photometric variables in set~1 (resp. set~2); for RV-variable giant stars, this rate decreases to 0\% (resp. 11\%) for set~1 (resp. set~2). Among the sets~1 and 2, only one cross-match exists with the General Catalogue of Variable Stars\footnote{When the full GES iDR5 is compared with the GCVS,  about 400 cross-matches are obtained, corresponding to less than 0.5\% of the GES iDR5 targets.}  \citep[GCVS;][]{samus2017} and remains after the cleaning for photometrically induced variability, namely V406~CMa (GES 06292710-3118285; red-star symbol in the left panels of Figs.~\ref{fig:sigmav_sigmam} and \ref{fig:sigmav_sigmam_set2}). This star is classified in GCVS as `$\gamma$~Dor:' variable. Its estimated photometric period is 0.768~d \citep[VSX catalogue,][]{watson2006}. Nevertheless, its SB1 nature seems well established because the maximum RV difference is of the order of 100~\kms\ over only 1~d. Another SB1 candidate, GES 12394853-3653483, is reported in the VSX catalogue\footnote{There are  1746 stars in common between the VSX catalogue and GES iDR5 when adopting a cone-search radius of 10 arcsec. The investigation of those photometric variables is however beyond the scope of this paper.} ---originally from the Catalina Sky Survey \citep{drake2017} -- as an eclipsing system of type EA ($\beta$ Persei-type, \emph{i.e.} Algol) with a photometric period of 1.438~d and an amplitude of 0.58 mag. 

There are less than 100 GES targets flagged as variable in the \emph{Gaia} DR2: 4 Cepheids, 40 RR~Lyr, 18 long period variables, 2 short-timescale and  36 rotation-modulation. Among the SB1 candidates, only a handful are flagged as variables in \emph{Gaia} DR2, as shown by the red dots in Figs.~\ref{fig:sigmav_sigmam} and \ref{fig:sigmav_sigmam_set2}.

When the confidence level of the statistical $\chi^2$ test is degraded from $5\sigma$ (set~1) to $3\sigma$ (set~2), the number of RV variables almost doubles (Table~\ref{tab:selection}). Nevertheless, when comparing Fig.~\ref{fig:sigmav_sigmam_set2} with Fig.~\ref{fig:sigmav_sigmam}, it turns out that the additional RV variables are mainly dwarfs falling in the $\delta$~Sct region and giants falling in the photometric variability band. Moreover, these additional RV variables mainly have a short-time sampling ($<1$~d, represented by black dots).

In summary, we find a contamination of the SB1s by RV variables supposedly due to $\delta$~Sct pulsations, jitter, and spots which amounts to about 18\% ($= 1 - 607 / 738)$ for set~1 and to 45\% ($= 1 - 718 / 1310)$ for set~2 (Table~\ref{tab:selection}). The evaluation of the contamination for set 3 was performed as for sets 1 and 2 using similar  figures and thresholds (separately for dwarfs and giants), and results in a correction factor of 91\% ($= 1 - 975 / 10479)$. Therefore, the 26\% ($=11316/43421$) of RV variables obtained in Sect.~\ref{Sect:RVvariables} for set~3 leads, after cleaning from photometric variability, to $\sim2.4\%$ ($=975/40551)$ as the raw (\emph{i.e.} not corrected for the detection efficiency) global fraction of SB1 detectable in the GES. The much larger actual SB1 fraction ($14.1\pm3.1$\%) will be obtained in Sect.~\ref{Sect:PDMF}, after evaluating the detection efficiency of SB1 by the GES.

\begin{figure*}
 \includegraphics[width=0.33\linewidth]{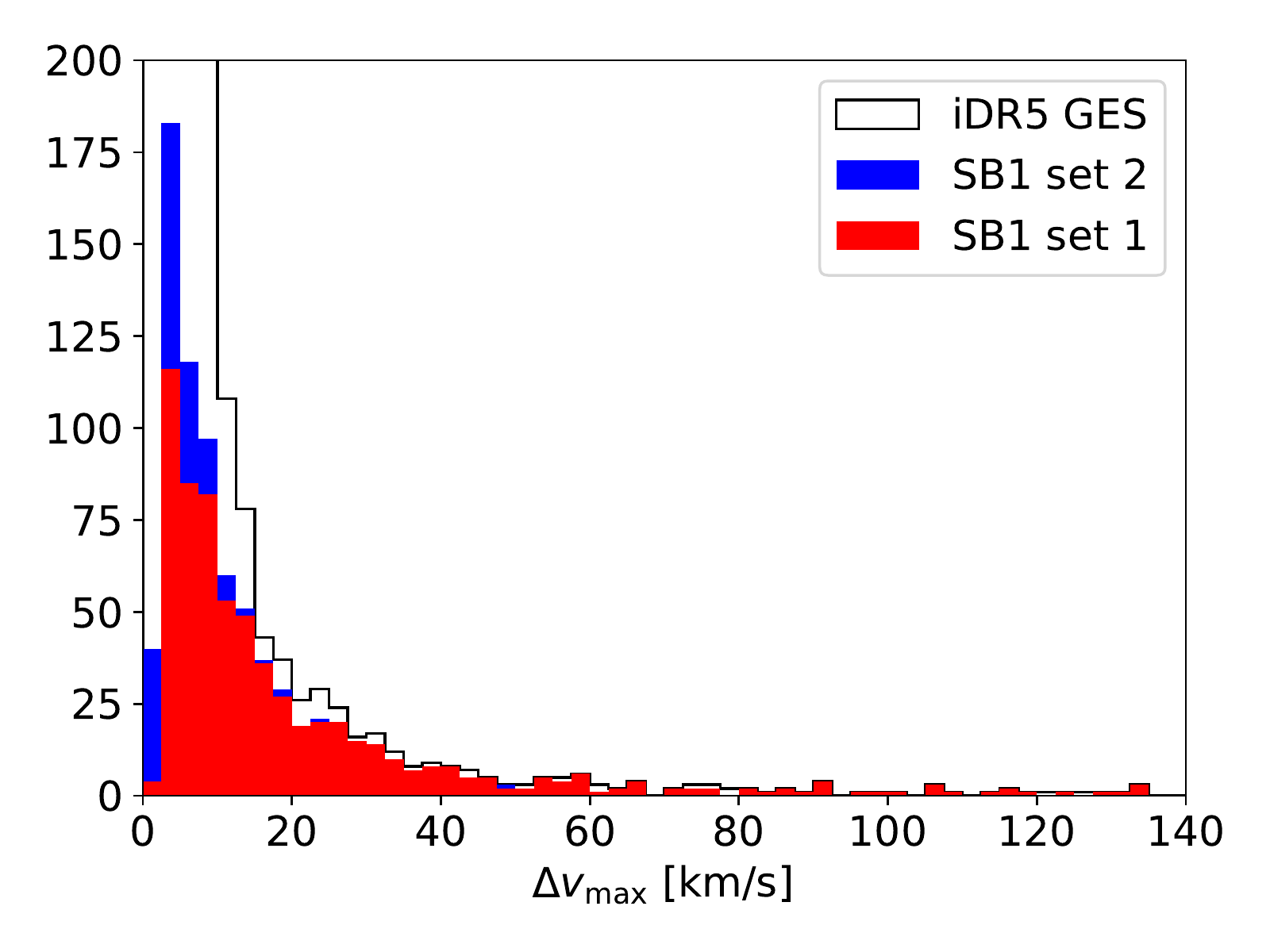}
 \includegraphics[width=0.33\linewidth]{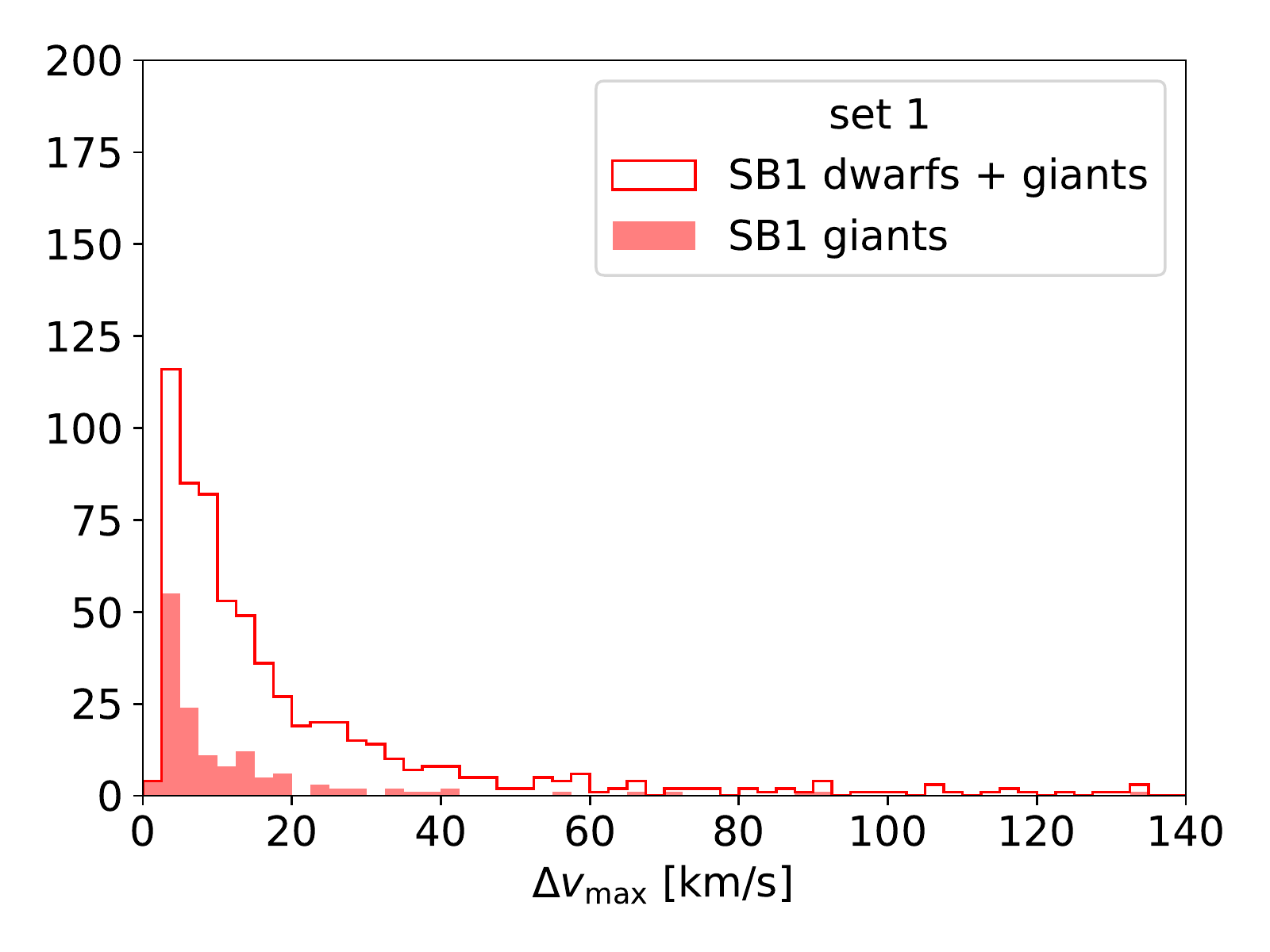}
 \includegraphics[width=0.33\linewidth]{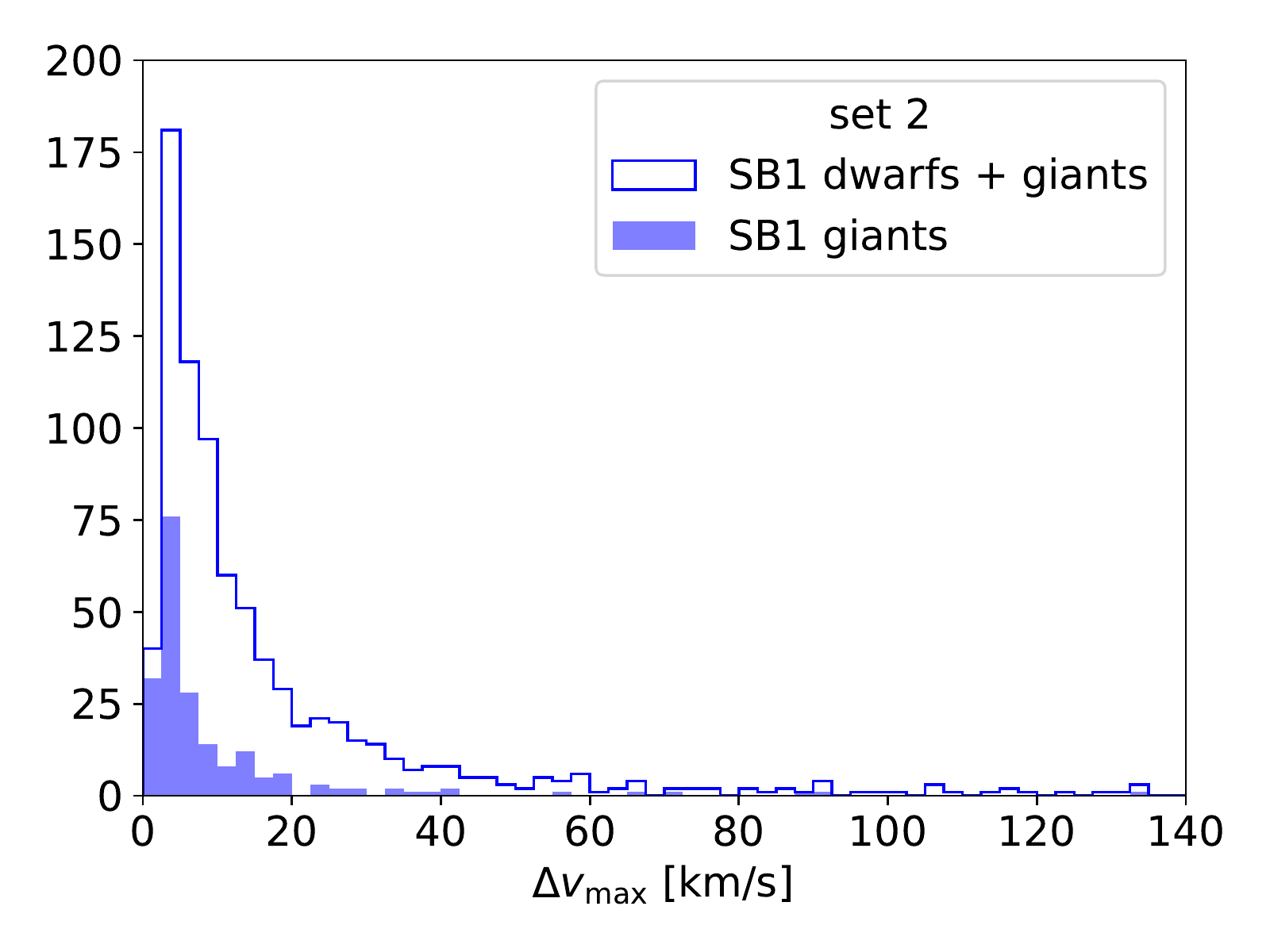}
 \caption{Left: RV range histograms for the analysed GES iDR5 sample (white) and for SB1 candidates at $\sim5\sigma$ (red, set~1) and at $\sim3\sigma$ (blue,  set~2). Middle: %\textbf{cumulative} 
 Corresponding histograms for SB1 set~1 separately for giants (filled red) and dwarfs+giants (red line). Right: Same as middle panel but for set~2.}
 \label{fig:histo_drvmax}
\end{figure*}

\begin{sidewaystable*}
\caption{Catalogue of SB1 candidates (set~1) ordered by increasing GES ID. $V$ is the visual magnitude, the `type' column is defined by the luminosity criterion (Eq.~\ref{eq:lc}), $\Delta v_{\max}$ is the difference between the highest and the lowest RV values, $\Delta t(\Delta v_{\max})$ is the corresponding time difference, $\bar{v}$ and $\sigma_{\bar{v}}$ are the mean and standard deviation of the $N$ RVs, $F2$ statistics is related to the $\chi^2$ and the degree of freedom number as given by Eq.~\ref{eq:F2},  $\sigma_v$ is the intrinsic RV dispersion as defined by Eq.~\ref{eq:sigma_v} and $\sigma_G$ is the intrinsic $G$ dispersion as defined by Eq.~\ref{eq:sigma_G}. The three last columns are the GES recommended parameters: effective temperature \Teff, surface gravity $\log{g}$, and metallicity [Fe/H]. The full table is available online.}
\footnotesize      

\begin{tabular}{rrrrrrrrrrrrrrr}
\hline\hline
GES ID & field & $V$ & type & $\Delta v_{\max}$ &  $\Delta t(\Delta v_{\max})$ & $\bar{v}$ & $\sigma_{\bar{v}}$  & $N$ & $F2$ & $\sigma_v$ & $\sigma_G$ & \Teff & $\log{g}$ & [Fe/H] \\
       &       &   &      & [\kms]            & [d]                          &     [\kms] & [\kms]             &     &      & [\kms]     &  [mag]          &   [K]      &           & \\
\hline
$00012323-5956412$ & -            & 16.80 & dwarf & $    24.8$ & $    0.092$ & $    74.1$ &   12.4 &  2 &  12.461  &   12.3 &    0.020 &- &- & $ 0.04\pm0.22$ \\
$00033430-0108236$ & -            & 13.20 & dwarf & $     9.2$ & $ -379.070$ & $     2.3$ &    3.7 &  6 &  16.140  &    3.7 &    0.006 & $4660\pm  57$ & $4.64\pm0.12$ & $-1.15\pm0.02$ \\
$00034629-0045387$ & -            & 14.30 & dwarf & $     5.6$ & $  371.015$ & $    -9.0$ &    2.7 &  4 &   9.460  &    2.7 &    0.016 &- &- & - \\
$00035138-0108589$ & -            & 16.10 & dwarf & $    10.7$ & $  379.070$ & $    56.3$ &    4.0 &  5 &   8.304  &    3.9 &    0.019 & $5177\pm 233$ & $4.74\pm0.24$ & $ 0.19\pm0.29$ \\
$00035355-4701511$ & -            & 16.00 & dwarf & $    27.3$ & $   31.921$ & $   -79.0$ &   13.3 &  4 &  29.533  &   13.2 &    0.014 &- &- & - \\
$00035356-0102018$ & -            & 14.40 & dwarf & $     8.2$ & $ -379.070$ & $  -193.3$ &    3.8 &  5 &  13.880  &    3.8 &    0.010 & $5880\pm 134$ & $4.98\pm0.18$ & - \\
$00040670-4710273$ & -            & 14.20 & dwarf & $     9.0$ & $  -31.921$ & $    -0.4$ &    4.1 &  5 &  16.274  &    4.1 &    0.008 & $4706\pm 163$ & $4.67\pm0.20$ & - \\
$00092680-5001133$ & -            & 15.60 & dwarf & $    24.8$ & $ -264.263$ & $    -4.0$ &    9.7 &  5 &  32.832  &    9.7 &    0.015 &- &- & - \\
$00092897-4953026$ & -            & 16.60 & dwarf & $    15.4$ & $  264.263$ & $   130.3$ &    7.3 &  4 &  12.027  &    7.2 &    0.023 & $4854\pm 264$ & $4.14\pm0.27$ & - \\
$00094700-0326395$ & -            & 14.80 & dwarf & $     4.2$ & $ -263.179$ & $    34.7$ &    1.8 &  4 &   7.293  &    1.8 &    0.011 & $6182\pm 154$ & $4.12\pm0.20$ & $-1.39\pm0.32$ \\
$00221531-7207097$ & NGC104       & 13.87 & giant & $    14.9$ & $-2474.212$ & $    -8.5$ &    7.4 &  2 &  15.541  &    7.4 &    0.003 &- &- & - \\
$00224888-7205257$ & NGC104       & 14.88 & giant & $     9.4$ & $ 3539.305$ & $    -8.6$ &    2.5 & 11 &  13.274  &    2.4 &    0.005 & $4923\pm  38$ & $2.82\pm0.09$ & $-0.48\pm0.40$ \\
$00230185-7203068$ & NGC104       & 14.28 & giant & $     9.0$ & $-2474.212$ & $   -20.8$ &    4.5 &  2 &  10.953  &    4.5 &    0.004 &- &- & - \\
$00231071-7211351$ & NGC104       & 13.70 & giant & $    16.6$ & $-3668.893$ & $   -16.9$ &    4.6 & 10 &  21.546  &    4.5 &    0.003 & $4670\pm  33$ & $2.51\pm0.08$ & - \\
$00231325-7209375$ & NGC104       & 14.63 & giant & $     7.9$ & $ 3565.329$ & $   -10.3$ &    2.3 & 11 &  16.242  &    2.2 &    0.005 & $4806\pm  75$ & $2.75\pm0.16$ & $-0.76\pm0.28$ \\
\dots &\dots &\dots &\dots &\dots &\dots &\dots &\dots &\dots &\dots &\dots &\dots &\dots &\dots &\dots \\
\hline
\end{tabular}  
\label{tab:results1}
\vspace{1cm}
\caption{Catalogue of the additional SB1 candidates that make up set~2, along with the ones in Table~\ref{tab:results1} ordered by increasing GES ID. Headers are detailed in the caption of Table~\ref{tab:results1}. The full table is available online.}
\footnotesize      
\begin{tabular}{rrrrrrrrrrrrrrr}
\hline\hline
GES ID & field & $V$ & type & $\Delta v_{\max}$ &  $\Delta t(\Delta v_{\max})$ & $\bar{v}$ & $\sigma_{\bar{v}}$  & $N$ & $F2$ & $\sigma_v$ & $\sigma_G$ & \Teff & $\log{g}$ & [Fe/H] \\
       &       &   &      & [\kms]            & [d]                          &     [\kms] & [\kms]             &     &      & [\kms]     &   [mag]         &   [K]      &           &                                \\
\hline
$00001461-5451588$ & -            & 14.70 &       & $     2.0$ & $   -0.018$ & $    70.1$ &    0.8 &  4 &   3.679  &    0.7 &    0.083 & $5989\pm  61$ & $4.43\pm0.12$ & $ 0.04\pm0.22$ \\
$00032437-3702589$ & -            & 14.90 &       & $     4.2$ & $    0.073$ & $    12.7$ &    1.9 &  3 &   3.463  &    1.7 &    0.072 &- &- & $-1.15\pm0.02$ \\
$00040478-3656202$ & -            & 13.20 & dwarf & $     2.6$ & $   -0.113$ & $   -20.5$ &    1.0 &  4 &   3.924  &    0.8 &    0.005 & $5685\pm  49$ & $4.12\pm0.11$ & - \\
$00040489-0056005$ & -            & 13.30 & dwarf & $     2.5$ & $  371.033$ & $   -31.9$ &    1.0 &  6 &   4.467  &    0.9 &    0.005 & $5106\pm  36$ & $3.38\pm0.09$ & $ 0.19\pm0.29$ \\
$00223198-7206489$ & NGC104       & 12.88 & giant & $     2.9$ & $ 3988.090$ & $   -14.6$ &    1.5 &  2 &   4.125  &    1.4 &    0.004 & $4282\pm  57$ & $1.76\pm0.11$ & - \\
$00244100-7158187$ & NGC104       & 13.30 & giant & $     2.2$ & $-1513.778$ & $   -19.2$ &    0.9 &  3 &   3.219  &    0.8 &    0.006 & $4505\pm 102$ & $2.31\pm0.14$ & - \\
$00250427-7206399$ & NGC104       & 12.25 & giant & $     2.2$ & $-2474.311$ & $   -16.0$ &    1.1 &  2 &   3.280  &    1.0 &    0.007 & $3996\pm 104$ & $1.27\pm0.16$ & - \\
$00250679-7203152$ & NGC104       & 13.43 & giant & $     2.4$ & $-2474.311$ & $    -9.5$ &    1.2 &  2 &   3.841  &    1.1 &    0.005 &- &- & - \\
$00263499-7206444$ & NGC104       & 13.39 & giant & $     1.9$ & $-3988.090$ & $   -11.8$ &    0.8 &  3 &   3.057  &    0.7 &    0.006 &- &- & - \\
$00294333-5002041$ & -            & 17.40 &       & $     5.0$ & $    0.025$ & $    90.5$ &    2.1 &  3 &   3.508  &    0.1 &    0.036 & $4994\pm 212$ & $3.68\pm0.28$ & $-1.39\pm0.32$ \\
$00382912-6000310$ & -            & 16.10 &       & $     3.5$ & $   23.911$ & $   231.8$ &    1.4 &  4 &   3.856  &    1.3 &    0.014 & $5100\pm 151$ & $2.70\pm0.20$ & - \\
$00400135-3708024$ & -            & 15.80 & dwarf & $     7.6$ & $    0.018$ & $    48.0$ &    2.8 &  4 &   4.537  &    2.5 &    0.015 & $5537\pm 187$ & $3.92\pm0.24$ & $-0.48\pm0.40$ \\
$01012143-7053118$ & NGC362       & 15.48 & giant & $     2.3$ & $   -0.933$ & $   220.5$ &    1.2 &  2 &   3.128  &    1.1 &    0.007 &- &- & - \\
$01031990-7046247$ & NGC362       & 15.47 & giant & $     2.8$ & $   -0.933$ & $   220.4$ &    1.4 &  2 &   3.977  &    1.3 &    0.006 &- &- & - \\
$01042164-7047555$ & NGC362       & 15.58 & giant & $     5.1$ & $   40.009$ & $   221.8$ &    1.6 &  7 &   3.576  &    1.5 &    0.007 & $5073\pm  41$ & $2.75\pm0.10$ & $-0.76\pm0.28$ \\
\dots &\dots &\dots &\dots &\dots &\dots &\dots &\dots &\dots &\dots &\dots &\dots &\dots &\dots &\dots \\
\hline
\end{tabular}  
\label{tab:results2}

\end{sidewaystable*}

\subsection{Properties of SB1 candidates}

\subsubsection{Orbital properties}

%We now characterise the SB1 candidates from sets~1 and 2 selected after the photometric cleaning described above, as a function of their dwarf or giant nature (Table~\ref{tab:selection}). 
We obtain 641 SB1 candidates in set~1 (Table~\ref{tab:results1}). This number is extended by 162 additional SB1 candidates when  considering set~2 (Tables~\ref{tab:results1} and \ref{tab:results2}).  
Our GES iDR5 subsample counts 43\,421 stars, thus leading to an SB1 detection rate of 1.5\% and 1.8\% at the $5\sigma$ and $3\sigma$ confidence levels, respectively. 
We note that this binary frequency, in both confidence levels, is much lower than that obtained by other studies. We therefore also evaluate our SB1 detection efficiency in Sect.~\ref{sec:efficiency}.

To distinguish between main sequence stars (dwarfs) and red or asymptotic giant branch stars (giants), we prefer to rely on the more precise luminosity criterion defined in Sect.~\ref{sec:photo} (Eq.~\ref{eq:lc}) and based on \emph{Gaia} DR2 compared to the less precise GES recommended $\log{g}$. The absolute $G$-magnitude estimate can be computed for 718 of the 803 SB1 candidates, allowing us to distinguish between giants and dwarfs. About a quarter of SB1 candidates are giant stars. The left panel of Fig.~\ref{fig:histo_drvmax} shows the distribution of the RV range for each target among the analysed subsample of the GES iDR5 (open histogram), and among the SB1 candidates in sets~1 and 2. We obtain the maximum RV range of 133~\kms~ for the GES ID 18503579-0620339 ($V=13.2$) belonging to the massive open cluster NGC~6705 (M~11) located in the Galactic plane. The two recorded velocities are $32.5\pm0.8$~\kms\ at MJD 56103.1096 and $-100.5\pm4.6$~\kms\ at MJD 56442.4004. \citet{cantat2014} found a probability of 92\% for this star to be a member of M~11 which has a cluster velocity of $34.8\pm0.4$~\kms. The 2MASS image shows another star $\sim2.2$~arcsec away. There are no recommended atmospheric parameters provided by the GES for this SB1 candidate, but we classified it as a giant according to its \emph{Gaia} absolute magnitude. This conclusion is puzzling though, since a giant cannot be hosted by a system close enough to exhibit such a large  RV amplitude except if this object hosts a compact invisible companion \citep[see, \emph{e.g.}][]{zwitter1993},  or a pair of low-mass stars in a A-(Ba,Bb) hierarchical triple system. We decided to keep this object among the SB1 candidates.

The minimum RV range among SB1 candidates is 2.2 and 1.6~\kms\ for sets~1 and 2, respectively. We clearly see on the left panel of Fig.~\ref{fig:histo_drvmax} that set~2 adds SB1 candidates with low RV ranges since the confidence level is lower for that set. The middle and right panels of Fig.~\ref{fig:histo_drvmax}  separate the distribution of RV range for SB1 candidate dwarfs and giants. The tail of the distribution for giant stars barely reaches 40~\kms, but exceeds 100~\kms\ for dwarfs. This expected behaviour is indicative of longer periods for SB1s with a giant primary component. This trend is also reported in \citet{badenes2018}, who thoroughly investigated the maximum RV range as a function of surface gravity and mass. They found a similar difference between dwarfs and giants.

\begin{figure*}
\centering
    \includegraphics[width=0.49\linewidth]{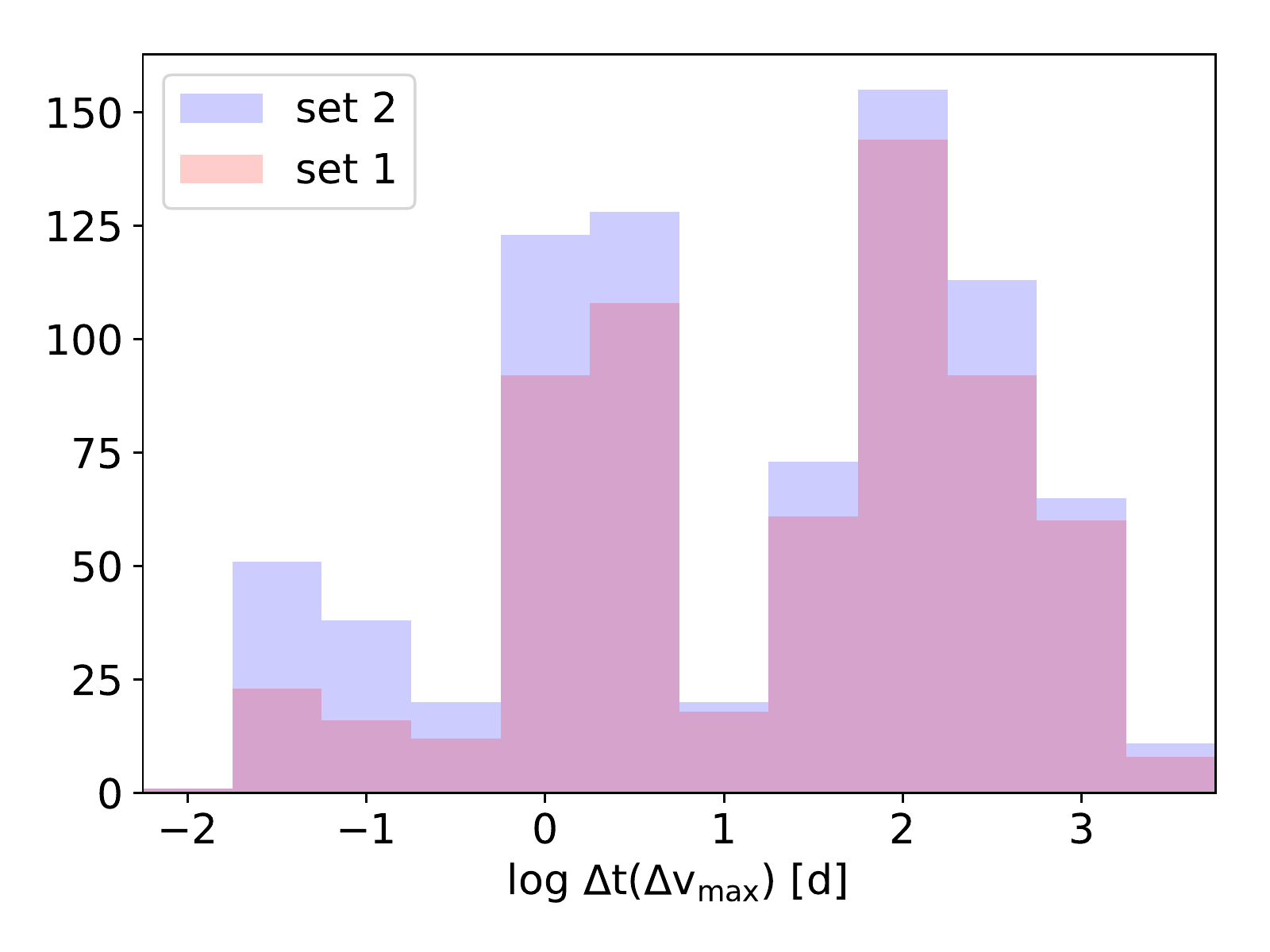}
    \includegraphics[width=0.49\linewidth]{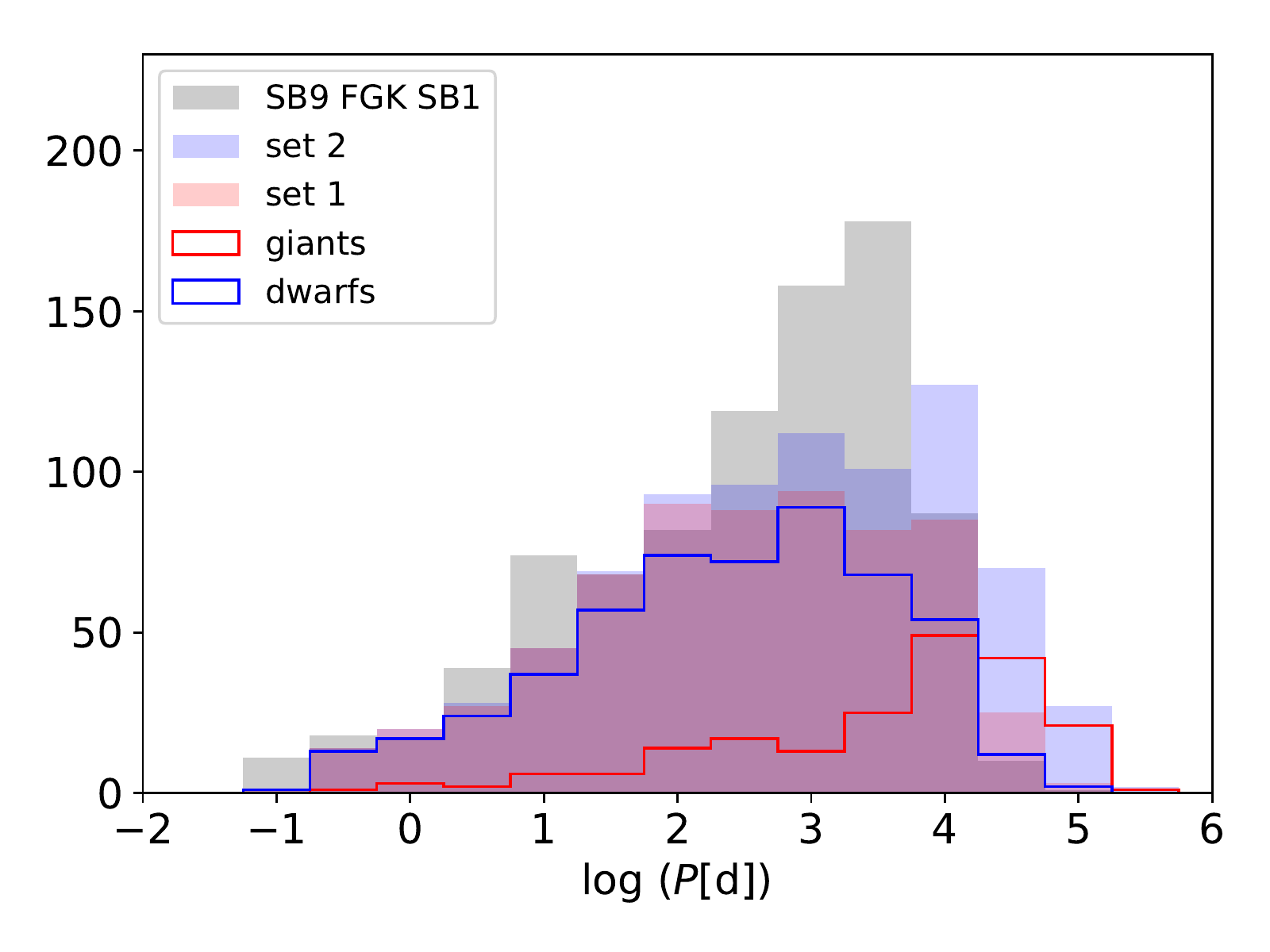}
    \caption{Left: Distribution of the time sampling corresponding to the maximum RV differences (RV range)  for SB1 candidates of the sets~1 (red) and 2 (blue). Right: Distribution of the SB1 orbital periods assuming that the amplitude $K$ of the RV curve may be derived from Eq.~\ref{eq:period} for sets~1 and~2 (red and blue filled histograms) and for SB1 dwarfs and giants (from set~2, blue and red empty histograms). The distribution of FGK-type SB1s from the SB9 catalogue \citep{pourbaix2004} is overplotted in grey.}
    \label{fig:histo_dtl}
\end{figure*}

\begin{table*}[]
    \centering
        \caption{Tentative orbital parameters and GES recommended atmospheric parameters for two SB1 dwarfs. 
        Masses are estimated from Eq.~(\ref{eq:m_teff}).}
    \begin{tabular}{rrr}
    \hline\hline
        GES ID      & 21303728+1202029$^a$ & 12575381-7053113 \\
        field & M~15 & NGC~4833\\
        \emph{Gaia} DR2 source ID & 1745931560476128000 & 5843793495793252992\\
        $G$ & 15.72 & 16.02 \\ 
        \hline
        $P$ [d] &$5.8$ & $3.9147\pm0.0002$\\
        $e$ & $0.3$& $0.26\pm0.02$\\
        $\omega$ [$^\circ$] & $82$ & $187.5\pm1.6$ \\
        $T_0$ - 2\,400\,000.5 [d] & $56241.2$& $56478.9\pm1.4$\\
        $v_0$ [\kms] & $17.7$& $-2.63\pm0.26$ \\
        $K_1$ [\kms] & $5.6$& $69.82\pm0.95$\\
        $f(M)$ [M$_\odot$] & $-$ & $0.124\pm0.006$\\
        $\sigma_1$ (O-C) [\kms] & 0.324& 0.410\\
        $N_\mathrm{obs}$ & 13& 10\\
        $N_\mathrm{epochs}$ & 4 & 5\\ 
        \hline
        \Teff\ [K]& $5754\pm58$ & $5318\pm170$  \\
        $\log{g}$  & $4.52\pm0.12$ & $4.13\pm0.21$ \\
         $\mathrm{[Fe/H]}$ & $-0.42\pm0.26$ & - \\
        \hline
        $M_1$ [M$_\odot$] & $1.04\pm0.03$ & $0.88\pm0.07$ \\
        \hline
    \end{tabular}
    
    $^a$ Only orbital parameters are given here as a tentative solution. More observations are needed for this star.
    \label{tab:orb}
\end{table*}

\begin{figure*}
    \centering
    \includegraphics[width=\linewidth]{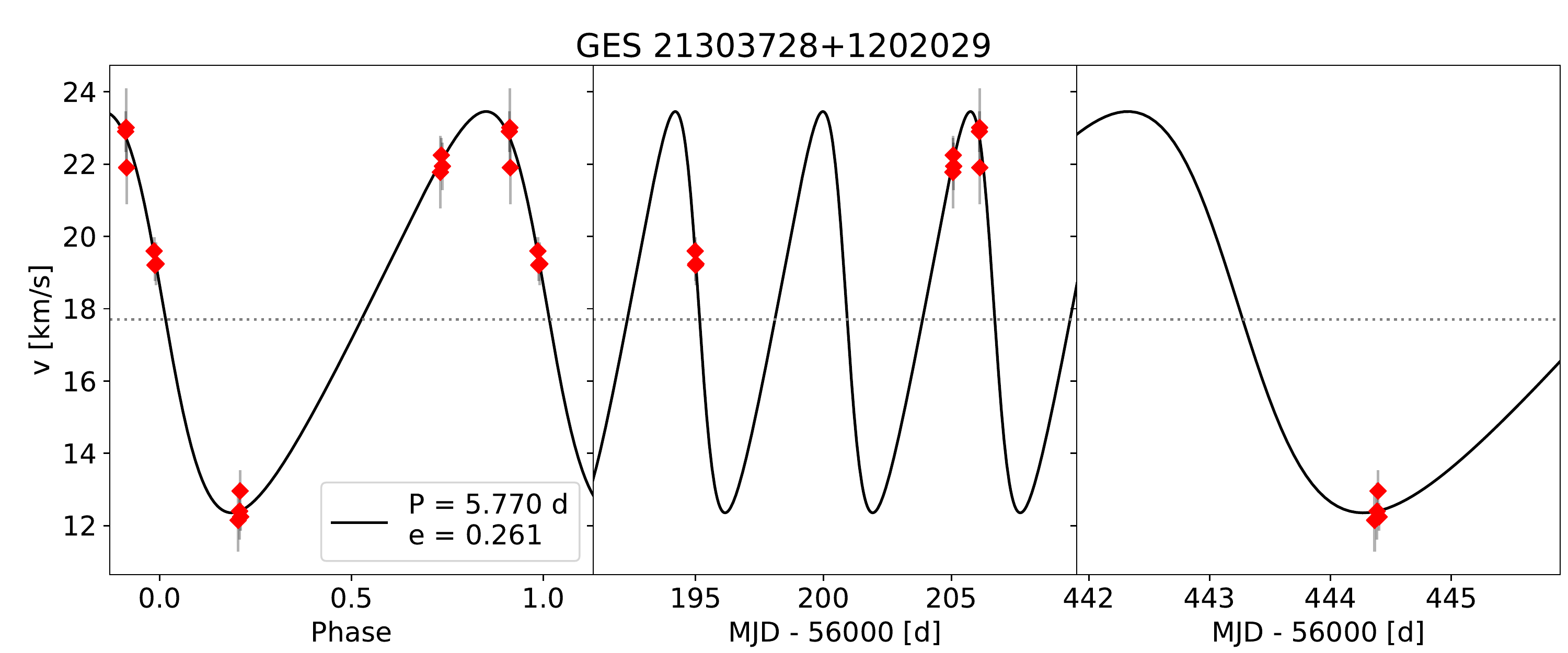}
    \includegraphics[width=\linewidth]{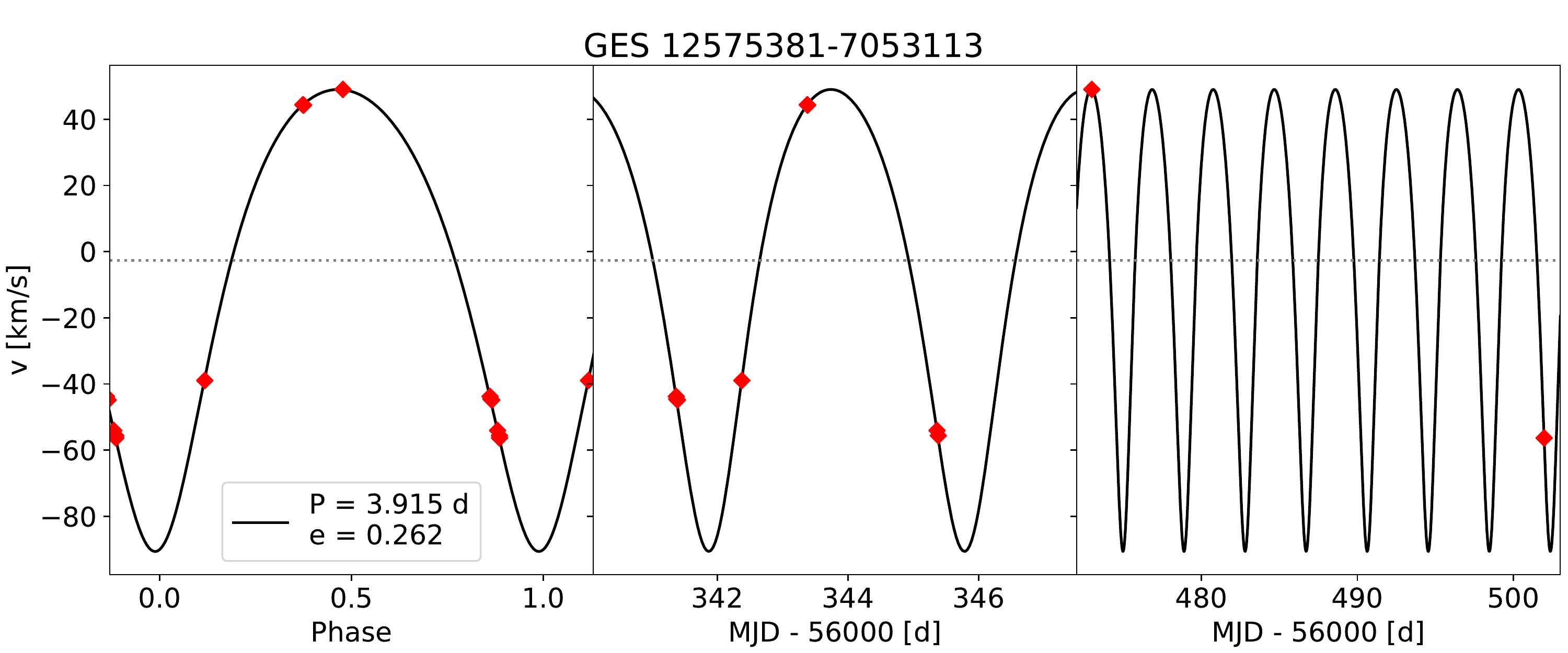}
    \caption{Two SB1s with tentative orbits. Radial velocity as a function of the phase (left panels) and as a function of MJD on the middle and right panels for GES ID 21303728+1202029 in M~15 (top panels) and GES ID 12575381-7053113 in NGC~4833 (bottom panels).}
    \label{fig:sb1_orbits}
\end{figure*}

The GES observation time sampling is irregular and suffers from a deficit between $\Delta t = 8$~d and $\Delta t = 20$~d as seen in the middle panel of Fig.~\ref{fig:dtmax}, which is reflected in the distribution of the time sampling for SB1 candidates shown in left panel of Fig.~\ref{fig:histo_dtl}. Indeed, the distribution of the sampling is spread from 1 to 1000~d with a significant  deficit around 10~d. More quantitatively, about 20\% of the detected SB1s have a maximum time span lower than 1~d, 20\% have a maximum time spac of between 1 and 10 d, and 60\% have a maximum time span larger than 10~d. 

Though the GES time sampling is too scarce to derive orbital periods, we can estimate their distribution from the RV amplitude $K$. Indeed, assuming a sinusoidal RV curve and a dense and uniform sampling, the standard deviation of the RV measurements is related to the amplitude by $K = \sqrt{2} \sigma_v$. While these conditions do not hold in the case of the GES, $ \sqrt{2} \sigma_v$ can still be used as a proxy for $K$, allowing a relative statistical comparison of periods between SB1s with dwarf and giant primaries. Assuming a mass of $M_1=1$~M$_\odot$ for the visible component and a mass ratio $q=0.25$, we then have:
\begin{equation}
\label{eq:period}
    P = 9.650\times 10^4 \frac{1}{K^3} \frac{\sin^3 i}{(1-e^2)^{3/2}} 
\end{equation}
where $P$ is expressed in days, $K$ in \kms~ and $i$ is the orbital inclination. Adopting the most probable inclination of $68^\circ$ (corresponding to random inclinations on the sky) and an average eccentricity $e=0.2$ (corresponding to the median eccentricity of FGK-type SB1s in \citealt{pourbaix2004}), we get the distributions represented in the right panel of Fig.~\ref{fig:histo_dtl}. In this figure, the blue (resp., red) filled histograms correspond to sets 1 (resp., 2) while the blue (resp., red) lines correspond to the distributions of dwarfs (resp., giants) from set~2. 

Among systems with giant primaries (red line), there is a clear lack of short-period systems with respect to systems with dwarf primaries (blue line). Indeed, giant primaries cannot fit into orbits with too short a period (typically $P \la 100$~d) without overflowing their Roche lobe. In addition we notice that the maximum period of giants is longer than that of dwarfs. This bias might be caused by our procedure to remove false positives (Sect.~\ref{sec:photo}, Figs.~\ref{fig:sigmav_sigmam} and \ref{fig:sigmav_sigmam_set2}), leading us to keep SB1 giant candidates with smaller $\sigma_v$ (hence longer periods) than SB1 dwarf candidates.

For comparison purposes we overplot in Fig.~\ref{fig:histo_dtl} (right panel) the period distribution of FGK-type SB1 of the SB9 catalogue \citep{pourbaix2004}. Although the global shapes of the SB9 and GES period distributions are similar, there is a shift of the histogram of GES estimated periods towards larger values. This might be due to the quite irregular and scarce sampling of GES RV, inducing too small a $K$ when derived from $\sigma_v$ and thus producing overly large periods. Moreover, when the sampling is scarce and the orbit is eccentric, the estimated orbital period (assuming a circular orbit) tends to be further overestimated. Correcting for these effects would be delicate, given the irregular GES RV sampling. With these effects in mind, we can give an upper limit of the probed orbital periods of $\log{P[\text{d}]} \lessapprox 4$.

Finally, we report on our attempts to derive orbital parameters for some of the SB1s by fitting Keplerian orbits to their RV time series. The SB1 candidates counting the largest number of observations were observed in stellar clusters and in the CoRoT field. We identified about 20 SB1 candidates with at least five epochs, but we could find reasonable orbital solutions for only two of them (both belonging to set~1) with periods shorter than one week. GES ID 21303728+1202029 is located in M~15 (but is not considered as a member, \citealt{drukier1998}) and GES ID 12575381-7053113 in  NGC~4833; both have \emph{Gaia} DR2 identifiers (see Table~\ref{tab:orb}). We show in Fig.~\ref{fig:sb1_orbits} the observed RVs along with the calculated orbits as a function of either orbital phase (left panels) or time (middle and right panels). The orbital parameters for these two SB1s are listed in Table~\ref{tab:orb} together with the mass function $f(M)$, the standard deviation of the residuals $\sigma_1(\mathrm{O-C})$, the number $N_\mathrm{obs}$ of GES RVs used in the fit of the orbit and the corresponding number of epochs $N_\mathrm{epochs}$. The periastron time is given in MJD. The uncertainties on the orbital parameters of GES ID 21303728+1202029 are quite large due to poor phase coverage, indicating that the orbit must be considered as preliminary. The only reported RV for this latter source is $12.58\pm0.46$~km/s \citep{drukier1998}, in good agreement with the obtained orbital solution. In both cases, observations span over two seasons more than 150 days apart, which corresponds to a large number of orbits and can produce aliasing effects. The Lomb-Scargle periodogram of GES 12575381-7053113 shows that the orbital period is $\sim4\pm2$~d. This uncertainty is more realistic than the one given in Table~\ref{tab:orb} which is the intrinsic uncertainty on the fitted parameter.

\begin{figure*}
  \includegraphics[width=0.5\linewidth]{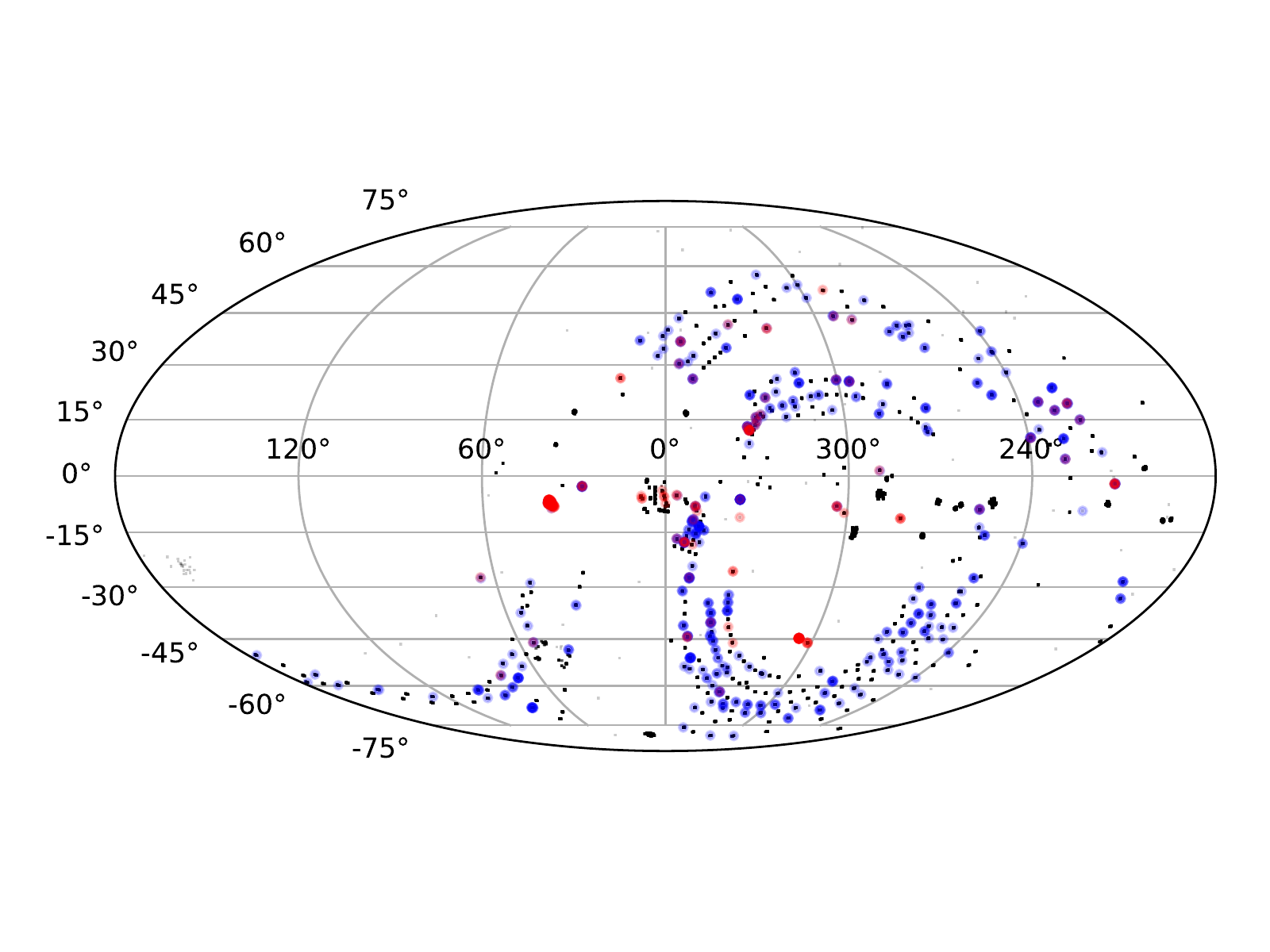}
  \includegraphics[width=0.5\linewidth]{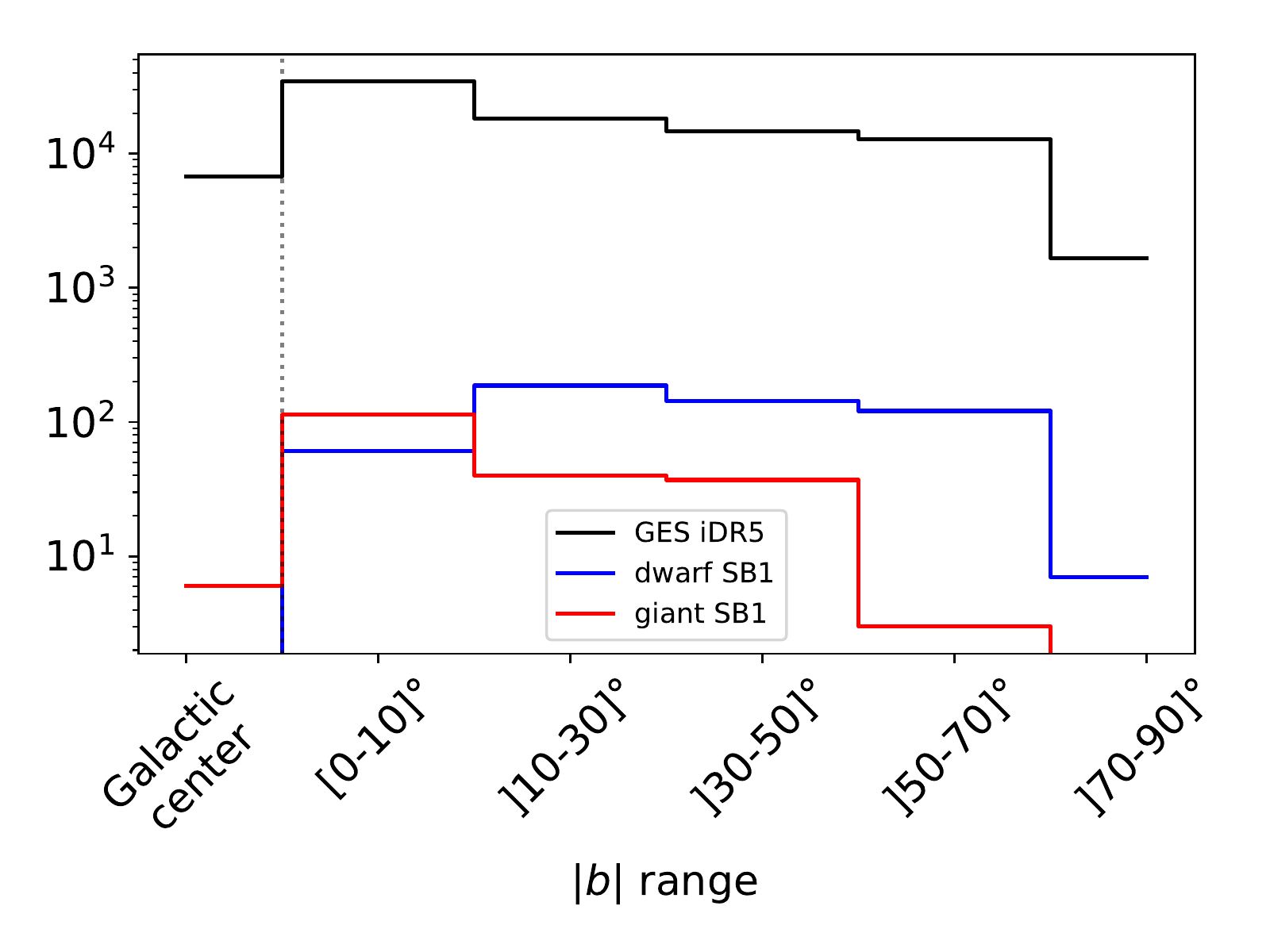}
\caption{
Left: Mollweide projection of the GES iDR5 subsample (small grey pixels), with SB1 dwarf (large blue dots) and giant (large red dots) candidates from set~2 in Galactic coordinates. The darker a given region, the denser the observations in this region. Right: Number of SB1 dwarfs (blue) and SB1 giants (red) within the specified Galactic latitude range excluding the Galactic centre. The first bin shows the same statistics but within 20$^\circ$ around the Galactic centre. GES iDR5 targets are also shown (black line).}
\label{fig:mollweide_dist}
\end{figure*}

\begin{figure}
    \centering
 \includegraphics[width=\linewidth]{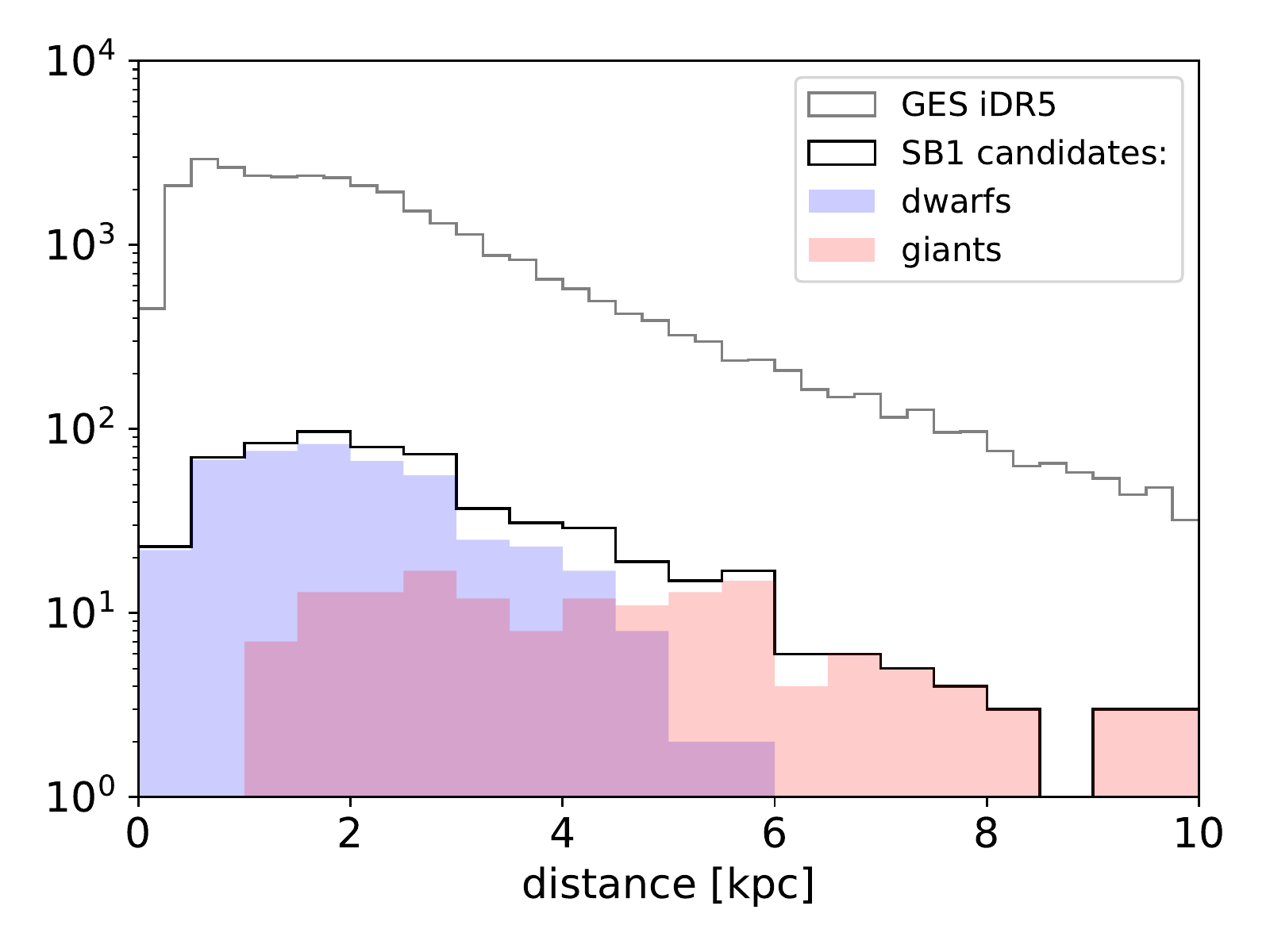}
\caption{Right: \emph{Gaia} distances for stars in the analysed GES iDR5 subsample (grey),  and those classified as dwarf (blue) and giant (red) SB1 candidates from set 2 (black).}
    \label{fig:sb1_dist}
\end{figure}

\begin{figure*}
\includegraphics[width=0.33\linewidth]{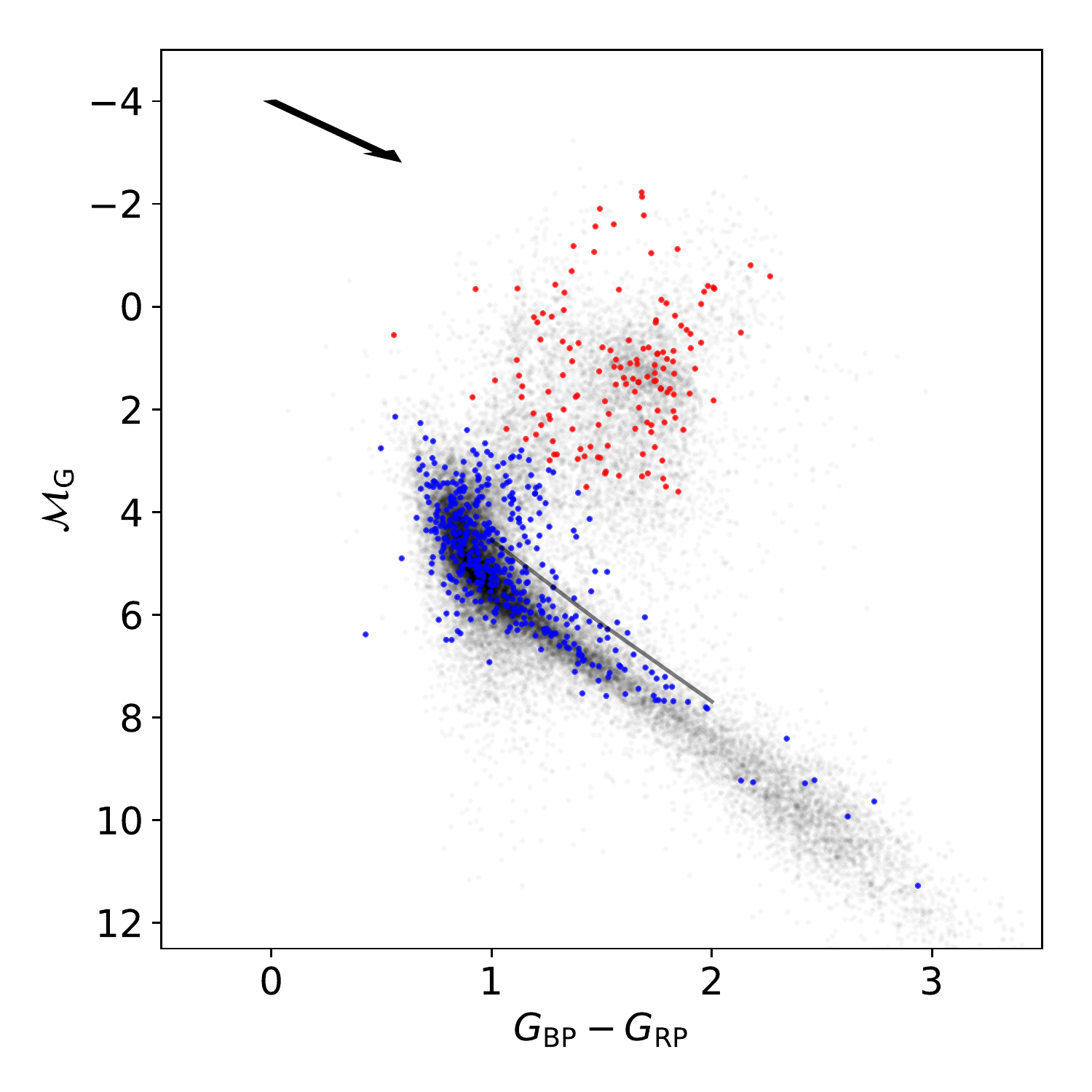}
\includegraphics[width=0.33\linewidth]{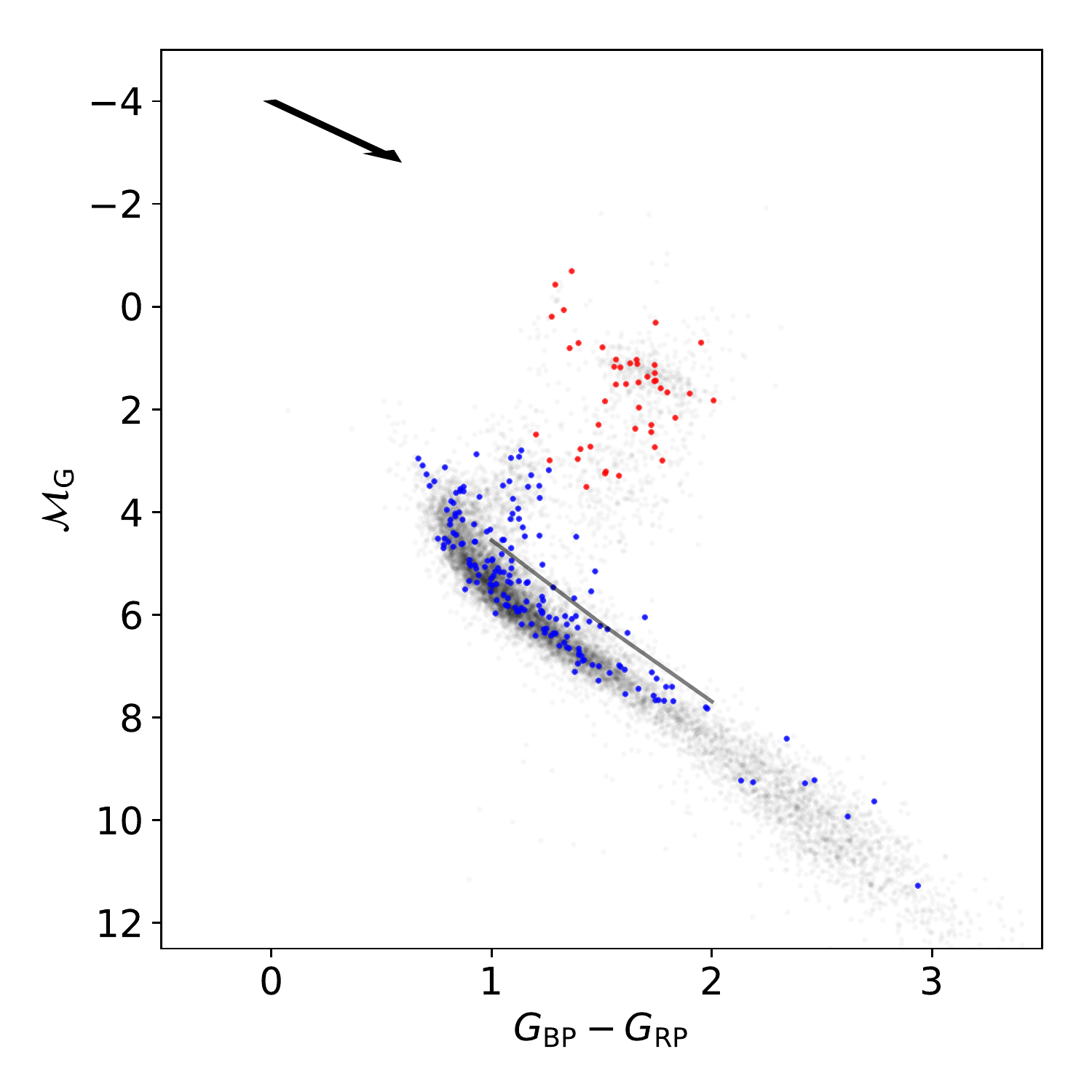}
\includegraphics[width=0.33\linewidth]{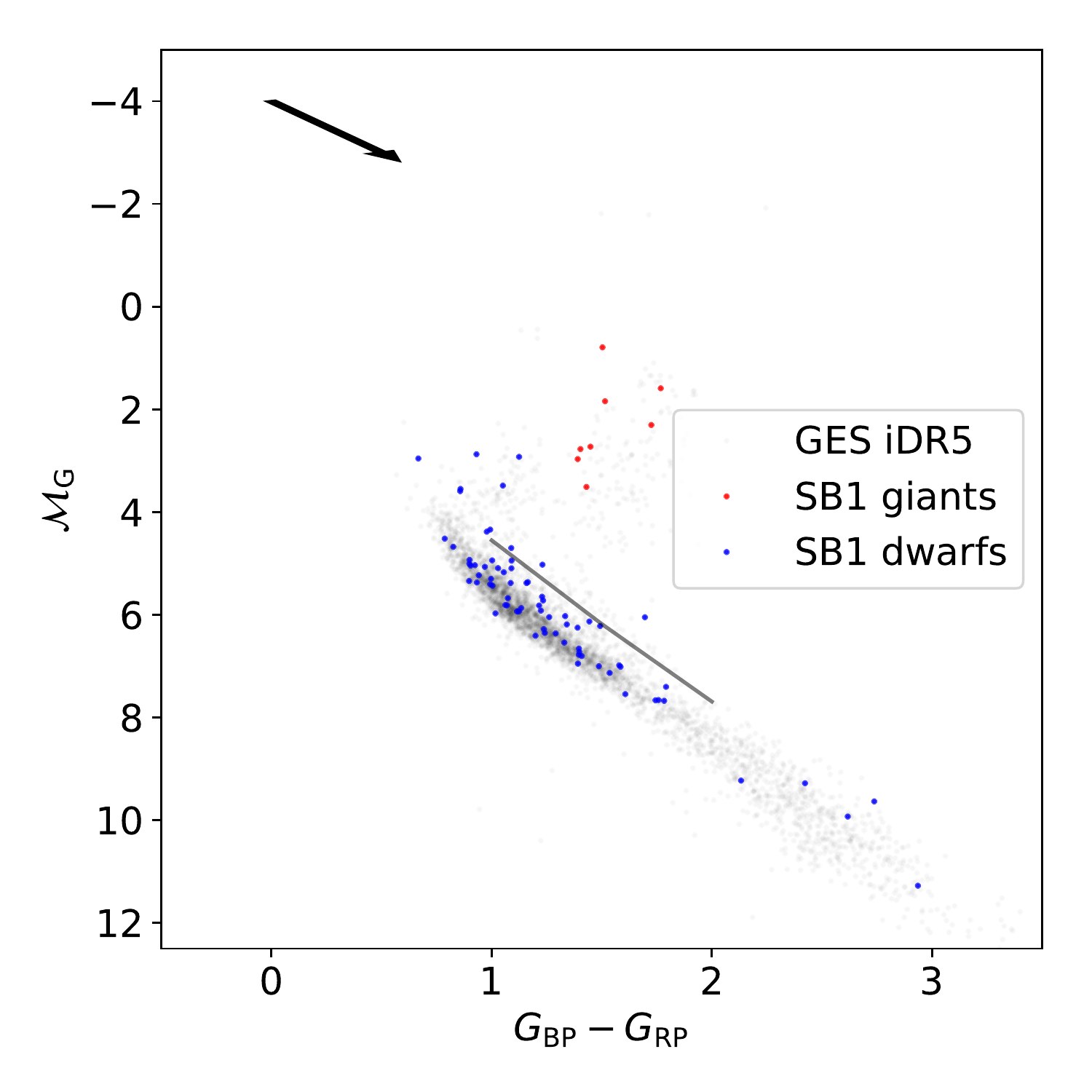}
\caption{Colour -- absolute magnitude diagrams (non-dereddened) of GES iDR5 SB1 candidates using \emph{Gaia} DR2 parallaxes and photometry. \emph{Gaia} constraints on relative parallaxes and photometry errors amount to  50\% (left), 10\% (middle), and 5\% (right). The grey line depicts the main sequence for twin SB2 with $q=1$, \emph{i.e.} 0.75 mag above the main sequence. The top left arrow shows the direction of the reddening vector. }
\label{fig:cmd_sb1}
\end{figure*}

\subsubsection{Spatial distribution}
The left panel of Fig.~\ref{fig:mollweide_dist} shows the GES iDR5 targets  on a  Mollweide projection\footnote{Pseudo-cylindrical projection that locally preserves the area.} in Galactic coordinates. In the Galactic plane, mainly the bulge, the CoRoT field, and open clusters have been observed. At this scale, the individual stars are not resolved, so that each dot corresponds to a GIRAFFE line of sight that can simultaneously acquire up to  132 targets. We superimposed the SB1 dwarf (in blue) and giant (in red) candidates. The SB1 giant candidates are mainly located in the Galactic plane whereas SB1 dwarfs candidates are more numerous in the galactic-latitude range ]10$-$30]$^\circ$ (right panel of Fig.~\ref{fig:mollweide_dist}). Fig.~\ref{fig:sb1_dist} shows the distance distribution of the GES targets (grey),  reaching up to 20~kpc (not displayed on Fig.~\ref{fig:sb1_dist}). We only considered targets whose relative parallax error is better than 0.5. The SB1 dwarf (blue) and giant (red) candidates from set~2 (black) are shown, with a maximum distance around 6~kpc for SB1 dwarfs and 10~kpc for SB1 giants. The number of SB1 candidates peaks around 2~kpc. The decrease of the number of SB1 candidates with increasing distance nicely follows the evolution of the number of GES iDR5 targets with distance, confirming that the SB1 detection efficiency is not affected by the target distance, as expected for spectroscopic binaries.

\subsubsection{Gaia colour-magnitude diagram}
Figure~\ref{fig:cmd_sb1} locates the  GES SB1 candidates in the colour-absolute magnitude diagram. The background grey dots are the GES iDR5 subsample stars that are used in the $\chi^2$ test and that satisfy various
\emph{Gaia} parallax and photometry criteria described below. In all cases, we excluded stars that do not belong to the `gold'\footnote{Gaia sources for which the photometry  was produced by the full calibration process and used to establish the internal photometric system.} photometric dataset \citep{riello2018} as well as stars with negative parallaxes. We then constructed various subsamples by varying the relative error on (all four) parallaxes, $G_\mathrm{BP}$, $G_\mathrm{RP}$ and $G$ mean fluxes from 50 to 10\%, and finally 5\% (left to right panels of Fig.~\ref{fig:cmd_sb1}). The respective numbers of GES iDR5 stars are 31\,350, 11\,200, and 4\,300, while the numbers of SB1 candidates are 581, 222, and 81 when the relative error is set at 50, 10, and 5\%, respectively. The dispersion in the data sequences strongly decreases when reducing the relative error on \emph{Gaia} parallaxes and photometry, as expected.

The theoretical main sequence of twin SB2 (\emph{i.e.} with $q=1$) is overplot on Figure~\ref{fig:cmd_sb1} as the grey line, located 0.75 magnitudes above the single-star main sequence. We note that imposing a stronger constraint on relative error removes outlier SB1 candidates that are located below the main sequence. Those remaining above the main sequence might have a companion that affects the absolute magnitude and could correspond to SB2 candidates that are not detected as such because the two CCF peaks are strongly blended, that is, the velocity difference between the two components remains below $\sim25$~\kms, which corresponds to the GES SB2 detection threshold for GIRAFFE observations as reported by \citet{vdswaelmen2018a}. 

We now comment on specific outliers in the left panel of Fig.~\ref{fig:cmd_sb1}. The five SB1 candidates (4 dwarfs and 1 giant) with $G_\mathrm{BP}-G_\mathrm{RP} <0.6$ do not have GES recommended parameters and have no Simbad identifiers. Candidate SB1s below the main sequence and with $0.6 < G_\mathrm{BP}-G_\mathrm{RP} <1$ have GES recommended effective temperatures and gravities. One of them has a recommended $\log{g}=2.23\pm0.34$ (pointing at a giant star) but at the same time is the faintest SB1 dwarf candidate (GES ID 18133564-4224252, $\mathcal{M}_\mathrm{G}=6.92$) at  $G_\mathrm{BP}-G_\mathrm{RP} = 1$. These SB1 candidates located between main and white dwarf (WD) sequences overlap with the region occupied by cataclysmic variables (see Figures 7 and 11 of \citealt{eyer2018}). However, these candidates disappear when the relative uncertainties on both parallaxes and photometry are reduced from 50\% to 10\% or 5\%  (middle and right panels of Fig.~\ref{fig:cmd_sb1}), and therefore, they might be spurious detections.

Candidate SB1s with $2 < G_\mathrm{BP}-G_\mathrm{RP}$ and $\mathcal{M}_\mathrm{G}>8$ (\emph{i.e.} at the low-mass end of the main sequence) have no GES recommended parameters and no entries in the Simbad database. Interestingly, some of them show chromospheric activity as observed from the strong core emission in the \ion{Ca}{II} triplet in HR21, as identified with the t-SNE approach (Traven et al., in prep.), and probably correspond to eruptive variables \citep{eyer2018}.

\begin{figure*}
\includegraphics[width=0.247\linewidth]{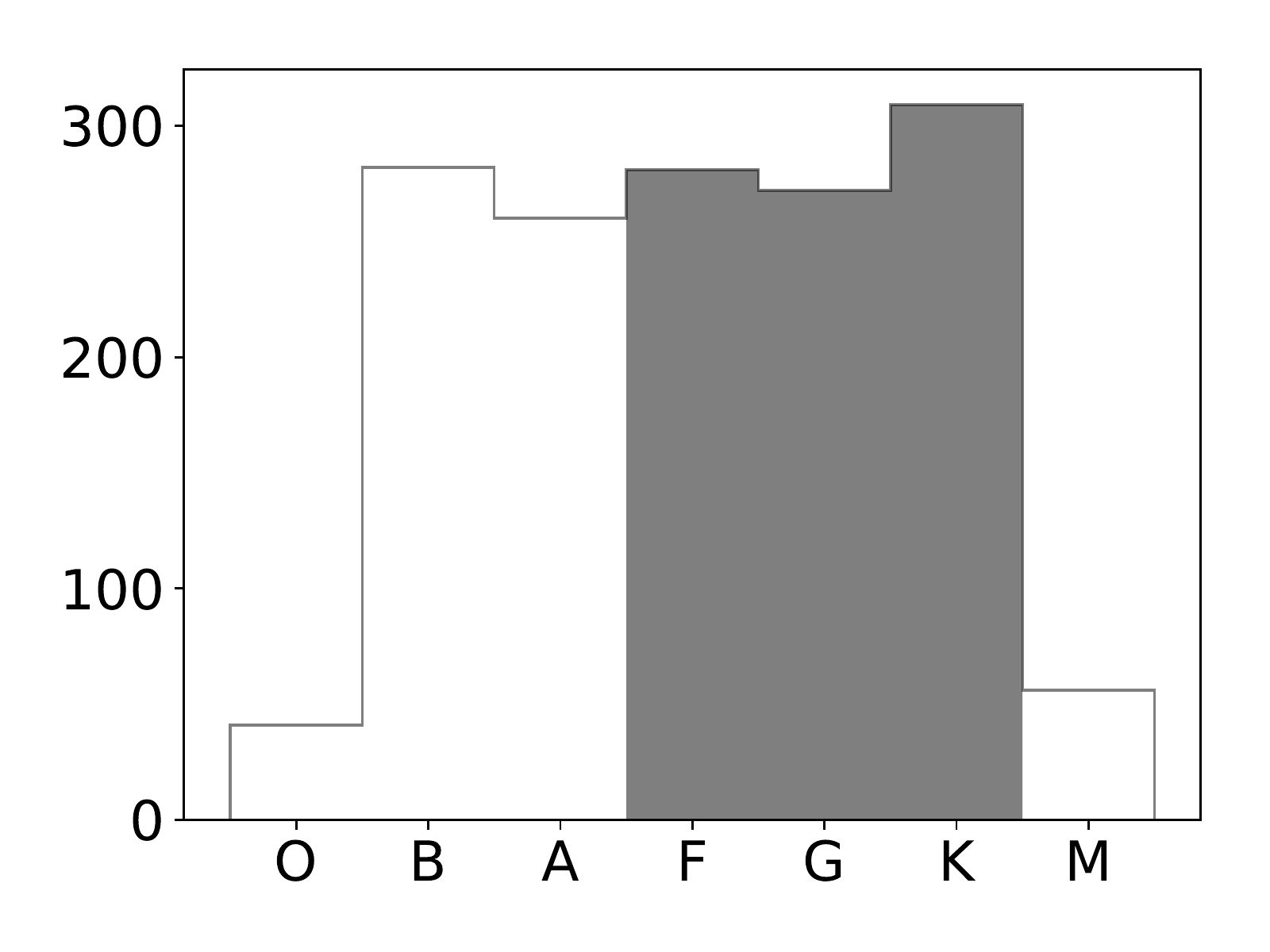}
\includegraphics[width=0.247\linewidth]{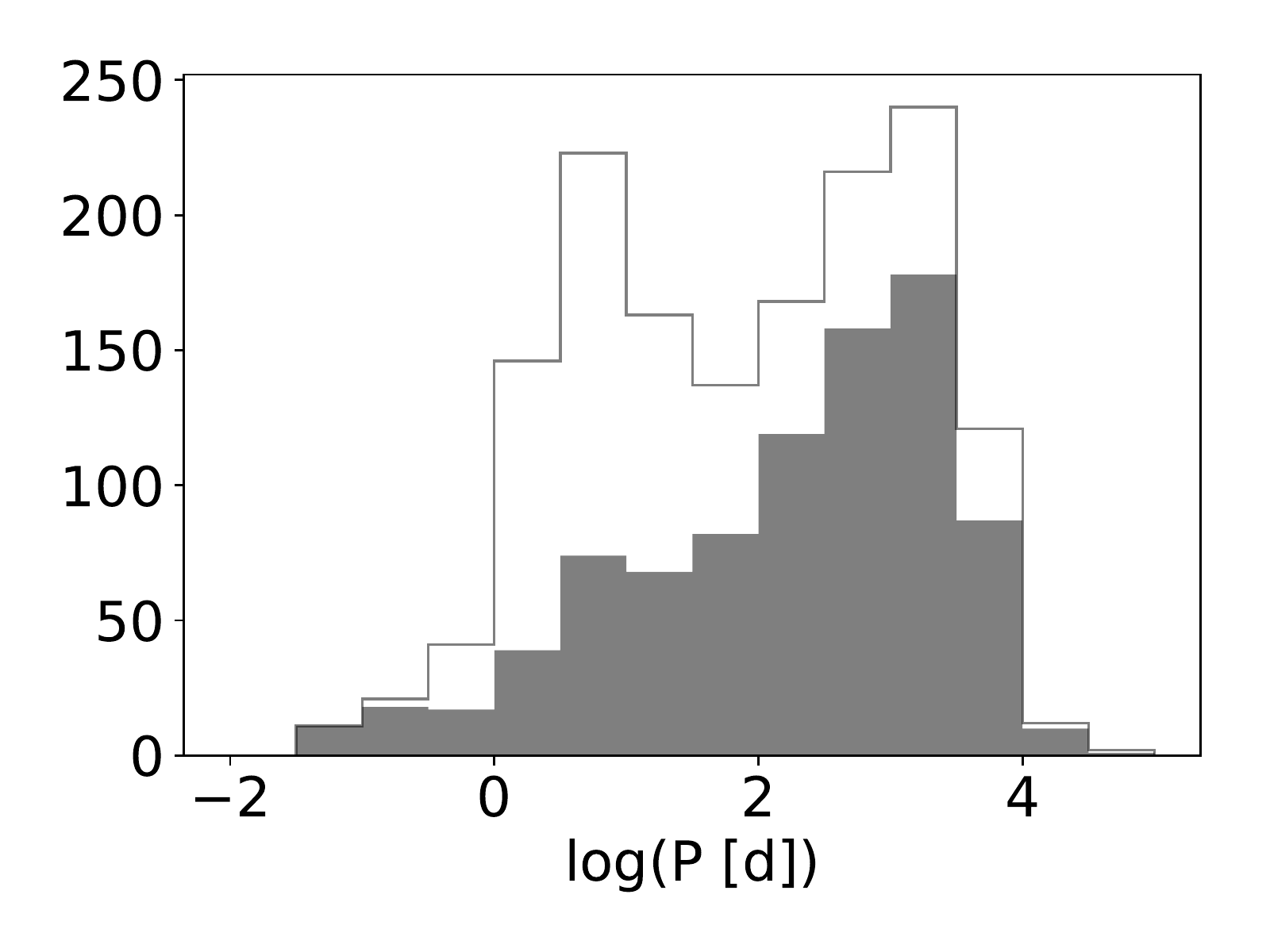}
\includegraphics[width=0.247\linewidth]{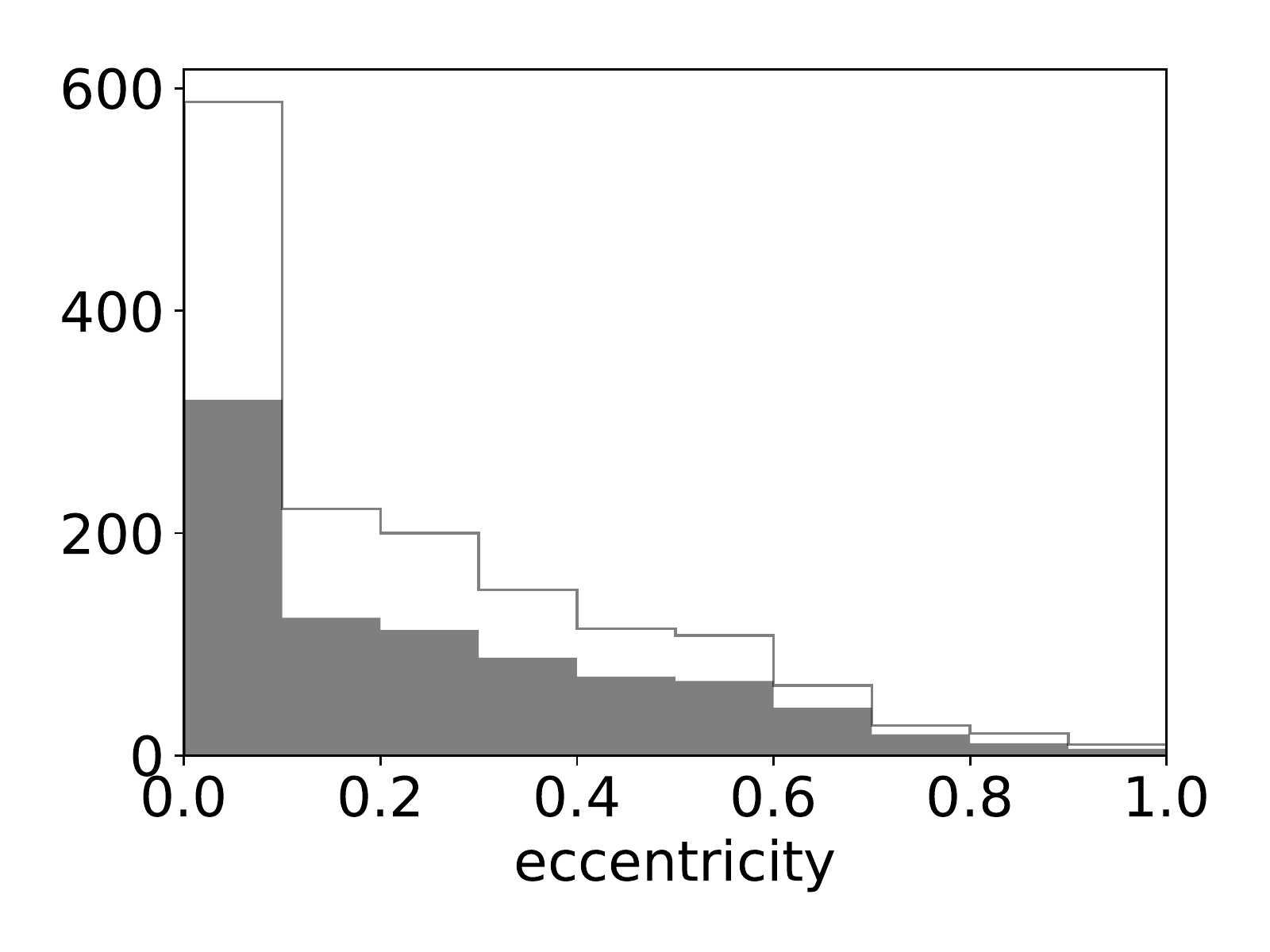}
\includegraphics[width=0.247\linewidth]{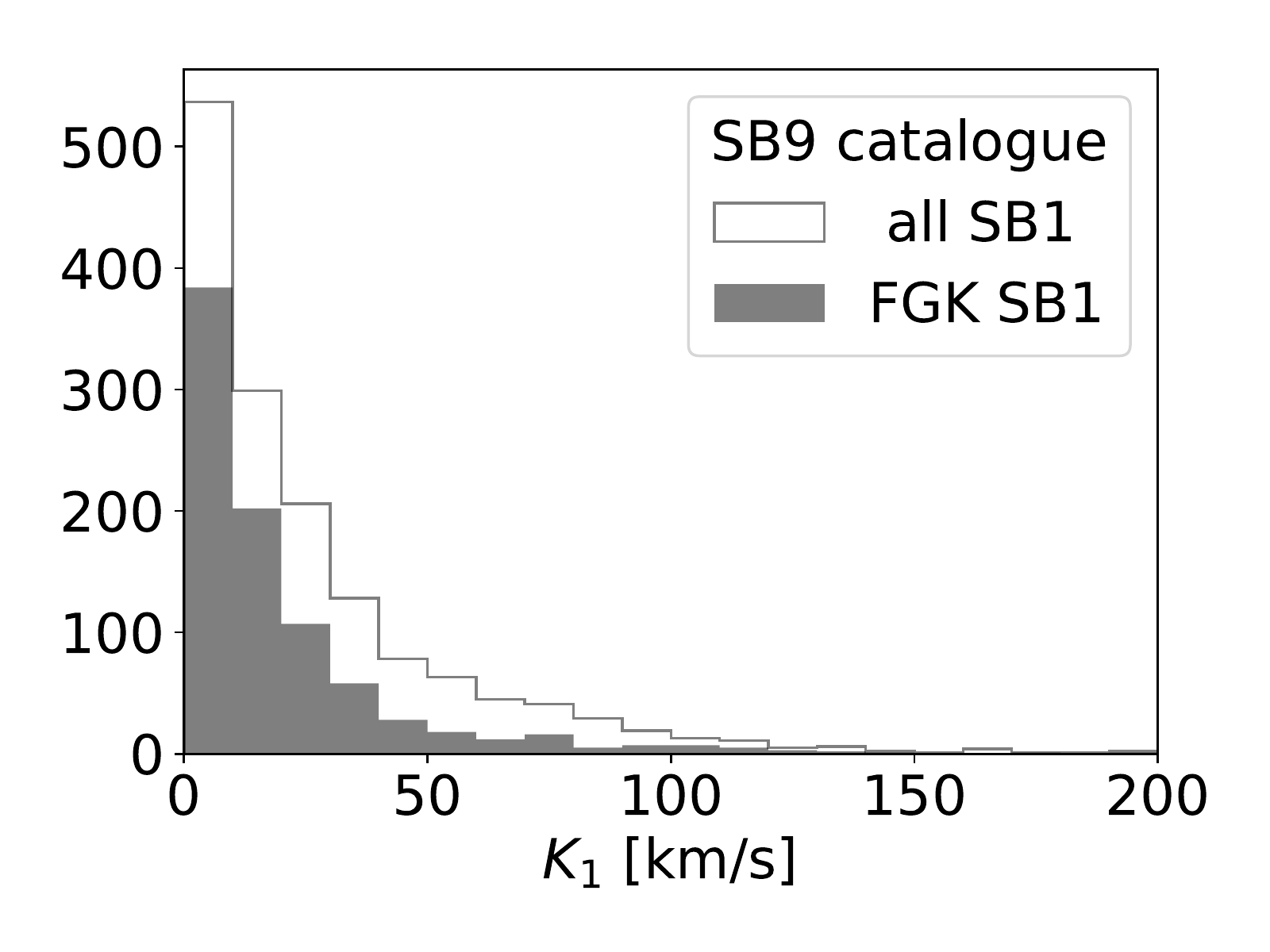}
\caption{Statistical properties of SB1s from the SB9 catalogue used in the Monte-Carlo simulations. From left to right: spectral-types, periods, eccentricities, and RV amplitude histograms.}
\label{fig:sb9_stats}
\end{figure*}

\subsubsection{Completeness correction}
\label{sec:efficiency}
In this section, we describe our attempt to evaluate the detection efficiency of our method.  To do so, we adopted the {\it Ninth Catalogue of Spectroscopic Binary Orbits}  \citep[SB9 catalogue;][]{pourbaix2004} as a (supposedly) complete sample. 
%This allowed us to correct the detection rate of GES SB1 before investigating the SB1 frequency as a function of  effective temperature (Sect.~\ref{sec:sb1_temp}), metallicity (Sect.~\ref{sec:sb1_met}), and both of them (Sect.~\ref{sec:2d}).
The SB9 catalogue is itself probably not free from bias because it is a compilation of orbits from the literature. However, the alternative approach based on a synthetic sample of binary systems is not easy to implement, as it would require building not only a family of pre-mass-transfer systems (which is relatively straightforward), but also post-mass-transfer systems (since the GES sample must contain pre- and post-mass-transfer systems altogether), which is currently unrealistic since the evolutionary channels followed by some post-mass-transfer systems like barium stars are not yet fully understood \citep[\emph{e.g.}][]{izzard2010,saladino2019}.
This is why we resort to the SB9 catalogue for our detection efficiency estimate.

\begin{figure*}
\includegraphics[clip, angle=-90, width=0.5\linewidth, trim={6cm 3.7cm 5cm 3.7cm}]{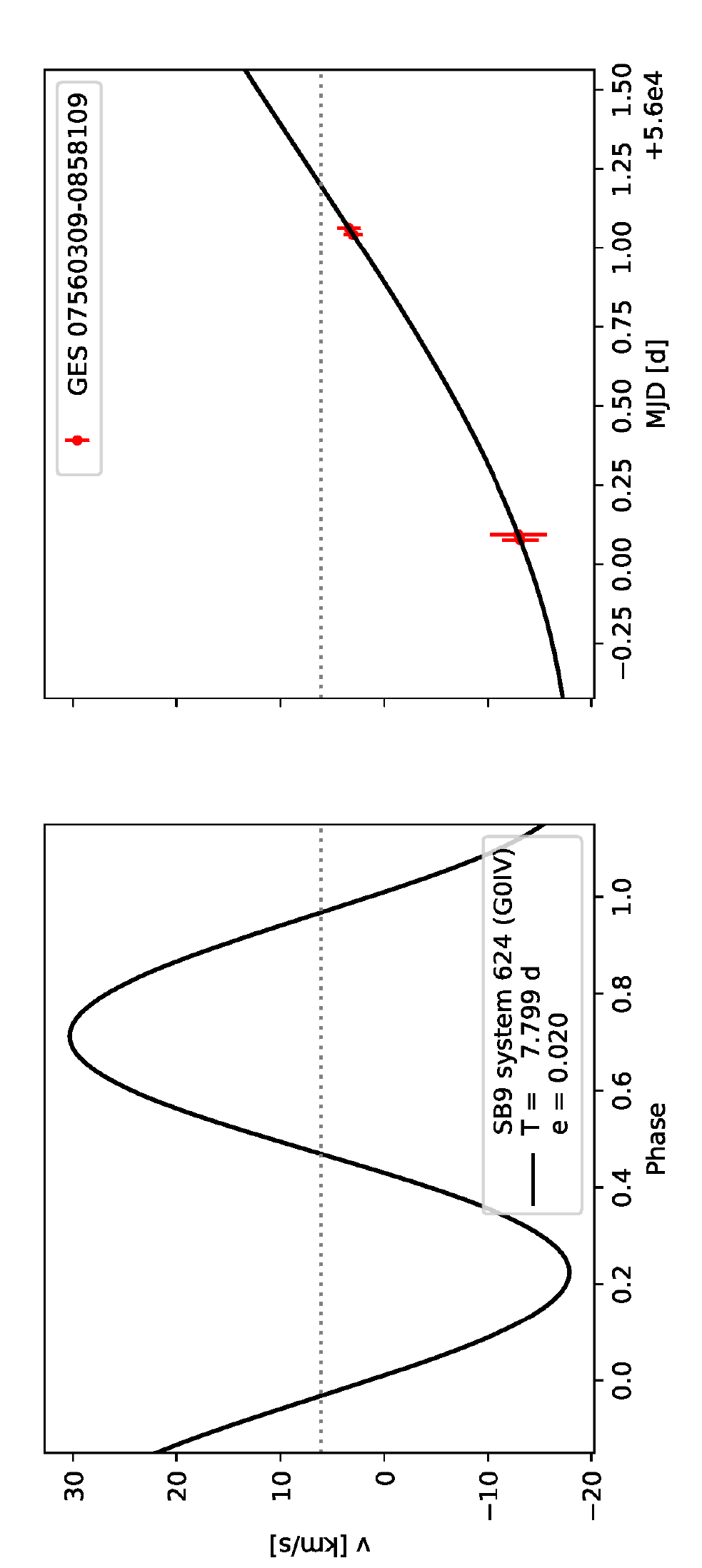}
\includegraphics[clip, angle=-90, width=0.5\linewidth, trim={6cm 3.7cm 5cm 3.7cm}]{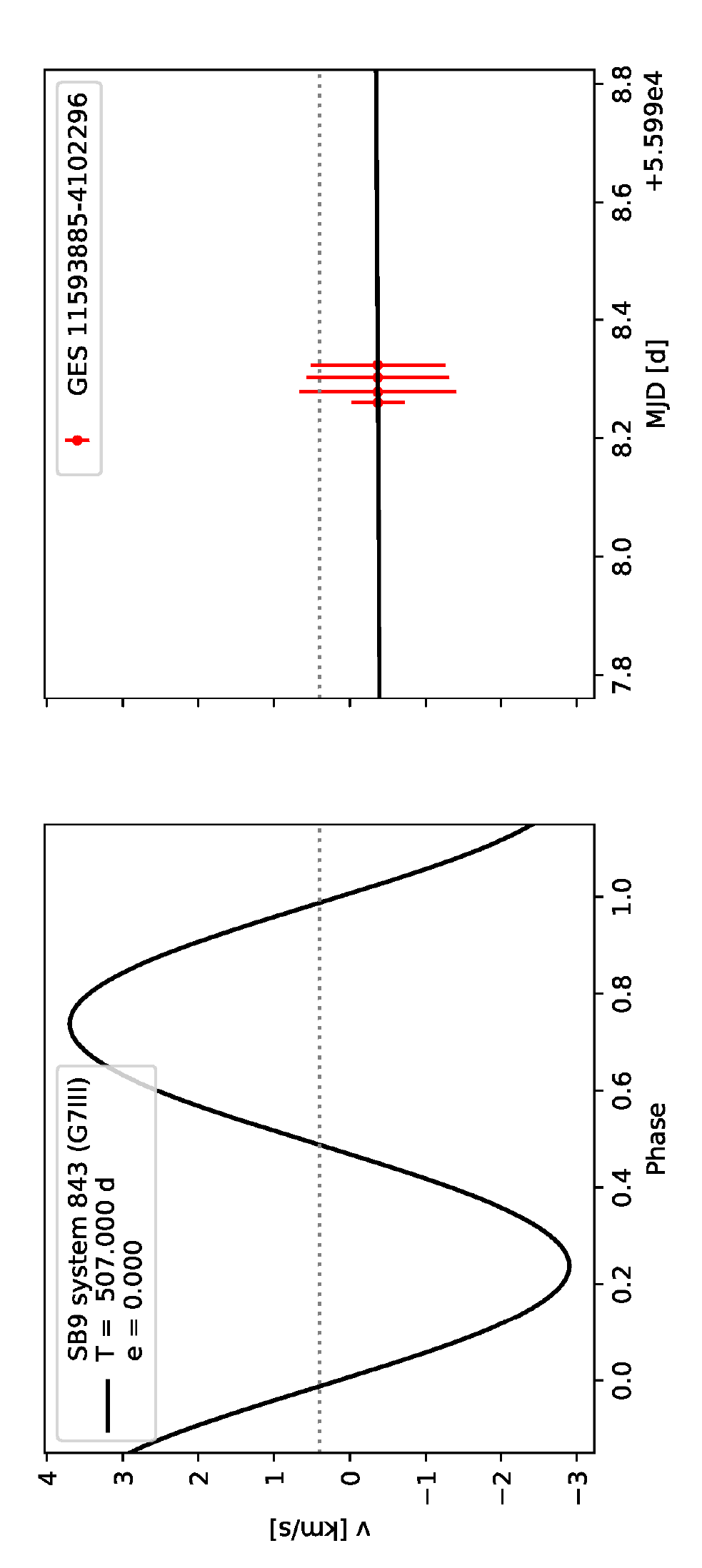}
\caption{Examples of Monte-Carlo simulations with (two left panels) and without (two right panels) detection of known circular SB1s from the SB9 catalogue. Red points represent simulated RV measurements at the epochs of a randomly chosen GES iDR5 target, with their estimated uncertainties. Spectral type, orbital period and eccentricity are indicated in the inset.}
\label{fig:mcs_sin}
\end{figure*}
\begin{figure*}
\includegraphics[clip, angle=-90, width=0.5\linewidth, trim={6cm 3.7cm 5cm 3.7cm}]{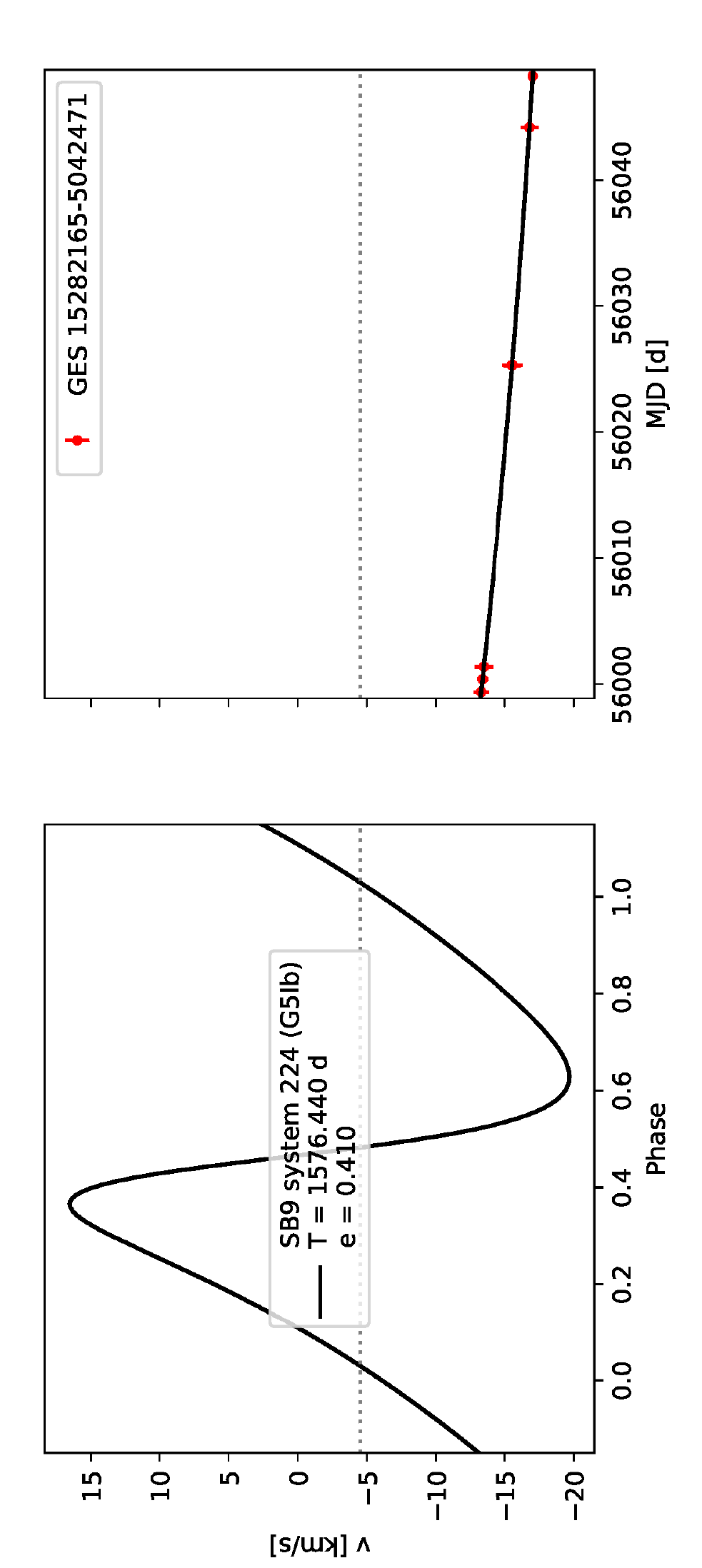}
\includegraphics[clip, angle=-90, width=0.5\linewidth, trim={6cm 3.7cm 5cm 3.7cm}]{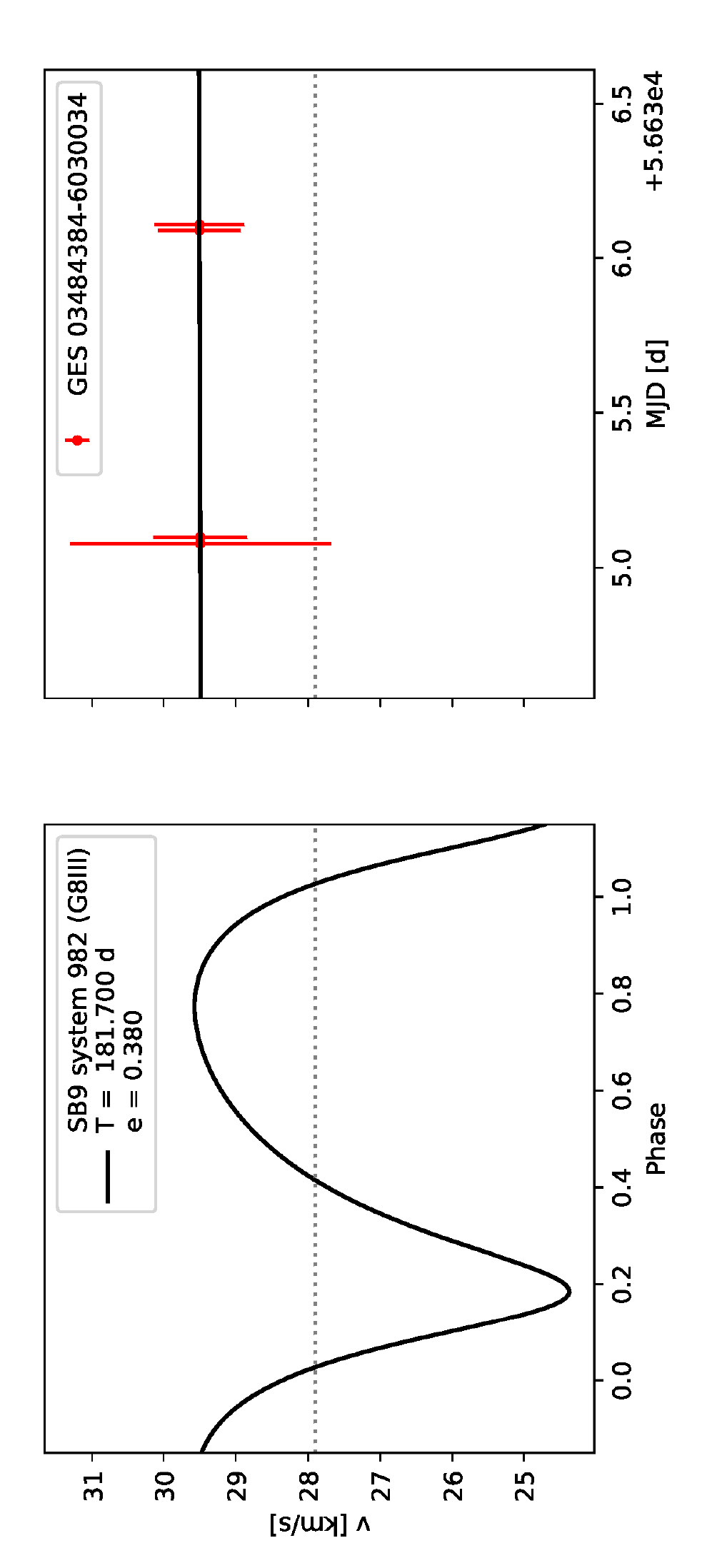}
\caption{Same as Fig.~\ref{fig:mcs_sin} but for eccentric orbits.}
\label{fig:mcs_exc}
\end{figure*}

{\it SB9 subsample selection}. 
We used the SB9 2018-04-11 version, listing 4\,544 (resp. 3\,045) orbits for 3\,611 stellar systems (resp. 2\,560 SB1 systems). In the SB9, 1\,512 SB1 have an associated spectral type. In total, SB9 therefore contains 283, 279, and 309 SB1 systems with primary spectral types F, G, and K, respectively, and with full orbital solutions (corresponding to a total of 871 systems). This subset, tagged as the `SB9 subsample' in the following, serves as our input unbiased catalogue and is characterised by the statistical distributions displayed in grey in Fig.~\ref{fig:sb9_stats} (to be compared to the full SB1 sample in the SB9 catalogue, in white). 

{\it Monte-Carlo simulations}.
In order to account for spectroscopic analysis failures ($S\!/N$ below our threshold but also defective spectra preventing the computation of CCFs), the association between SB9 systems and GES targets is performed using the full HR10+HR21 set of observations (\emph{i.e.} before any kind of cleaning), which corresponds to approximately $49\,000$ targets and about $210\,000$ observations.

To derive our detection efficiency, we computed the fraction of SB1 binaries in the SB9 subsample that would be detected if they were observed with the GES time sampling and processed with the same $\chi^2$ test as described in Sect.~\ref{Sect:chi2}. More precisely, to each system from the SB9 subsample, we associated a GES target with its specific time sampling, computed the corresponding RVs from the known SB9 orbit, and applied the  $\chi^2$ test to this RV dataset (with the RV uncertainties -- from NACRE CCFs -- and $S\!/N$ corresponding to the associated GES exposure). A specific detection efficiency could then be computed for this simulation. The process was repeated by randomly changing the association between the SB9 subsample stars and GES time samplings. We show some examples of this process in Figs.~\ref{fig:mcs_sin} (for circular orbits)  and \ref{fig:mcs_exc} (for eccentric orbits). The left panel of each pair in these figures depicts an orbital solution selected from the SB9 subsample and sampled as the GES target labelled in the right panel. 

\begin{figure*}
    \centering
    \includegraphics[width=1.1\linewidth, clip=true, trim=40 0 40 0]{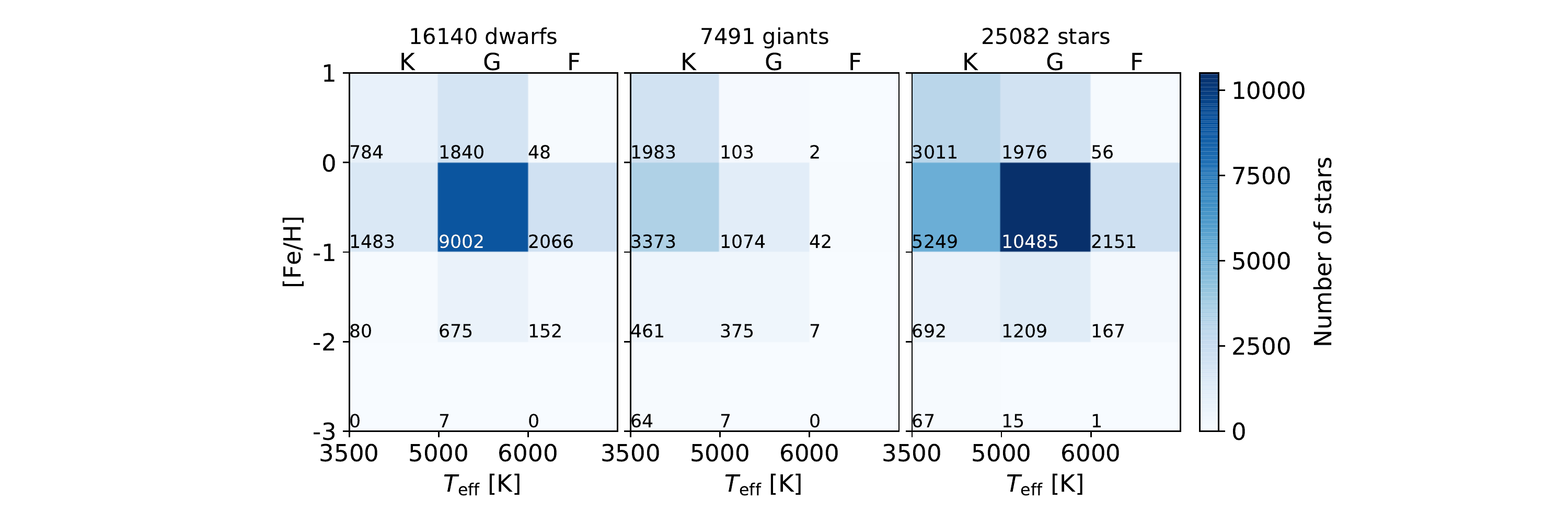}
    \includegraphics[width=1.1\linewidth, clip=true, trim=40 0 40 0]{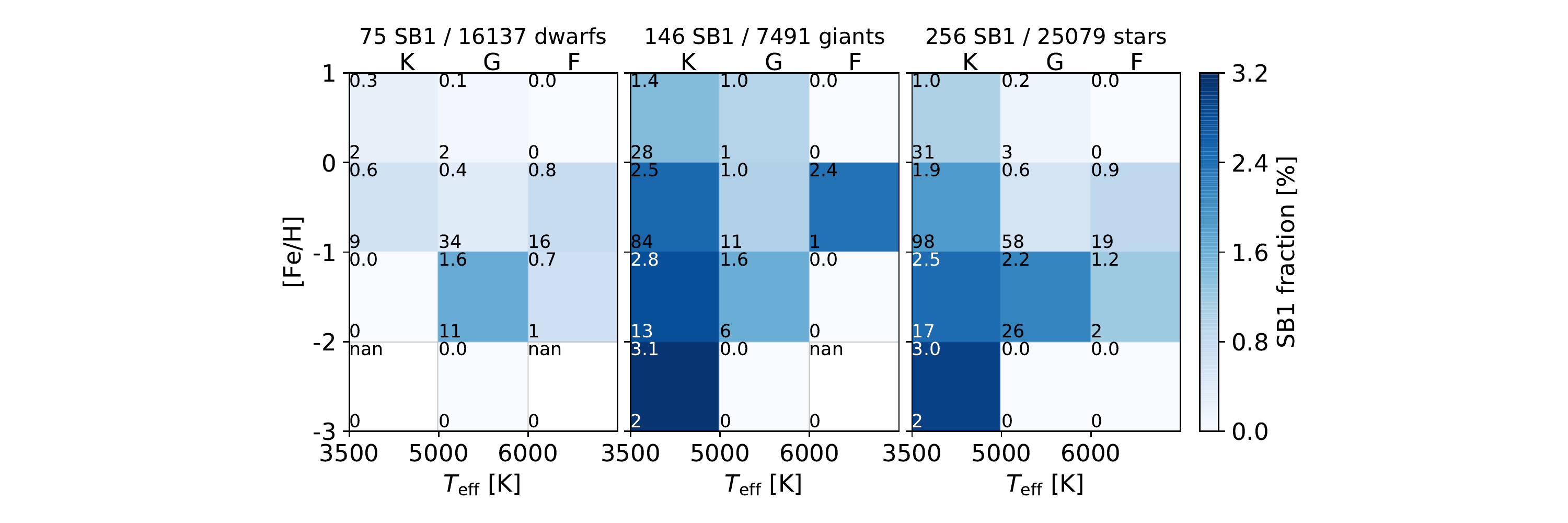}
    \includegraphics[width=1.1\linewidth, clip=true, trim=40 0 40 0]{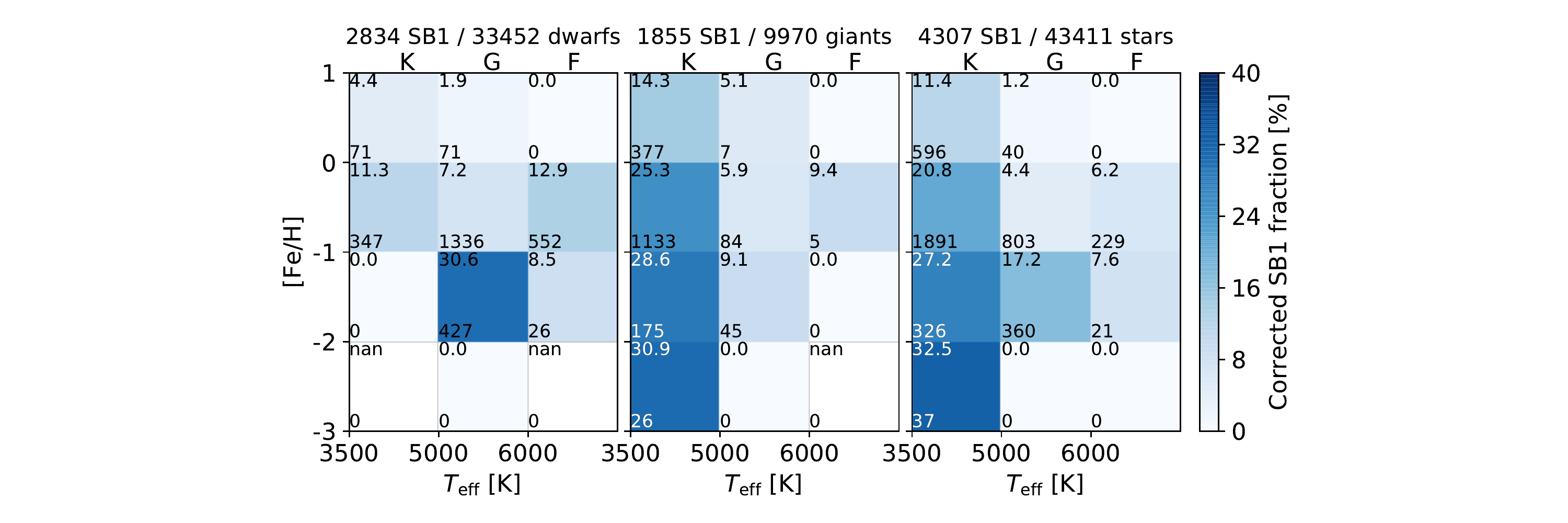}
    \caption{Two-dimensional (\Teff\ and [Fe/H]) dependence of the SB1 fraction using set~2. Left, middle and right panels refer to dwarfs, giants, and total GES iDR5 targets, for which we have recommended \Teff\ and [Fe/H] values. The colour scale corresponds to the number of stars per bin (top panels),  the uncorrected SB1 fractions per bin (middle panels), and the SB1 fraction corrected for the detection efficiency $\eta$ (Table~\ref{tab:m_teff_de}, bottom panels). The number of stars is indicated in the lower left corner of each bin, while the SB1 fraction is listed in the upper left corner.}
    \label{fig:2Ddepend}
\end{figure*}

\begin{figure*}
    \centering
    \includegraphics[width=\linewidth]{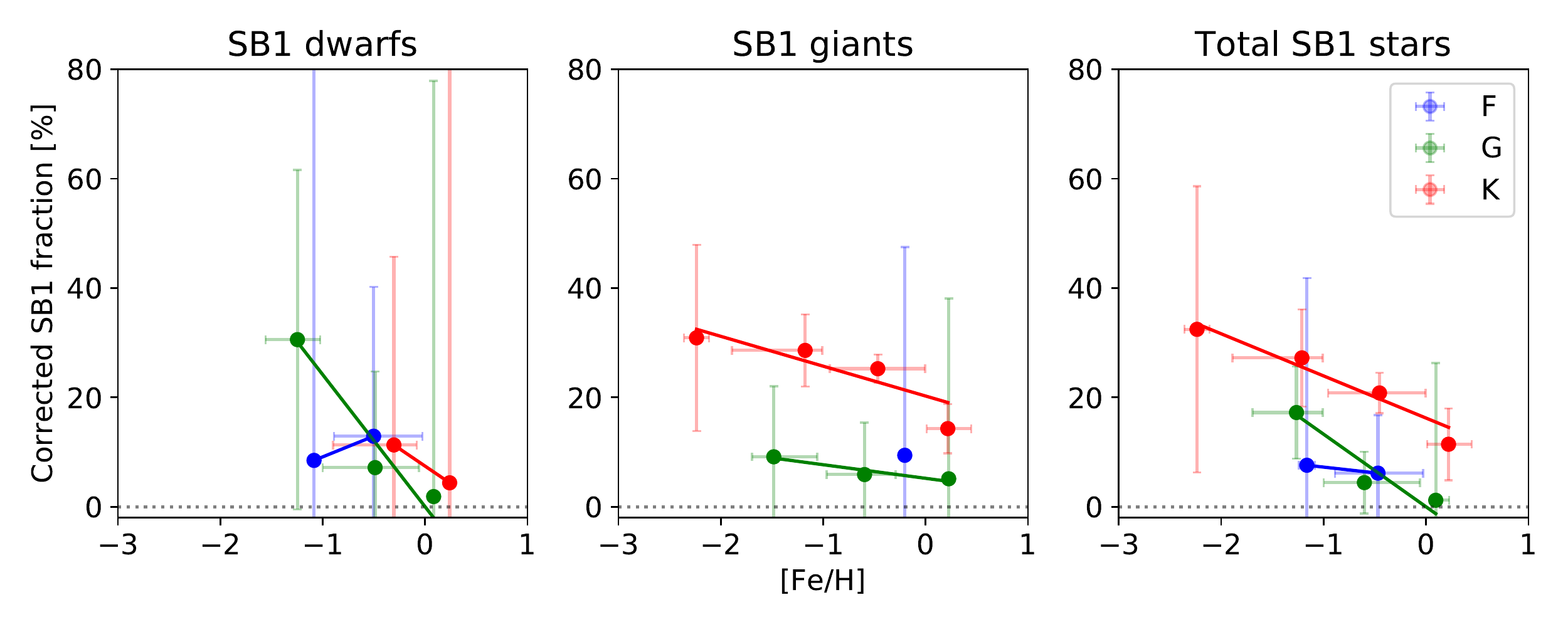}
    \caption{Fraction of SB1s as a function of the metallicity for SB1s with dwarf primaries (left), giant primaries (middle) and the whole SB1 sample (right) of set 2. The SB1 fraction is corrected for the detection efficiency per spectral type and for the different biases listed at the end of Sect.~\ref{sec:efficiency}. The SB1 fraction with K spectral type is the largest.}
    \label{fig:met_spt}
\end{figure*}

{\it Results.} The GES SB1 detection efficiency is evaluated from 100 Monte-Carlo simulations as described above. Each Monte-Carlo simulation includes 200 F, 200 G, and 200 K-type orbits from the SB9 catalogue, sampled with the GES epochs, RV uncertainties and $S\!/N$s. We derived the detection efficiency per spectral type ($\eta$) as listed in Table~\ref{tab:m_teff_de}. 
The detection efficiency is similar for F and G-type stars ($\sim 25$\%) whereas it drops significantly for K-type stars ($\sim10$\%). This intriguing drop of the detection efficiency between F-G and K spectral types by a factor of two may be traced back to the fact that  SB1 systems with K-giant primaries are over-represented in the SB9 catalogue with respect to the ones with dwarf primaries. This results from the presence in that catalogue of the many binaries involving K-giant primaries  discovered by \cite{mermilliod2007} in open clusters. Because of their longer periods, such systems involving giant stars are harder to detect using the GES. To avoid this bias, we decided to evaluate the detection efficiency separately for systems with K-dwarf and K-giant primaries, and applied to K dwarfs  the same detection efficiency as for  F-type stars.

Additional biases must also be taken into account because, as we show below, the SB1 frequency varies with effective temperature (Sect.~\ref{sec:sb1_temp}) and with metallicity (Sect.~\ref{sec:sb1_met}). In Sect.~\ref{sec:2d}, we first split the sample into 12 (\Teff, [Fe/H]) bins to separate the impact of these two parameters on the SB1 frequency. In this process, we also distinguish dwarfs from giant stars. Unfortunately, the required information (\Teff,  [Fe/H], as well as giant or dwarf classification) is only available for subsamples of the GES total sample. Restricting these subsamples to their intersection would dramatically reduce the number statistics. Therefore, in the remaining, `corrected' statistics refers to those that have been corrected for the availability of \Teff, [Fe/H], and dwarf or giant classification in the corresponding sample (assuming that the distribution of these parameters in the subsample where they are available is representative of their distribution in the whole sample).

In the following, we thus take into account:
\begin{itemize}
    \item the bias introduced by the differential availability of GES recommended \Teff\ and [Fe/H] for the whole sample and for the sets of SB1 candidates; 
    \item the bias introduced by the fact that about 10\% of SB1 from set~2 do not have dwarf/giant classification from \emph{Gaia} DR2 (due to filtering on the precision of parallaxes and photometry);
    \item the detection efficiency per spectral type. 
\end{itemize}
Among these corrections, the third one dominates. 

\subsubsection{Two-dimensional dependence of the SB1 fraction}
\label{sec:2d}

\begin{figure*}
    \centering
    \includegraphics[clip=true, width=\linewidth]{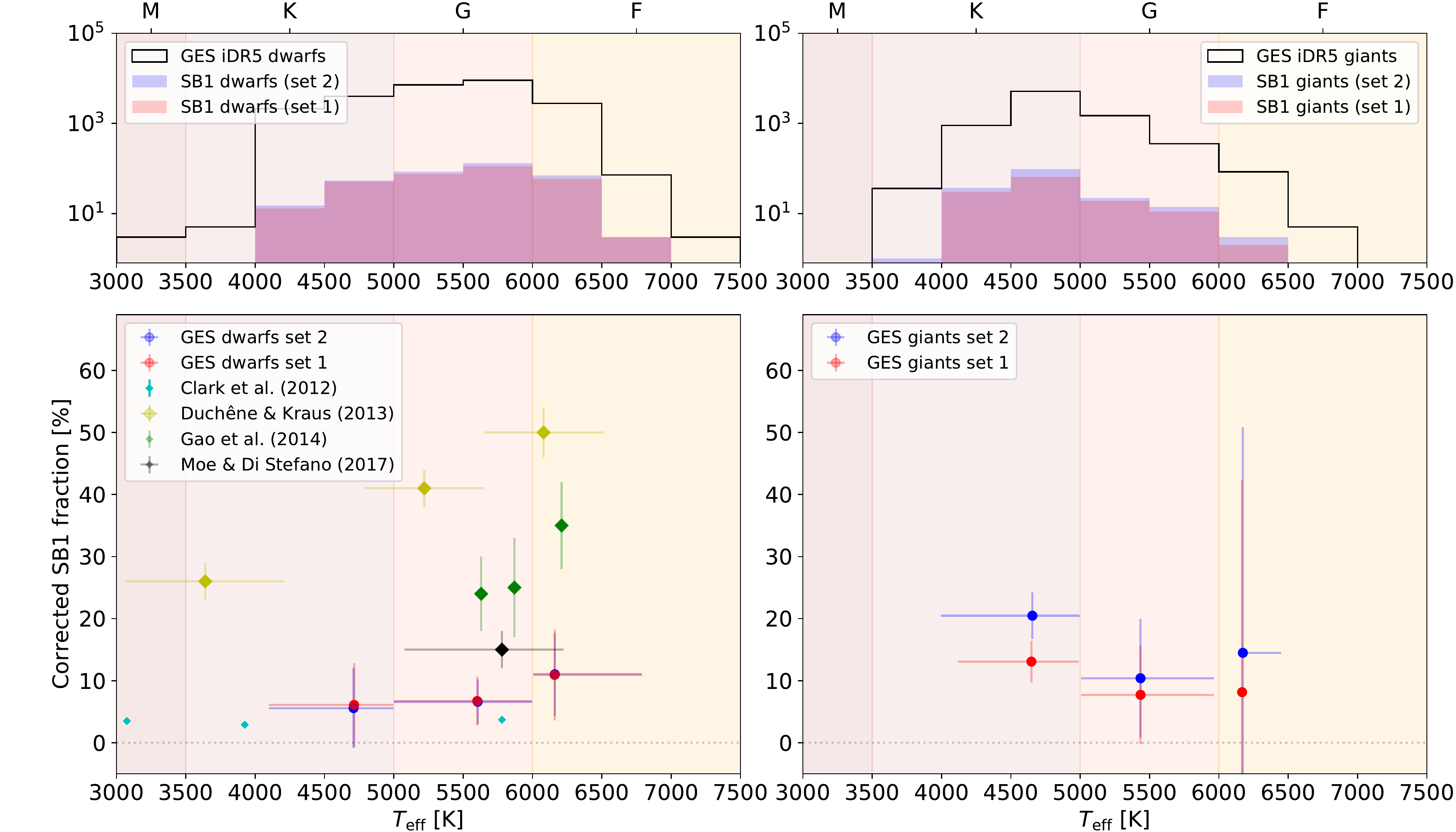}
    \caption{Top panels: Effective-temperature histograms for GES iDR5 stars (solid line), and SB1 candidates from sets~1 (red) and~2 (blue). Left panels depict main sequence stars and right panels correspond to giant stars. Bottom panels: Corrected SB1 fractions as a function of \Teff\ for SB1 candidates from sets~1 and 2 (red and blue dots, respectively). The  dots  are  centred  on the median \Teff\ in each spectral type and the \Teff\ range in each spectral type is  illustrated  by  the  horizontal  error bars. Yellow data points from \citet{duchene2013} correspond to the total binary rather than the SB1 fraction which is limited to shorter orbital periods (see text for details) and are not directly comparable with other datasets.}
    \label{fig:sb1_temp}
\end{figure*}

Figure~\ref{fig:2Ddepend} shows the colour-coded variation of the number of (analysed) stars (first row), the biased SB1 frequency (second row), and the corrected SB1 frequency (third row) as a function of temperature (horizontal axis) and metallicity (vertical axis). The first, second, and third columns provide colour maps for dwarfs only, giants only, and dwarfs+giants respectively. The first two rows are provided for comparison. Maps displayed on the third row are based on SB1 fractions corrected for the detection efficiency per spectral type, for the dwarf or giant classification availability, and for the (\Teff, $\log{g}$) parameter availability (separately for dwarfs and giants when applicable). The SB1 fractions increase with decreasing metallicity, and this trend is observed among each spectral type, except for dwarf F-type stars. Analysing the evolution of the SB1 fraction with spectral type is less straightforward; however the SB1 fraction increases when going from warm metal-rich objects (top right corners) to cool metal-poor stars (bottom left corners). This diagonal trend is observed for the whole SB1 sample (lower right panel of Fig.~\ref{fig:2Ddepend}) as well as when considering dwarfs and giants separately (lower left and middle panels). This finding is consistent with the trend illustrated in figure~4 of \cite{gao2017}, who considered dwarf stars ($-1 < $~[Fe/H]~$< 0.5$), also showing a decreasing binary fraction with metallicity. These latter authors find a higher binary fraction among stars with \Teff\ $\ga 6500$~K, as we do in Sect.~\ref{sec:sb1_temp} (left panel of Fig.~\ref{fig:sb1_temp}).

\begin{table}
    \centering
    \caption{Detection efficiency ($\eta$) of the method as a function of spectral type tested on SB1 from the SB9 catalogue. The last column lists the present-day-mass-function probability-density function (PDMF PDF) for associated effective temperatures and masses (see Sect.~\ref{Sect:PDMF} for more details).}
    \begin{tabular}{ccccc}
    \hline \hline
       Spectral & $\eta$ & \Teff & $M$  & PDMF  \\
       type & [\%] & [K] & [M$_\odot$] & PDF \\
       \hline
        F & $24.7 \pm 2.3$ & 6750 & 1.38 & 0.04  \\    
        G & $21.5 \pm 2.7$ & 5500 & 0.95 & 0.05  \\      
        K dwarf & \emph{$21.5 \pm 2.7$} & 4250 & 0.51 & 0.07  \\  
        K giant & $12.3 \pm 2.4$ & 4250 & 1.3 & 0.001  \\
        \hline
    \end{tabular}

    \label{tab:m_teff_de}
\end{table}

We investigate  in Fig.~\ref{fig:met_spt} the combined effects of effective temperature (or spectral type) and metallicity on the SB1 frequency, also separating dwarfs from giants. We provide in Table~\ref{tab:sb1_frac_fits}, the coefficients of the linear fits of the SB1 fraction with metallicity in the form $f=a~\mathrm{[Fe/H]} + b$, with their [Fe/H] validity range, separately for dwarf and giant stars, as well as for different spectral types. The trend of increasing SB1 frequency with decreasing metallicity is observed for two (G and K-type) out of the three considered spectral types for both dwarfs and giants. For F giants, only one data point is available and therefore it is not possible to draw conclusions, while for F dwarfs, the only two data points define a decreasing trend with decreasing metallicity. Given the error bar on the frequency at [Fe/H$] = -1$, this trend is probably spurious. We note however that \cite{hettinger2015} also obtained a {\it positive} slope with metallicity at the 2$\sigma$ level for field F-type main sequence stars from SDSS low-resolution spectra. 

We note that the SB1 fraction of K giants is higher than that of earlier spectral types but within the errorbars, as seen in the following section (Fig.~\ref{fig:sb1_temp}). The SB1 frequency is higher among K giants than among K dwarfs, but this difference has to be interpreted by considering that these two samples do not originate from stars of similar initial masses. Indeed, K giants typically evolve from A-type stars characterised by a much higher binary frequency than that of K dwarfs.

\subsubsection{Dependence of SB1 frequency with effective temperature}
\label{sec:sb1_temp}
Recommended effective temperatures are available for 82\% of our GES iDR5 subsample and for $\sim74\%$ of SB1 candidates (for both sets 1 and 2, see Table~\ref{tab:selection}). The \Teff\ range is [$3400-7500$]~K and its distribution is shown in Fig.~\ref{fig:sb1_temp}, with the dwarf SB1 candidates in the top left panel and the giant SB1 candidates in the top right panel. The distributions of dwarf and giant stars peaks at early G and early K spectral types, respectively. We then computed the SB1 frequency per spectral type for dwarf and giant stars. This frequency takes into account the detection efficiency per spectral type ($\eta$, Table~\ref{tab:m_teff_de}), the bias introduced by the availability of \Teff\ and  the availability of dwarf/giant classification (Table~\ref{tab:selection}). The SB1 frequencies corresponding to sets~1 (red) and 2 (blue) are shown in the bottom panels of Fig.~\ref{fig:sb1_temp} along with their Poissonian errors. 

There is a positive correlation of the SB1 fraction containing dwarf primaries with temperature, and hence with the mass of the primary while no significant correlation is visible for the fraction of SB1 with giant primaries, probably due to the large errorbars. The fact that the frequency of binary systems decreases towards the lowest stellar masses or the coolest temperatures is already well established in the literature \citep{raghavan2010, duchene2013, moe2017}. It should nevertheless be mentioned that these studies generally refer to FGK stars as a whole under the denomination `solar-type stars', defined as encompassing types F6 to K3, and usually they provide a unique binary fraction for FGK stars, unlike the present study that provides binary fractions separately for F-, G-, and K-type stars.

\begin{table}[]
    \centering
        \caption{SB1 frequency--metallicity relation: parameters of linear fits obtained from set 2 as illustrated on Figs.~\ref{fig:met_spt} and \ref{fig:sb1_met_new}. The $a$ and $b$ parameters correspond to the slope and the $y$-intercept, respectively.}
    \begin{tabular}{cccc}
    \hline\hline
    Type     & $a$  & $b$  & [Fe/H] \\
    & [\% dex$^{-1}$] & [\%] & valid range \\
    \hline
    SB1 dwarfs \\
    F & $7.6$ & $16.7$  &$[-1.1, +0.5]$\\
    G & $-24.2\pm5.7$ & $0.0\pm6.3$ & $[-1.3, +0.0]$ \\
    K & $-12.7$ & $7.4$ & $[-0.3, +0.2]$\\
    \hline
    SB1 giants\\
    G & $-2.5\pm0.7$ & $5.2\pm0.8$ & $[-1.5, +0.2]$ \\
    K & $-5.5\pm2.1$ & $20.3\pm3.0$ & $[-2.2, +0.2]$\\
    \hline
    Total SB1\\
    F & $-2.1$ & $5.2$ & $[-1.2, -0.5]$\\
    G & $-13.3\pm4.5$ & $0.0\pm5.1$ &$[-1.3, +0.0]$\\
    K & $-7.7\pm1.3$ & $16.2\pm1.2$ & $[-2.2, +0.2]$\\
    \hline
    Total SB1\\
    all spectral types & $-8.8\pm3.0$ & $7.1\pm2.0$ &$[-2.4, +0.4]$\\
    \hline
    
    \end{tabular}

    \label{tab:sb1_frac_fits}
\end{table}

Figure~\ref{fig:sb1_temp} also displays the binary fraction from \citet{duchene2013}, which exhibits a monotonically decreasing binary fraction from 6000 to 3600~K. This fraction is however very different from the SB1 frequencies reported by both the present study as well as by \cite{clark2012}, \citet{gao2014}, and \citet{moe2017}. The reason is related to the fact that \citet{duchene2013} include long-period systems whereas all the other studies are restricted to shorter-period systems. This is further confirmed by the agreement between the binary frequencies found by the present GES study and by \citet{moe2017} (the black diamond in the lower left panel of Fig.~\ref{fig:sb1_temp}). These authors found a frequency of {\it close binaries} of $15\pm3$\% for `solar-type stars' (actually covering the F and G spectral types, or stellar masses between 0.8~M$_\odot$ (\Teff~$=5075$~K) and 1.2~M$_\odot$ (\Teff~$=6225$~K)); they define `close binaries' as those with $\log P({\mathrm d}) < 3.7$ (\emph{i.e.} $P < 14$~yr) and $q > 0.1$. Here we cannot derive the orbital periods of most GES SB1 candidates (due to poor time sampling), but Eq.~\ref{eq:period} allows us to obtain period estimates, resulting in the period distribution displayed in  Fig.~\ref{fig:histo_dtl}. Most of the GES SB1 systems comply with the definition of `close binaries' given by Moe \& Di Stefano. The two SB1 fractions may thus be compared and, as illustrated in the bottom left panel of Fig.~\ref{fig:sb1_temp}, they are in rough agreement. 

\begin{figure}
    \centering  
    \includegraphics[width=\linewidth]{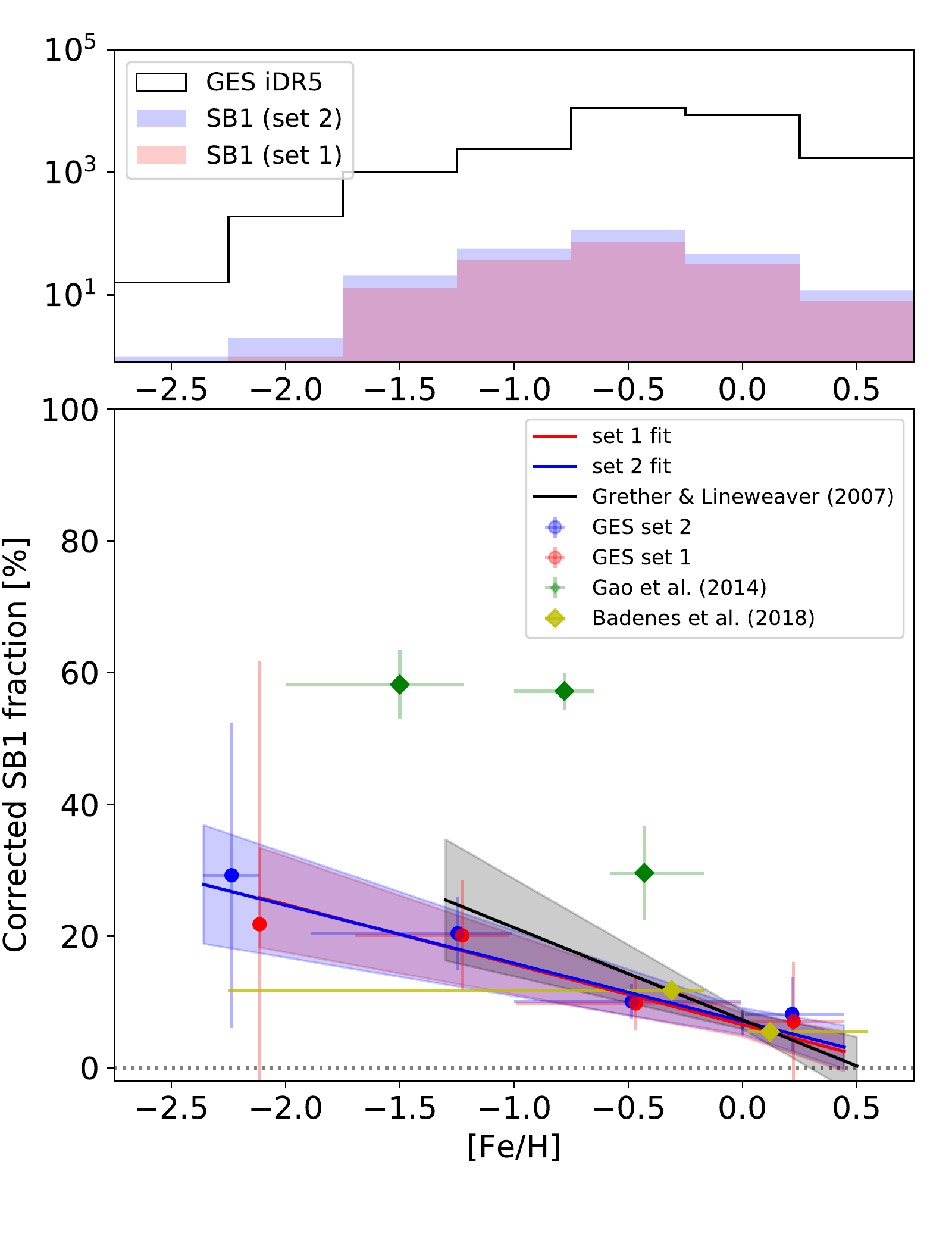}      
    \caption{Top panel:  Metallicity histograms for GES iDR5 stars (solid line), and SB1 candidates from sets~1 (red) and~2 (blue). Bottom panel: Sensitivity of the global GES SB1 fraction to the metallicity combining the different spectral types of the right panel of Figure~\ref{fig:met_spt} and comparison with literature results. Red and blue dots correspond to SB1 fractions of GES sets 1 and 2; the dots are centred on the median metallicity of each bin (of width 1 dex) and the metallicity range in each bin is illustrated by the horizontal error bars. Red and blue trends are almost superimposed. Yellow diamonds correspond to the SB1 fractions of \citet{badenes2018} (see discussion in Sect.~\ref{sec:sb1_met}) centred on the median metallicities of their two samples covering a metallicity range illustrated by the yellow error bars.}
    \label{fig:sb1_met_new}
\end{figure}

\subsubsection{Dependence of GES SB1 frequency with metallicity}
\label{sec:sb1_met}
The GES-recommended metallicities are available for 58\% of our GES iDR5 subsample and for $26$\% and $32$\% of SB1 candidates (for sets~1 and 2, respectively). In addition, we noticed that we have twice the number of giants with GES-recommended metallicities than dwarfs. The number of stars with [Fe/H] determination is 25\,082, among which 166 are detected as SB1 according to set 1 confidence level ($\sim 5 \sigma$) and 256 according to set 2 confidence level ($\sim 3 \sigma$, see Table~\ref{tab:selection}). The metallicity ranges from [Fe/H]~=~$-2.7$ to $+0.6$, and its distribution is shown in the top panel of Fig.~\ref{fig:sb1_met_new}. The number of stars and the SB1 candidates are maximum in the [Fe/H]$=-0.5$~dex bin. We computed the SB1 frequency in four bins by taking into account both (i) the detection efficiency per spectral type (Table~\ref{tab:m_teff_de}) and (ii) the bias introduced by the availability of the recommended metallicity. The frequencies corresponding to sets 1 (red) and 2 (blue) are shown in the bottom panel of Fig.~\ref{fig:sb1_met_new} along with their Poissonian errors. As expected, these errors reveal that the most precise SB1 frequency is the one with the largest number of SB1 detections, namely $f_1(\mathrm{[Fe/H]}=-0.5)=9.8\pm4.1$\% and $f_2(\mathrm{[Fe/H]}=-0.5)=10.1\pm2.7$\% (for sets 1 and 2, respectively). 

A linear fit to both sets 1 and 2 (taking the uncertainties into account) results in:
\begin{equation}
\label{eq:f1}
f_1^{\mathrm{lin}} = a_1\mathrm{[Fe/H]} + b_1 \text{~~~~~with} \begin{cases} a_1=-9.1\pm2.7\%~\text{dex}^{-1} \\b_1=+6.6\pm1.8\%\end{cases}
\end{equation}
\begin{equation}
\label{eq:f2}
f_2^{\mathrm{lin}} = a_2\mathrm{[Fe/H]} + b_2 \text{~~~~~with} \begin{cases} a_2=-8.8\pm3.0\%~\text{dex}^{-1} \\b_2=+7.1\pm2.0\%\end{cases}   
\end{equation}
The slopes $a_1$ and $a_2$ and the y-intercepts $b_1$ and $b_2$ agree with each other within 1$\sigma$. The frequency-metallicity relation for set 2 is listed in Table~\ref{tab:sb1_frac_fits} (last row) along with its temperature dependence (see previous section). The GES SB1 frequency decreases with metallicity by about $9\pm3$\% per metallicity dex. 

However, we need to firmly establish the reality of this trend by performing a test of the null hypothesis that the binary frequency does not depend on metallicity against the alternative hypothesis that it depends linearly on metallicity.  A $F$-test is used to discriminate between these two models:
\begin{equation}
\label{eq:f}
F = \frac{N-2}{1} \frac{\chi_\mathrm{cst}^2- \chi_\mathrm{lin}^2}{\chi_\mathrm{lin}^2},
\end{equation}
where $N$ is the number of metallicity bins with SB1 frequencies available, and $\chi_{\mathrm{[cst,lin]}}^2$ are the $\chi^2$ resulting from the linear (`lin') or constant (`cst') fits, defined as:
\begin{equation}
\label{eq:chi2mod}
\chi_{\mathrm{[cst,lin]}}^2 = \sum_{i} \frac{(f_i^\mathrm{meas} - f_i^{\mathrm{[cst,lin]}})^2}{\epsilon_i^2}     
\end{equation}
where $\epsilon_i$ is the Poissonian error attached to each measured SB1 frequency $f_i^\mathrm{meas}$, and $f_i^{\mathrm{[cst,lin]}}$ is the model-predicted SB1 frequency. All these quantities are listed in Table~\ref{tab:ftest}. The $F$-test reveals that the linear fits better match the data sets, with a much lower reduced $\chi^2$. The null hypothesis of a binary frequency independent of metallicity may be rejected with a confidence level higher than 95\% (for set~1) or 93\% (for set~2). In other words, the first kind risk of rejecting the null hypothesis of no metallicity trend while it is true is below 5\% with data set~1 and below 7\% with data set~2. Nevertheless, this trend should be considered as a lower limit because the detection efficiency we have estimated in Sect.~\ref{sec:efficiency} does not account for a possible dependency on metallicity. This effect will be assessed in the forthcoming analysis of the final GES data release.

\begin{table}[]
    \centering
        \caption{$F$-test (see Eqs.~\ref{eq:f} and \ref{eq:chi2mod}) comparing the linear fits (null and non-null slope) of SB1 frequency with metallicity.}
    \begin{tabular}{cccccc}
    \hline \hline
               & $N$ &  $\chi^2_\mathrm{cst}$ & $\chi^2_\mathrm{lin}$ & $F$-value & Confidence level \\
         \hline\\
         set 1 & 4 & 0.388 & 0.059 & 11.1 & $>95$\%\\
         set 2 & 4 & 0.953 & 0.179 &  8.7 & $>93$\%\\
         \hline
    \end{tabular}

    \label{tab:ftest}
\end{table}

\citet{grether2007} found a similar trend (with a slightly larger slope of $-14\pm6$\%~dex$^{-1}$) at a 2$\sigma$ (95\%) confidence level for a volume-limited ($<25$~pc) sample of FGK binaries with orbital periods shorter than 5~yr (black line in the bottom panel of Fig.~\ref{fig:sb1_met_new}). However their sample is limited to metallicities above [Fe/H$]\sim -1.2$. We also note that our sample is more extended spatially, covering distances up to $10$~kpc (Fig.~\ref{fig:sb1_dist}). It is worth noting that in their second, more extended sample (from 25 to 50~pc), these latter authors no longer find such  a trend (of increasing SB1 fraction with decreasing metallicity), probably because this more distant sample relies on less accurate photometric metallicities. Their closer sample relies, like ours, on (intrinsically more precise) spectroscopic metallicities. The good agreement between the slopes obtained by \citet{grether2007} and our study (when accurate spectroscopic metallicities are used) might indicate the universality of the SB1 frequency--metallicity relation, independent of the volume sampled in the Galaxy.

Additional evidence for this conjecture comes from the recent analysis of the APOGEE data \citep{badenes2018}. In the large sample of main sequence stars analysed by these latter authors, the ratio of binary frequencies among low- and high-metallicity stars (\FeH$=-0.31$ and $+0.12$, respectively) is about 2.2 (their Fig.~13). If we assume a binary fraction of $5.5$\% (using Eq.~\ref{eq:f1}) at $\mathrm{[Fe/H]}=+0.12$, this corresponds to a slope of $-15$\%~dex$^{-1}$ for the \citet{badenes2018} SB1 frequency--metallicity relation. This value is in good agreement with our findings (see yellow diamonds in the bottom  panel of Fig.~\ref{fig:sb1_met_new}). As mentioned in Sect.~\ref{sec:comp}, there are no SB1s in common between GES and APOGEE, thus strengthening the universality of this SB frequency--metallicity relation. 

A re-analysis of the APOGEE data by \citet{moe2018}, combined with Kepler eclipsing binaries and several other surveys, led to the finding that the close-binary fraction increases from 10\% at [Fe/H]~=~+0.5~dex to 40\% at [Fe/H]~=~$-1.0$~dex (\emph{i.e.} a $-20$\%~dex$^{-1}$ slope across the range $-1.0 \le $~[Fe/H]~$ \le 0.5$), and then to $\sim55$\% at  [Fe/H]~=~$-3.0$ (\emph{i.e.} a somewhat less steep slope; see their Fig.~18). The slope obtained by \citet{moe2018} in their larger metallicity range is similar to that found by \citet{gao2014} from the LEGUE and SEGUE survey data for main sequence FG stars (see the green diamonds in our Fig.~\ref{fig:sb1_met_new}), and is larger than ours by a factor of two. A trend with a steep slope of $-50$\%~dex$^{-1}$ was found by \citet{gao2017} for  G-type stars covering the metallicity range [$-0.9$, +0.4] in the LAMOST survey \citep[a similar result using LAMOST data was presented by][where  the binary--metallicity trend is clearly visible only for G dwarfs, and not for F or K dwarfs]{Tian2018}. Somewhat older investigations, which show less clear trends between binary fraction and metallicity, can be found in \citet{raghavan2010} and \citet{rastegaev2010}.

These previous investigations did not include a correction for the bias introduced by the fact that the sampling of F, G and K-type stars favours the longest-lived family. This bias can be corrected for by taking into account the Present-day-mass-function Probability-density function (PDMF PDF), as we present in the following section for our GES sample in order to derive a `PDMF-corrected' binary fraction and per metallicity bin.

\begin{figure}
    \centering
    \includegraphics[width=\linewidth]{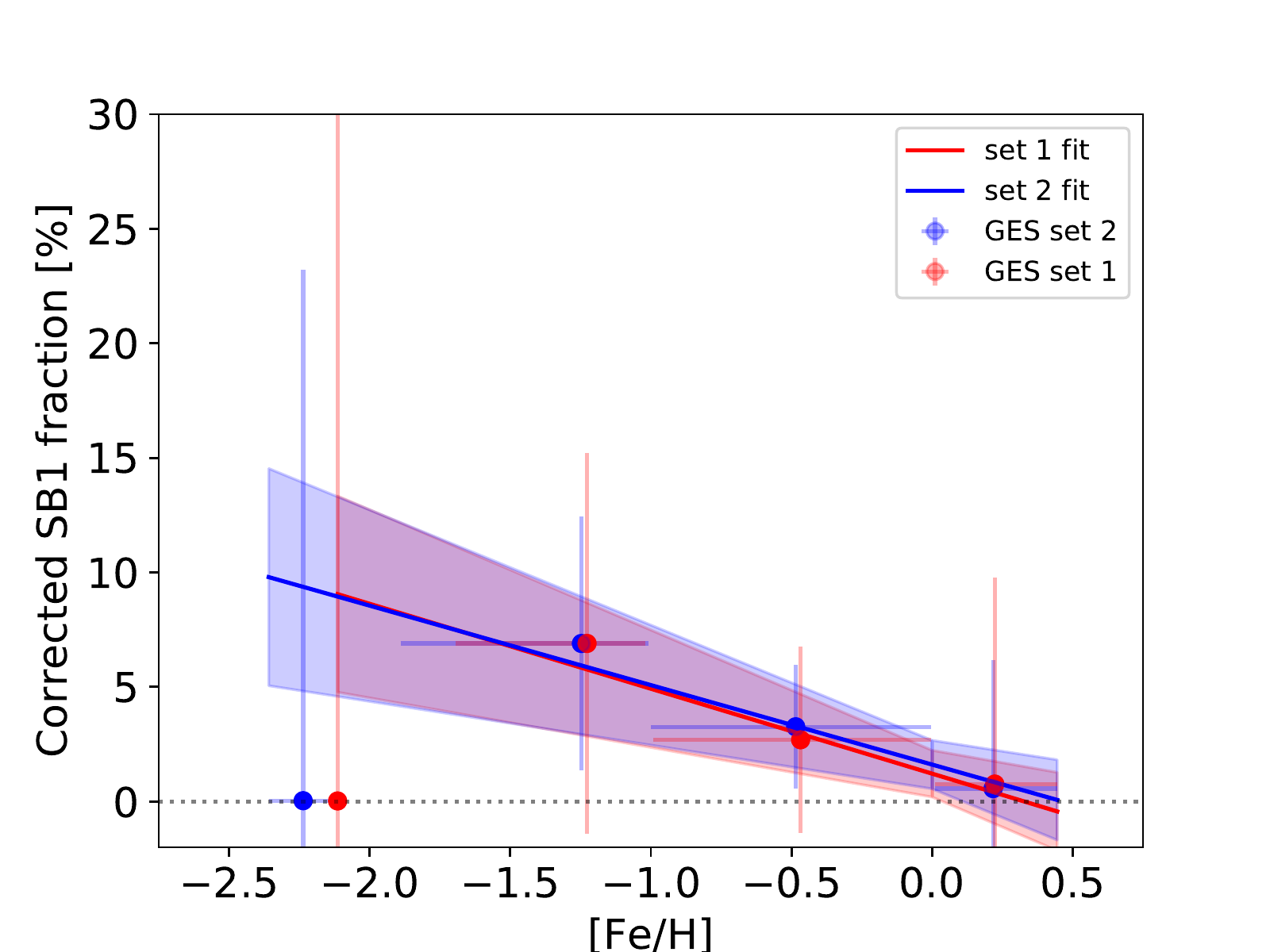}
    \caption{Sensitivity of the global GES SB1 fraction when the PDMF weighting is applied to the fraction per spectral type, distinguishing dwarfs and giants for K type.}
    \label{fig:sb1_frac_z2_w_pdmf}
\end{figure}

\subsubsection{The PDMF-corrected SB1 fraction and its metallicity dependence}
\label{Sect:PDMF}
To obtain the PDMF-corrected detection efficiency, the detection efficiency per spectral type was weighted by the PDMF PDF presented in Fig.~2(b) of \citet{tapiador2017}. These authors use a hierarchical Bayesian modelling from the masses derived by the FLAME\footnote{Final Luminosity, Age and Mass Estimator} module of the \emph{Gaia} data processing \citep{andrae2018}. The PDMF weighting strongly decreases with increasing mass following a multi-component power law  \citep{kroupa2002}. A mean effective temperature (\Teff) is associated to each spectral type, and the corresponding main sequence mass is then computed  from the mass -- \Teff\ relation of \citet{moya2018}:
\begin{equation}
M = (-0.964\pm0.004) + (3.475\pm0.006)\times10^{-4}\; T_\mathrm{eff}
\label{eq:m_teff}
\end{equation}
where $M$ is the stellar mass expressed in M$_{\odot}$. This empirical relation is  accurate to better than 10\% and is based on 125 stars among which 41\% are eclipsing binaries, whereas 57\% and 2\% of the masses are derived from   asteroseismic and interferometric observations, respectively. \citet{moya2018} investigated numerous other functional relations (\emph{e.g.} stellar mass as a function of \Teff\ and surface gravity) and, among those, some have better regression coefficients than the one we selected, which is nevertheless largely sufficient for our purpose. As a check, Eq.~\ref{eq:m_teff} predicts $M=1.04$~M$_\odot$ for the Sun (\Teff$=5780$~K) and appears to be valid in the range [4780 -- 11000]~K. 

\begin{table}[]
    \centering
        \caption{$F$-test when PDMF weighting is taken into account (see Eqs.~\ref{eq:f} and \ref{eq:chi2mod}) comparing the linear fits (null and non-null slope) of SB1 frequency with metallicity.}
    \begin{tabular}{cccccc}
    \hline \hline
               & $N$ &  $\chi^2_\mathrm{cst}$ & $\chi^2_\mathrm{lin}$ & $F$-value & Confidence level \\
         \hline\\
         set 1 & 4 & 0.073 & 0.019 & 5.77 & $>90$\%\\
         set 2 & 4 & 0.169 & 0.050 & 4.75 & $>88$\%\\
         \hline
    \end{tabular}

    \label{tab:ftest2}
\end{table}

With these estimated masses per spectral type, we weight the detection efficiencies per spectral type $\eta$ listed  in Table~\ref{tab:m_teff_de} with the  PDMF PDF from Figure 2(b) of \citet{tapiador2017}. However, for K giants we have to proceed differently because they are evolved objects. The progenitors of K giants are supposed to be F main sequence stars of approximately 1.3~M$_\odot$. To estimate the PDMF weighting for K giants, we therefore consider that they have similar birth rates than those of FV stars. The PDMF weighting is obtained by integrating the FV birth rate over the lifetime of that of a KIII giant. Assuming a time-independent stellar formation rate, we estimate the PDMF PDF (noted $w$) of K giants as to the PDMF PDF of F dwarfs weighted by the lifetime ratio of a KIII over an FV star. Using the STAREVOL isochrones \citep{siess2006}, we have $w$(KIII) = $2\times10^{-2}$~$w$(FV), giving $w$(KIII)$~\approx 0.001$.

We finally obtain a PDMF-corrected SB1 detection efficiency  of $\eta_t = 22.2\pm1.6$\% (as compared to $20.0\pm4.6$\% when the PDF weighting is not considered).
We note that our detection efficiency corrections did not account for a possible metallicity dependence: the RV uncertainties are expected to be larger for metal-poor stars, which have weaker absorption lines leading to an under-detection of metal-poor binaries.

In summary, we present in Table~\ref{tab:sb1_freq} the total SB1 fraction as well as the giant and the dwarf SB1 fractions for sets~1, 2 and 3 before and after corrections for the total detection efficiency $\eta_t$ that includes the weighting by the PDMF. The uncorrected SB1 fractions and their associated Poissonian errors can be retrieved from the numbers provided in Table~\ref{tab:selection}. The PDMF-corrected fraction of SB1 with dwarf primaries $f_\mathrm{d}$ (defined as the corrected ratio of the number of SB1 dwarfs over the number of dwarfs) is in the range 7-8\% with a typical uncertainty of 4\%. Similarly, the PDMF-corrected fraction of SB1 with giant primaries $f_\mathrm{g}$ is in the range 7-19\% but with a larger typical uncertainty of 7\%. We note that we do not expect to find consistent binary frequencies between dwarfs and giants because they do not probe stellar samples of similar initial mass. Combining these two fractions ($f_\mathrm{d}$ and $f_\mathrm{g}$) leads to a global PDMF-corrected SB1 fraction $f_\mathrm{t}$ in the range 7-14\% with a typical uncertainty of 4\%.

We now investigate the SB1 fraction dependence with metallicity. If we do not weight by the PDMF, the SB1 fraction is given by:
\begin{equation}
    \frac{n_\text{b}}{n_\text{t}} = \frac{n_\text{b}^\text{F}+n_\text{b}^\text{G}+n_\text{b}^\text{K}}{n_\text{t}^\text{F}+n_\text{t}^\text{G}+n_\text{t}^\text{K}} 
\end{equation}
where $n_\text{t}^\text{F}$, $n_\text{t}^\text{G}$, and $n_\text{t}^\text{K}$ are the numbers of F, G, and K-type stars, and $n_\text{b}^\text{F}$, $n_\text{b}^\text{G}$, and $n_\text{b}^\text{K}$ are the corresponding numbers or binaries.
The previous relation can also be expanded in the following way to account for the PDMF:
\begin{equation}
\frac{n_\text{b}}{n_\text{t}} = \frac{n_\text{b}^\text{F}}{n_\text{t}} + \frac{n_\text{b}^\text{G}}{n_\text{t}} + \frac{n_\text{b}^\text{K}}{n_\text{t}} = \frac{n_\text{t}^\text{F}}{n_\text{t}}\frac{n_\text{b}^\text{F}}{n_\text{t}^\text{F}} + \frac{n_\text{t}^\text{G}}{n_\text{t}}\frac{n_\text{b}^\text{G}}{n_\text{t}^\text{G}} + \left( \frac{n_\text{t}^\text{dK}}{n_\text{t}}\frac{n_\text{b}^\text{dK}}{n_\text{t}^\text{dK}} + \frac{n_\text{t}^\text{gK}}{n_\text{t}}\frac{n_\text{b}^\text{gK}}{n_\text{t}^\text{gK}}\right)
\end{equation}
where dK and gK refer to K dwarfs and giants, respectively. In the right member of the second equality, we can identify the first ratio of each term as the PDMF per spectral type $w$ (as given in Table~\ref{tab:m_teff_de}):
\begin{equation}
\label{eq:sb1_pdmf}
    \frac{n_\text{b}}{n_\text{t}} = w(\text{F})\frac{n_\text{b}^\text{F}}{n_\text{t}^\text{F}} + w(\text{G})\frac{n_\text{b}^\text{G}}{n_\text{t}^\text{G}} + \left(w(\text{dK})\frac{n_\text{b}^\text{dK}}{n_\text{t}^\text{dK}} + w(\text{gK}) \frac{n_\text{b}^\text{gK}}{n_\text{t}^\text{gK}}\right)
\end{equation}
with:
\begin{equation}
 w(\text{F}) + w(\text{G}) + w(\text{dK}) + w(\text{gK}) = 1
\end{equation}
 The SB1 fraction (weighted by the PDMF) per metallicity bin is illustrated in Fig.~\ref{fig:sb1_frac_z2_w_pdmf}. The most metal-poor bin is made of only two SB1 K giants with a very low PDMF $w$ factor, explaining the decrease of the SB1 fraction in this bin compared to the higher metallicity bins. The linear fit on set 2 is:
\begin{equation}
f(\mathrm{[Fe/H]}) = (-3.5\pm1.6)\%~\text{dex}^{-1} \mathrm{[Fe/H]} + (1.6\pm1.1) \% 
\label{Eq:trend-metallicity}
\end{equation}
with a confidence level higher than 88\% as shown in Table~\ref{tab:ftest2}. The effect of the PDMF weighting is to decrease the slope of the anticorrelation by a factor larger than two.

The trend presented by Eq.~\ref{Eq:trend-metallicity} cannot be compared to the literature because a PDMF-weighting accounting for the different spectral-type contributions is usually not performed. The present result can be important for stellar population syntheses and Galactic chemical evolution models that take into account the binary frequency as a function of metallicity. We note that it is not necessary to correct for the Malmquist bias (a bias leading to an over-representation of bright giants compared to dwarfs in a magnitude-limited sample) because this bias affects the numerators and denominators of Eq.~\ref{eq:sb1_pdmf} in a similar manner.

\begin{table} 
\centering
\caption{SB1 fraction of FGK-type stars before and after correction for the total detection efficiency $\eta_t$ which includes the weighting by the PDF PDMF. $f_\mathrm{d}$, $f_\mathrm{g}$  are the SB1 frequencies with dwarf primaries and giant primaries, respectively, and $f_\mathrm{t}$ is the total GES SB1 fraction.}
\begin{tabular}{crrr}
\hline\hline
SB1 fraction     &  set 1 & set 2 & set 3 \\
     \hline
Uncorrected &    & \\
$f_\mathrm{d}$ & $1.5\pm 4.6\%$ & $1.7\pm4.4\%$ & $1.8\pm4.2\%$\\
$f_\mathrm{g}$ & $1.5\pm 8.4\%$ & $2.1\pm7.1\%$ & $4.3\pm5.0\%$\\
$f_\mathrm{t}$ & $1.5\pm 4.1\%$ & $1.8\pm3.7\%$ & $3.1\pm3.2\%$\\
\hline
PDMF-corrected & & \\
$f_\mathrm{d}$ &  $6.7\pm4.6\%$ & $7.5\pm4.4\%$ & $8.3\pm4.2\%$\\
$f_\mathrm{g}$ & $6.9\pm8.4\%$ & $9.6\pm7.1\%$  & $19.3\pm5.0\%$\\
$f_\mathrm{t}$ &  $6.7\pm4.1\%$ & $8.0\pm3.7\%$ & $14.1\pm3.1\%$\\

    \hline
\end{tabular}
\label{tab:sb1_freq}
\end{table}

\subsection{Comparison with other surveys}
\label{sec:comp}
All cross matches were performed using the CDS X-match service\footnote{\url{http://cdsxmatch.u-strasbg.fr/\#tab=xmatch&}} with a large search cone radius of 10 arcsec. The following results are obtained when cross-matching with the GES iDR5:
\begin{itemize}
 \item 43 stars in common with SB9 \citep{pourbaix2004}, none of which identified as SB1 in the present study. Conversely, none of our SB1 candidates are in SB9.
\item 291 stars in common with RAVE DR5. \citet{matijevic2011} identified 1333 RV variables among 20\,000 southern stars (in vDR3\footnote{RAVE internal data release})  with magnitudes $I$ in the range [9--12]. None of our SB1 candidates are among these 1333 RV variables (Matijevi{\v c}, priv. comm.).
\item 292 stars in common with APOGEE \citep{badenes2018}. Since APOGEE (DR14) is a northern-hemisphere survey covering declinations above $\sim-32^\circ$, the overlap is necessarily small (the GES maximum declination is $\sim12^\circ$). None of our SB1 candidates are among their RV variables.
\end{itemize}
The null intersection of these cross-matching attempts allows us to conclude that our SB1 candidates are all new ones. Four were nevertheless already known to be photometric variables:
\begin{itemize}
    \item GES 06292710-311828  (V406 CMa), known as a $\gamma$~Dor variable (see Sect.~\ref{sec:photo});
    \item GES 12394853-365348 (SSS\_J123948.3-365349), known as an eclipsing binary of type EA (see Sect.~\ref{sec:photo});
    \item GES 11045767-1752043 and GES 15430381-4419571, classified as variables in the \emph{Gaia} DR2. These latter two are shown as red dots (dwarf stars) on the left panels of Figs.~\ref{fig:sigmav_sigmam} and \ref{fig:sigmav_sigmam_set2}. 
\end{itemize}

\section{Summary}
We investigated the GES iDR5 to detect and characterise spectroscopic binaries with one visible component (SB1) using single exposures from the GIRAFFE HR10 and HR21 setups. After a careful estimate of the RV uncertainties, we applied a $\chi^2$-test on a sample of approximately $43\,500$ stars that were observed at least twice and for which the $S\!/N$ of each spectrum is higher than three. In addition, we discarded about 100 stars with a RV amplitude larger than 177~\kms\ or with $(\Delta v/\Delta t)_{\max} \ge 160$~(\kms)/h, all of them including at least one outlying RV. The  $\chi^2$-test requires a correct evaluation of the RV uncertainties, which includes (i) inter-setup bias between HR10 and HR21, (ii) uncertainties associated with the Gaussian fit on the CCF peak, (iii) uncertainties due to template \Teff, rotational velocity and $S\!/N$, and (iv) uncertainties due to changes in the configuration of the GIRAFFE spectrograph with time. Overall,  the RV uncertainty may be estimated as $0.55\pm0.24$~\kms\ (right panel of Fig.~\ref{fig:drvs_errs}). 

The $\chi^2$-test identifies 641 or 803 SB1 candidates, depending on the adopted confidence level (respectively $5\sigma$ for set~1, or $3\sigma$  for set~2), and after removing RV variables associated with photometric variability and probably due to stellar pulsation, rotation, or spots (the latter cleaning was performed with the help of \emph{Gaia} DR2 photometry). Among SB1 candidates, 11\%  are located in stellar clusters (although it is beyond the scope of this paper to check for their cluster membership).  The raw binary frequency is just below 2\%, but when corrected for the detection efficiency, selection biases and PDMF weighting (Sects.~\ref{sec:efficiency} and \ref{Sect:PDMF}), the global GES SB1 fraction is in the range 7-14\% depending on the confidence level of the $\chi^2$ test, with a typical uncertainty of 4\%. 

Separating giants from dwarfs using absolute magnitudes derived from \emph{Gaia} DR2 parallaxes and  photometry (Eq.~\ref{eq:lc}), we found that about a quarter of the SB1 candidates are giant stars, are mainly located in the Galactic plane and can be as far as 10~kpc away (Fig.~\ref{fig:mollweide_dist}). The maximum RV variations reach 40~\kms\ for SB1 giants whereas for dwarfs, this may even exceed 100~\kms\  (Fig.~\ref{fig:histo_drvmax}). 

Tentative orbits are provided for two short-period binaries (orbital periods of 4 and 6~d) involving dwarf primaries from set~1   (Table~\ref{tab:orb} and Fig.~\ref{fig:sb1_orbits}). The other SB1 candidates need follow-up RV observations  to confirm their binary nature, and to derive their orbit. Nevertheless, the orbital period  distribution is estimated from the RV standard deviations (right panel of Fig.~\ref{fig:histo_dtl}). The detected SB1s have estimated orbital periods lower than $\log{P[\text{d}]} \lessapprox 4$. SB1s with dwarf primaries extend towards shorter orbital periods than SB1s with giant primaries, whose short-period tail is indeed expected to be depleted by  binary interaction (\emph{e.g.} Roche lobe overflow). 

Furthermore, we analysed the dependence of  SB1 frequency with \Teff\ and metallicity. A small increase of the SB1 frequency is observed from K- towards F-type stars, in agreement with previous studies. With a confidence level higher than 93\%, we show that the SB1 frequency decreases  with increasing metallicity with a slope of $-9\pm3$\%~dex$^{-1}$ (Eqs.~\ref{eq:f1} and \ref{eq:f2}) in the metallicity range $-2.7\le\mathrm{[Fe/H]}\le+0.6$. If we include the PDMF weighting, the slope turns to $-4\pm2$\% with a confidence level higher than 88\%. We also provide the SB1 fraction drifts with metallicity per spectral-type (Fig.~\ref{fig:sb1_temp} and Table~\ref{tab:sb1_frac_fits}). 

Since the observational biases impacting the SB1 detection efficiency are mostly due to the sparsity of the RV coverage, which is independent of metallicity, we do not expect them to be the cause of the observed SB1 frequency--metallicity relation. This relation is expected to have important implications \citep{moe2018, bate2019} for the formation scenarios of binary stars. 

\begin{acknowledgement}
We thank the referee for useful suggestions that improve the quality and the readibility of the manuscript.

T.M., M.V.d.S. and S.V.E. are supported by a grant from the Fondation ULB. 

We thank G. Matijevi{\v c} for providing RAVE data for SB1 detection.

T.B. was funded by the project grant ’The New Milky Way’ from the Knut and Alice Wallenberg Foundation.

This work has made use of data from the European Space Agency (ESA) mission {\it Gaia} (\url{https://www.cosmos.esa.int/gaia}), processed by the {\it Gaia} Data Processing and Analysis Consortium (DPAC, \mbox{\url{https://www.cosmos.esa.int/web/gaia/dpac/consortium}}). Funding for the DPAC has been provided by national institutions, in particular the institutions participating in the {\it Gaia} Multilateral Agreement.

This work was partly supported by the European Union FP7 programme through ERC grant number 320360 and by the Leverhulme Trust through grant RPG-2012-541. We acknowledge the support from INAF and Ministero dell'Istruzione, dell'Universit\`a e della Ricerca (MIUR) in the form of the grant ``Premiale VLT 2012''. The results presented here benefitted from discussions held during the \emph{Gaia}-ESO workshops and conferences supported by the ESF (European Science Foundation) through the GREAT Research Network Programme.

This research has made use of the SIMBAD database, the VizieR catalogue access tool and the cross-match service provided and operated at CDS, Strasbourg, France.

\end{acknowledgement}

\bibliographystyle{aa}
\bibliography{biblio}

\end{document}